\documentclass[11pt,a4paper,english]{aastex631}

\usepackage{xcolor,colortbl} 
\definecolor{ocre}{RGB}{52,177,201} 

\usepackage{avant} 
\usepackage{mathptmx} 

\usepackage{fancyvrb}
\usepackage{fvextra}
\usepackage{gensymb}
\usepackage{microtype} 
\usepackage[utf8]{inputenc} 
\usepackage[T1]{fontenc} 

\usepackage{natbib}
\bibpunct{(}{)}{;}{a}{}{,} 
\input{bib.def}
\definecolor{MyRed}{rgb}{0.9,0.0,0.0} 
\definecolor{MyDarkRed}{rgb}{0.6,0.0,0.0} 
\definecolor{MyLightRed}{rgb}{1.0,0.8,0.8} 
\definecolor{MyLightPink}{rgb}{1.0,0.9,0.9} 
\definecolor{MyLightOrange}{rgb}{1.0,0.8,0.5} 
\definecolor{MyPink}{rgb}{1.0,0.08,0.45} 
\definecolor{MyDarkBlue}{rgb}{0,0.08,0.45} 
\definecolor{MyDarkGreen}{rgb}{0,0.5,0.0} 
\definecolor{MyLightBlue}{rgb}{0.97,0.97,1.0} 
\definecolor{MyLightGreen}{rgb}{0.8,1.0,0.8} 
\definecolor{MyWarmYellow}{rgb}{1.0,0.95,0.85} 
\definecolor{MyMediumBlue}{rgb}{0.7,0.72,1.0} 

\newcommand{\emphbf}[1]{{\textbf{#1}}}
\newcommand{\param}[1]{\texttt{#1}}
\newcommand{\code}[1]{\textsc{#1}}
\newcommand{\tool}[1]{\textsc{#1}}
\newcommand{\task}[1]{\textbf{\textit{#1}}}
\newcommand{\function}[1]{\textbf{\textit{#1}}}
\newcommand{\techdoc}[0]{\textit{``Technical documentation of the developed software packages''}}
\newcommand{\techdocshort}[0]{\textit{Tech.\,Doc.}}

\begin{document}
\vspace{-2cm}
 \title{ALMA Memo 628\\[4mm]
 \large 
 High-cadence observations of the Sun} 

\author[0000-0002-5006-7540]{Sven Wedemeyer}
\author{Mikołaj Szydlarski}
\affiliation{Rosseland Centre for Solar Physics, University of Oslo, PO Box 1029, Blindern 0315 Oslo, Norway} 
\affiliation{Institute of Theoretical Astrophysics, University of Oslo, PO Box 1029, Blindern 0315 Oslo, Norway}

\author{M. Carmen Toribio}
\author{Tobia Carozzi}
\affiliation{Department of Space, Earth and Environment, Chalmers University of Technology, Onsala Space Observatory, 439 92 Onsala, Sweden} 

\author{Daniel Jakobsson}
\author{Juan Camilo Guevara G\'omez}
\affiliation{Rosseland Centre for Solar Physics, University of Oslo, PO Box 1029, Blindern 0315 Oslo, Norway} 
\affiliation{Institute of Theoretical Astrophysics, University of Oslo, PO Box 1029, Blindern 0315 Oslo, Norway}

\author{Henrik Eklund}
\affiliation{Rosseland Centre for Solar Physics, University of Oslo, PO Box 1029, Blindern 0315 Oslo, Norway} 
\affiliation{Institute of Theoretical Astrophysics, University of Oslo, PO Box 1029, Blindern 0315 Oslo, Norway}
\affiliation{European Space Agency (ESA), European Research and Technology Centre (ESTEC), Keplerlaan 1, 2201 AZ Noordwijk, The Netherlands}

\author{Vasco M. J. Henriques}
\affiliation{Rosseland Centre for Solar Physics, University of Oslo, PO Box 1029, Blindern 0315 Oslo, Norway} 
\affiliation{Institute of Theoretical Astrophysics, University of Oslo, PO Box 1029, Blindern 0315 Oslo, Norway}

\author{Shahin Jafarzadeh}
\affiliation{Rosseland Centre for Solar Physics, University of Oslo, PO Box 1029, Blindern 0315 Oslo, Norway} 
\affiliation{Institute of Theoretical Astrophysics, University of Oslo, PO Box 1029, Blindern 0315 Oslo, Norway}
\affiliation{
Niels Bohr International Academy, Niels Bohr Institute, Blegdamsvej 17, DK-2100 Copenhagen, Denmark}

\author{Jaime de la Cruz Rodriguez}
\affiliation{Institute for Solar Physics, Dept. of Astronomy, Stockholm University, Albanova University Center, 10691 Stockholm, Sweden} 
\begin{abstract}
The Atacama Large Millimeter/submillimeter Array (ALMA) offers new diagnostic capabilities for studying the Sun, providing complementary insights through high spatial and temporal resolution at millimeter wavelengths. ALMA acts as a linear thermometer for atmospheric gas, aiding in understanding the solar atmosphere's structure, dynamics, and energy balance.
Given the Sun's complex emission patterns and rapid evolution, high-cadence imaging is essential for solar observations. Snapshot imaging is required, though it limits available visibility data, making full exploitation of ALMA's capabilities non-trivial. 
Challenges in processing solar ALMA data highlight the need for revising and enhancing the solar observing mode. The ALMA development study High-Cadence Imaging of the Sun demonstrated the potential benefits of high cadence observations through a forward modelling approach. The resulting report provides initial recommendations for improved post-processing solar ALMA data and explores increasing the observing cadence to sub-second intervals to improve image reliability.
\end{abstract}


\section{Introduction} 
\label{sec:intro}

The Atacama Large Millimeter/submillimeter Array (ALMA) offers diagnostic possibilities for the study of our Sun that have not been available before. 
At millimeter wavelengths, the continuum radiation of the Sun emerges from its chromosphere --- an atmospheric layer sandwiched between the photosphere and the corona. Despite its fundamental importance for the energy transport in the solar atmosphere, observations of the chromosphere are hampered by the small number of available diagnostics and the difficulties with interpreting them. 
The radiation continuum at millimeter wavelengths offers diagnostic possibilities for the study of our Sun that are complementary to commonly used  diagnostics but which only now can be exploited thanks to ALMA's high spatial and temporal resolution. 
In particular, ALMA's ability to serve as an essentially linear thermometer of the atmospheric gas at unprecedented spatial and temporal resolution in the millimeter wavelength range has great scientific potential. Consequently, solar ALMA observations will contribute significantly to answering long-standing questions about the structure, dynamics and energy balance of the outer layers of the solar atmosphere and thus promise high-impact results \citep[see][ and references therein]{2016SSRv..200....1W,2018Msngr.171...25B}. 

In contrast to many other astronomical sources, the Sun does fill and actually exceeds the primary beam of ALMA with a complex emission pattern that covers a large range of spatial scales \emph{and} evolves on extremely short time scales of only seconds and even below. Fully exploiting ALMA's possibilities therefore requires imaging at very high cadence, at least for the majority of solar science cases. This high cadence essentially requires snapshot imaging, which limits the visibility data available for imaging at any given time. 
Exploiting the time domain to its full extent within the instrumental limitations is therefore crucial for reliable imaging of solar ALMA data. 
The currently offered cadence of 1\,s is already remarkable but an even higher cadence is technically possible.

The resulting challenges with processing solar ALMA data during the past years have revealed the need for a thorough and systematic review and further development of the current solar observing mode and processing of the resulting data, which is quite different from the standard processing of other ALMA data. Please see \citet{2017SoPh..292...87S} and \citet{2017SoPh..292...88W} for a technical introduction to solar observing with ALMA. The experimental study presented here explores the potential of increasing the cadence of solar observing even further towards sub-second time resolution in order to increase the reliability of the resulting image time series. 

Similar strategies that exploit an ultra-high cadence have already been routinely applied for a long time for observations of the Sun at visible wavelengths. There, it is common to take rapid bursts of 50-100 images with high cadence and exposures of less than 0.1\,s. Each burst is then combined into one science-ready image by using MOMFBD \citep[Multi-Object Multi-Field Blind Deconvolution,][]{2005SoPh..228..191V} or Speckle techniques \citep{1970A&A.....6...85L,2008A&A...488..375W,2011A&A...533A..21P}. 
The advantage is that the atmosphere of the Earth above the telescope at the time of the observation does not vary notably during such a short exposure. This way the influence of the varying atmosphere above the telescope can (at least partially) be corrected for and leads usually to a substantial increase in image quality.
ALMA observations are affected in a similar way, most notably in the form of phase corruptions caused in the Earth's troposphere. Further exploiting ALMA's high cadence for a possible reduction and correction of such phase errors is therefore promising.  
It should be mentioned that ultra-high time resolution imaging (at 20\,ms) is already routinely performed at the Very Large Array (VLA), demonstrating that a similar capability would be worth introducing for ALMA, too.

This report aims at demonstrating the potential of exploiting a higher cadence of solar observing with ALMA for the future improvement and extension of the solar observing mode and more reliable imaging products. The first part of the study aimed to provide first recommendations for optimal post-processing of solar ALMA data. This report should be understood as a first demonstration of the potential of employed techniques for the future improvement and extension of the solar observing mode. 
Using the same methodology, the aim of the second part of the study is to investigate whether combining visibility data from ultra-high-cadence (i.e., sub-second) observations across sufficiently short time windows can increase the precision with which brightness temperature\footnote{Please note that only interferometric observations are simulated for this report. No Total Power (TP) offsets are considered. The term brightness temperature is in this study used synonymously to the true brightness temperature minus a TP offset.} maps can be reconstructed in the imaging process. Based on these experiments, first recommendations for potential future development steps are given. 
The memo combines the following documents that were deliverables of the ESO study: 
\textit{Recommendations for optimal post-processing of solar ALMA data} and  
\textit{The potential of a high-cadence imaging mode for ALMA observations of the Sun}. 
Please note that the 
\textit{``Technical documentation of the developed software packages''} \citep[hereafter referred to as \techdocshort,][]{techdoc}  is not included but can be found online\footnote{\url{https://www.eso.org/sci/facilities/alma/development-studies/oso1.html}}.

\section{Methodology} 
\label{sec:method}

\subsection{Outline} 

\begin{figure}[t!]
    \centering  \includegraphics[width=12.5cm]{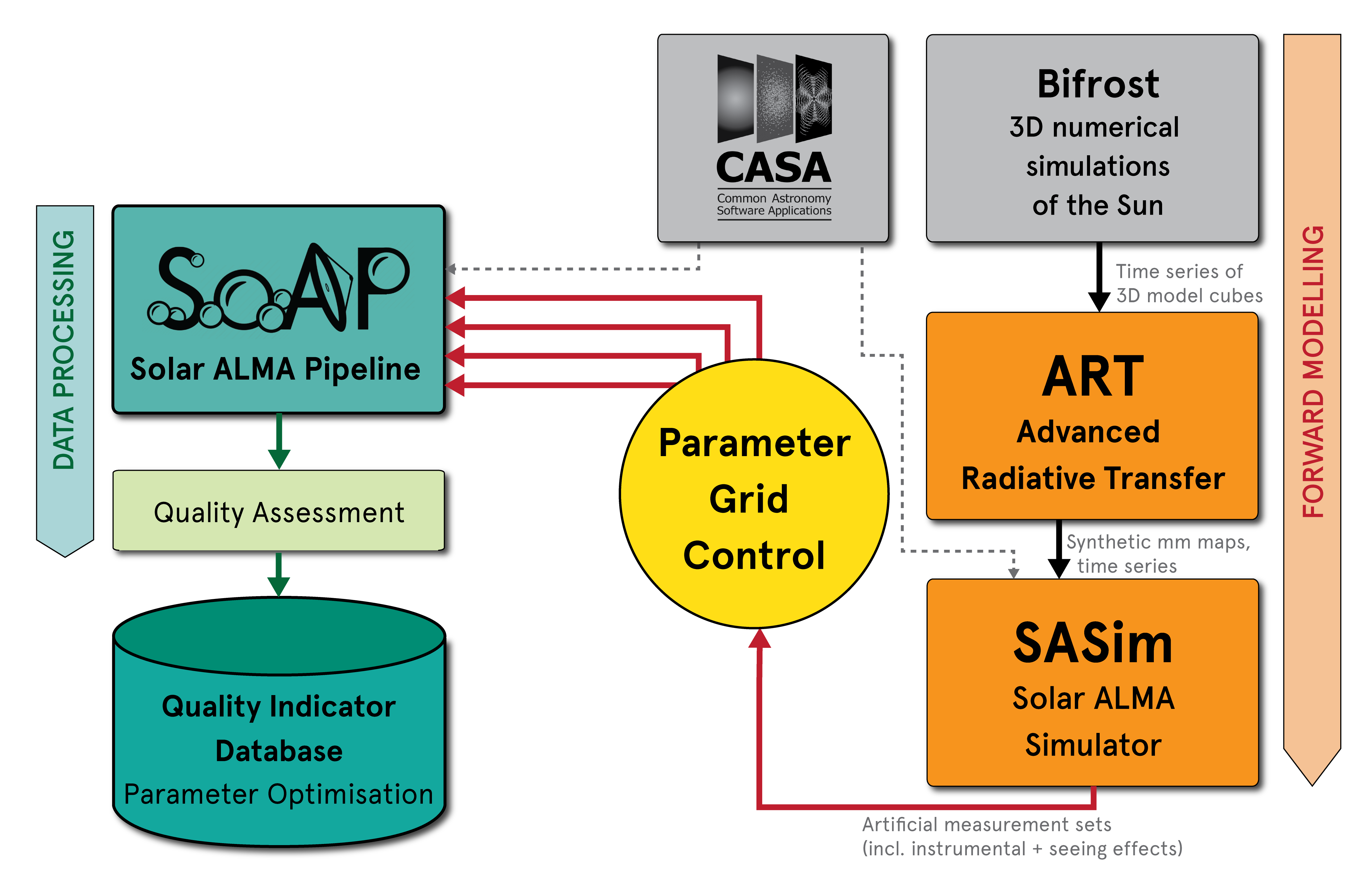}
    \caption{Simplified outline of the methodology. See Sect.~\ref{sec:method} for details.}
    \label{fig:method}
\end{figure}

The underlying principle is to use artificial observations based on a numerical model of the Sun, whose properties are known in detail, and to then apply post-processing methods with default and optimised parameters to these test data sets. Comparisons between the original and processed artificial ALMA observations then reveal which aspects of the input data can successfully be recovered (and which not) and how these procedures can be optimized accordingly. See Fig.~\ref{fig:method} for a flowchart illustrating the method. 
The following software packages and codes are used for this study: 

\begin{itemize}
    \item The stellar atmosphere simulation code \textbf{\tool{Bifrost}} \citep{2011A&A...531A.154G} is used for the production of time  sequences of 3D solar model atmospheres. 
    \item The \textbf{Advanced Radiative Transfer (\tool{ART}) code} \citep{2021zndo...4604825D} calculates synthetic brightness temperature maps that are used as input for \tool{SASim}.  
    \item  The \textbf{Solar ALMA Simulator (\tool{SASim})} is the main software package developed for this study. \tool{SASim} takes time series of synthetic brightness temperature maps as input and applies instrumental and weather (or ``seeing'') effects. The output is a simulated measurement set.  
    \item The \textbf{Solar ALMA Pipeline (\tool{SoAP})} is the main tool used for the  reconstruction of time series of brightness temperature maps. \tool{SoAP} accepts as input measurement sets  as downloaded from the ALMA Science Archive but also simulated measurement sets produced with \tool{SASim}.
\end{itemize}
\vspace{-1mm}\noindent
Please see also \citet{2022FrASS...9.7878W} for further information.

The forward modelling steps produce artificial ALMA observations, which serve as input for the post-processing steps. 
The input (sky) model is derived from radiative transfer calculations  for time series from a state-of-the-art 3D numerical simulation of the solar atmosphere (see Sects.~\ref{sec:solarsim}-\ref{sec:artcalc}). The produced brightness temperature maps for all spectral channels in the current setup of Band~3 and Band~6 for the solar observing mode represent an ideal artificial observation at the angular resolution of the computational grid of the 3D simulation. These ideal test data, which present a realistic time-dependent sky model with precisely known properties, are then input in the Solar ALMA Simulator (\code{SASim}, see Sect.~\ref{sec:sasim_desc}). This tool produces artificial visibilities in the form of a measurement set (MS) that closely resembles real observational MSs as produced by ALMA. \tool{SASim} facilitates studying the influence of different instrumental effects and set-ups and of the Earth’s atmosphere on the resulting measurement set. 

It is important to emphasize that the post-processing steps including the imaging are exactly the same that are applied to data from real ALMA observations (see below). 
A parameter grid was explored that includes the most important CLEAN parameters for relevant value ranges for different weather scenarios. 
Comparisons of the images output by \tool{SoAP} with the corresponding uncorrupted reference images reveal how well the imaging for a given choice of parameters performs. A set of quality indicators, which is calculated for each tested case, then allows for determining the optimal parameter combination and approach for the two receiver bands under different weather conditions. The quality indicators make use of brightness temperature differences, spatial power spectra, and temporal variations. Different ranges of spatial scales and variations as a function of radius in the maps are considered.  
In the following, a brief overview over the relevant components is given. A more detailed technical description is provided online in the technical documentation\footnote{\citet{techdoc}, \techdoc.  \url{https://www.eso.org/sci/facilities/alma/development-studies/oso1.html}}.
Please note that \code{CASA}~5.7 was used for the production of the artificial MSs and the imaging with SoAP.

\subsection{Construction of the solar test cases}

In this section, the construction of the solar test data are described. The data are derived from 3D simulations of the solar atmosphere (Sect.~\ref{sec:solarsim}) and radiative transfer calculations  (Sect.~\ref{sec:artcalc}) that result in a time-dependent \emph{sky model}. The latter is then used as input for the production of simulated measurement sets with the Solar ALMA Simulator (Sect.~\ref{sec:sasim_desc}). 

\subsubsection{Numerical simulations of the Sun} 
\label{sec:solarsim}

\begin{figure}[t!]
    \centering\includegraphics[width=\textwidth]{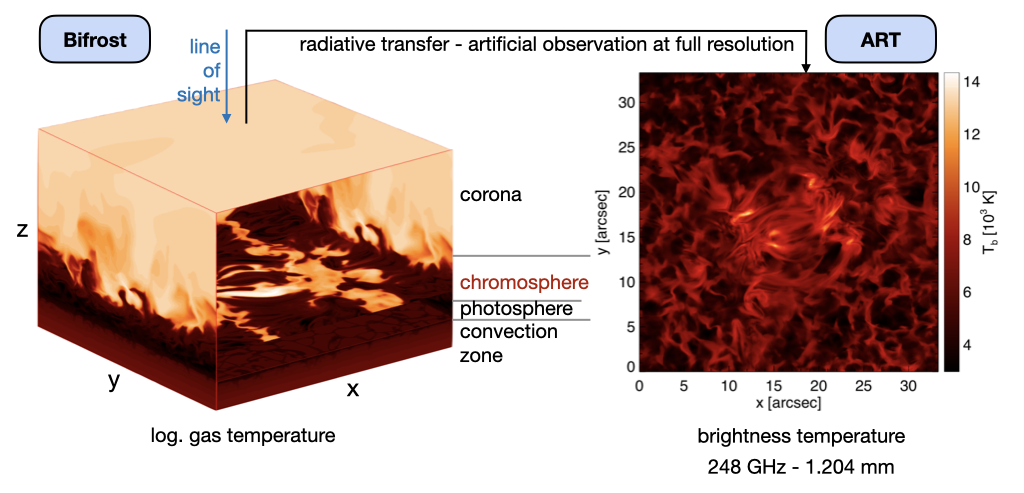}
    \vspace{-5mm}
    \caption{Numerical simulation of the Sun. Left: 3D snapshot of the RMHD simulation with \tool{Bifrost} showing the logarithmic gas temperature. The horizontal layers show the photosphere, chromosphere, and top of the simulation box. Right: The resulting brightness temperature as calculated with the Advanced Radiative Transfer (\tool{ART}) code for a frequency of 248 GHz (wavelength of 1.204 mm). Credits: Wedemeyer \& Szydlarski, UiO, 2023.}
    \label{fig:bifrostmodel}
\end{figure}

The  model of the Sun used in this report is produced with the state-of-the-art 3D radiation magnetohydrodynamics (RMHD) code \tool{Bifrost} \citep{2011A&A...531A.154G,2016A&A...585A...4C}. It should be emphasised that this model is state-of-the-art and thus as realistic as currently possible. 
The computational domain includes the top of the convection zone, the photosphere, the chromosphere and the lower parts of the corona, and thus all layers that are relevant for the formation of radiation at (sub)millimeter wavelengths. For this study, a 3D simulation is chosen that is representative for a large part of the Sun. It includes an enhanced network region with stronger magnetic fields and overarching coronal loops surrounded by Quiet Sun patches. The computational grid has a horizontal extent of 24.0\,Mm\,$\times$\,24.0\,Mm with a grid resolution of 47.6\,km, which at the distance Sun-Earth corresponds to 33.1”\,$\times$\,33.1” and 0.066”, respectively. The simulation sequences used for this report have a duration of 1\,min at 1\,s and 0.1\,s cadence, respectively.  
See Fig.~\ref{fig:bifrostmodel} for an illustration. 

\subsubsection{Radiative Transfer calculations} 
\label{sec:artcalc}

The resulting time series of 3D model snapshots is used as input for the Advanced Radiative Transfer (\tool{ART}) code (de la Cruz Rodriguez et al., in prep.), which then simulates how the model Sun would be seen at full resolution from Earth at frequencies observed with ALMA (see Fig.~\ref{fig:bifrostmodel}). 
These maps thus correspond to perfect artificial observations with a single solid surface telescope with a diameter of $\sim$5\,km (Band~6) and $\sim$11\,km (Band~3), respectively, and are thus superior to what ALMA could achieve even under ideal conditions. \tool{ART} is here used to calculate time series for all spectral channels for both Band~3 and Band~6. The frequencies for which \tool{ART} maps are calculated are chosen to be identical to the actual spectral setup of ALMA’s Band~3 and Band~6 solar observing modes in order to produce a test case that is as close to actual observations as possible. Please see the next section for details on the resulting data and the \techdoc\ for details on \tool{ART}.

\subsubsection{Simulated ALMA observations at 1\,s cadence} 
\label{sec:sasim_desc}
\label{sec:simseq_1s}

The next step is to convert the \tool{ART} maps into input files for the Solar ALMA Simulator (\tool{SASim}, see the  \techdocshort). A detail that should be mentioned is that the horizontal extent of the 3D \tool{Bifrost} model and the corresponding \tool{ART} map are too small to fill up the full field of view of ALMA at the frequencies of Band~3 and Band~6. The maps are therefore periodically repeated in all directions so that the final map consists of tiles filled with the initial \tool{ART} map. This is possible due to the periodic lateral boundary conditions of the \tool{Bifrost} model. This procedure is carried out for all time steps and all spectral channels. The resulting Band~3 and Band~6 input data sets are described in Table~\ref{tab:artdata}. 
In addition, one frequency-averaged map is calculated for each time step for both Band~3 and Band~6, which includes all spectral windows with a total bandwidth of 8\,GHz for each band. The resulting 1\,min long sequences are later used for the construction of reference models. 

\begin{table}[b]
    \centering
    \vspace{5mm}
    \begin{tabular}{|l|c|c|}
\hline
&\textbf{Band 3} & \textbf{Band 6}\\
\hline
Spectral set-up & 4 spectral windows & 4 spectral windows \\
                &with 128 spectral channels each& with 128 spectral channels each\\
\hline
Central frequencies of & [ 93.0,  95.0, 105.0, 107.0]&[230.0, 232.0, 246.0, 248.0]\\[-1mm]
spectral windows [GHz]&&\\
\hline
Bandwidth of spectral & [2.0, 2.0, 2.0, 2.0]&[2.0, 2.0, 2.0, 2.0]\\[-1mm]
windows [GHz]&&\\
\hline
Spatial extent&
165.5” x 165.5”&
99.3” x 99.3”\\
\hline
Spatial resolution (x,y)&
0.066”, 0.066”&
0.066”, 0.066”\\
\hline
Repetition of \tool{ART} maps& 5 $\times$ 5 &3 $\times$ 3 \\
\hline
Sequence duration&1 min (60 steps)&1 min (60 steps)\\
\hline
Time resolution&1s&1s\\
\hline
    \end{tabular}
    \caption{Properties of the brightness temperature maps produced with \tool{ART}. }
    \label{tab:artdata}
\end{table}

\begin{figure}[t]
    \centering
    \vspace{-3mm}
    \hspace{-3mm}
    \includegraphics[width=15cm]{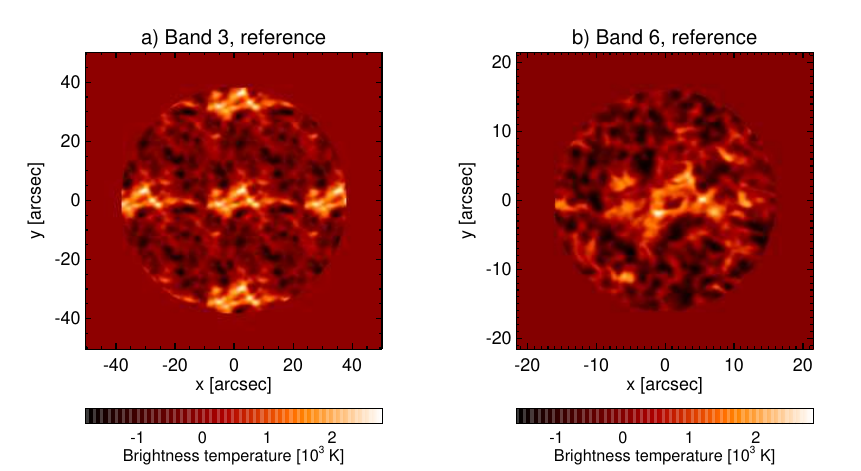}
    \caption{Reference images used for Band~3 (left) and Band~6 (right). 
    The first time step of each series is shown. 
    }
    \label{fig:tbmaps_ref}
\end{figure}
 
The sets of brightness temperature maps as summarised in Table~\ref{tab:artdata} are then used (separately) as input for \tool{SASim}. Time step by time step, each set of maps is used as an input sky model and then finally concatenated into one measurement set for Band~3 and one for Band~6. 
These uncorrupted MSs thus represent what an ideal ALMA could achieve under ideal weather conditions, i.e. in the absence of instrumental noise and in the  absence of Earth’s atmosphere. 
Please note that the date and time of the simulated observation was selected based on real ALMA observations of the Sun (see, e.g., the data on SALSA\footnote{\url{http://sdc.uio.no/salsa/}}). Consequently, the observations take place when the Sun is not close to the zenith, resulting in elongated synthetised beams as they are commonly seen for real ALMA observations of the Sun (see Table~\ref{tab:synbeam}).

In the next step, the impact of the Earth's atmosphere on the data is simulated by corrupting the phase information in the MS. 
For this purpose,  a time-dependent phase screen with fluctuating precipitable  water vapour (PWV) level generated with fractional Brownian Motion is used, which moves at a user-specified wind speed over the antenna array and thus decorrelates the interferometric signal. See the \techdoc{} for more details. 

It turned out that the phase corruption as implemented in \function{settrop} function in \tool{CASA} does not produce realistic results for daytime ALMA simulations at high cadence as required for solar observations. Consequently, the weather (i.e. phase corruption) scenarios are not selected by simply specifying the PWV level but rather through a detailed study of the phase corruption properties in real ALMA data~(see~Sect.~\ref{sec:obstestcase}). 

It should be mentioned that neither the brightness temperature maps for the spectral channels (see Sect.~\ref{sec:artcalc}) nor the reference model maps (see Sect.~\ref{sec:refmodel}) are  multiplied by the primary beam prior to being used as input for \code{SASim}. Application of the primary beam occurs as part of \task{simobserve()} when it is called during the execution of \code{SASim}. Please note that a mask (based on the primary beam response) is applied within \code{SoAP} and also to the reference model maps (see examples in~Fig.~\ref{fig:tbmaps_ref}).

\begin{table}[bp!]
    \centering
\vspace{5mm}
    \begin{tabular}{|c|c|c|c|c|c|c|}
\hline
robust&\multicolumn{3}{c|}{Band 3}&\multicolumn{3}{c|}{Band 6}\\
\cline{2-5}
&major&minor&angle&major&minor&angle\\
\hline
 -2.0  &     1.700  &     1.240  &     80.85 & 
             0.745  &     0.537  &     81.98\\
\hline
 -1.5  &     1.700  &     1.240  &     80.85 &
             0.745  &     0.537  &     81.98\\ 
\hline
 -1.0  &     1.700  &     1.241  &     80.85 & 
             0.746  &     0.537  &     81.98 \\
\hline
 -0.5  &     1.703  &     1.242  &     80.87 & 
             0.746  &     0.538  &     81.96 \\
\hline
0.0& & & & & & \\
\cline{1-1}
 0.5  &     1.865  &     1.334  &     80.99 & 
            0.805  &     0.571  &     81.63 \\
\hline
 1.0  &     2.147  &     1.465  &     81.11 & 
            0.905  &     0.617  &     81.41 \\ 
\hline
 1.5  &     2.267  &     1.516  &     81.30 & 
            0.947  &     0.635  &     81.39 \\ 
\hline
 2.0  &     2.282  &     1.523  &     81.30 & 
            0.956  &     0.637  &     81.42 \\
\hline
\end{tabular}
    \caption{Synthesised beam sizes as produced during the imaging process.}
    \label{tab:synbeam}
\end{table}

\subsubsection{Simulated ALMA observations at ultra-high cadence}
\label{sec:uncorruptms}
\label{sec:simseq_100ms}

Next to the 1s-cadence case described in Sect.~\ref{sec:simseq_1s}, a sequence with a duration of 60\,s at 0.1\,s cadence (i.e. 600~time steps) is used to simulate observations at ultra-high cadence. 
Each of these 600 3D~models is then used as input for the  Advanced Radiative Transfer (\tool{ART}) code that produces intensity maps at frequencies corresponding to ALMA's receiver bands 3 and 6 as offered for solar observations. These time series of maps are converted into flux and then used as input for the Solar ALMA Simulator (\tool{SASim}), which generates corresponding measurement sets. 
The result is a (uncorrupted) measurement set for a solar science target for each Band 3 and 6. In addition, corresponding measurement sets for a calibrator source are produced for both bands. As for the 1s-cadence sequence, the calibrator maps have a central pixel with a constant flux of 7\,Jy and all other pixels are set to zero. These maps are processed with \tool{SASim} in the same way as for the science target. Apart from the higher cadence, the same simulation setup as for the 1\,s is used, i.e. observations around local noon in array configuration~3 with baselines up to 500\,m. 
Please see Fig.~\ref{fig:Tbmaps_100ms_allcases}c-d for examples of images as produced with  \tool{SoAP} parameters  with standard imaging parameters (as described in Sect.~\ref{sec:imaging}) from  the uncorrupted MSs. 

\subsection{Reference models}
\label{sec:refmodel} 

\begin{figure}[t]
    \centering
    \includegraphics[width=14cm]{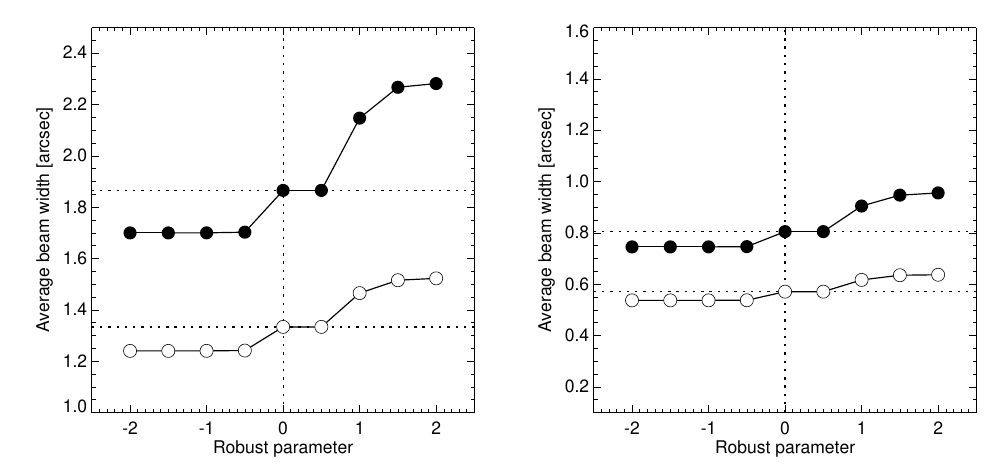}
    \vspace*{-4mm}
    \caption{Major (black symbols) and minor (white symbols) axes of synthesised beams as function of the robust parameter for Band~3 (left) and Band~6 (right).} 
    \label{fig:refbeamsizes}
\end{figure}

\subsubsection{Reference models for the 1s-cadence case}

The original \tool{ART} maps, which correspond to the \tool{SASim} input, are averaged over frequency before they are convolved with a  synthesised beam. This results in reference models with a time series of 60~maps at 1\,s cadence. 
The synthesised beam widths for the major and minor axes slightly increase with increasing value of the \param{robust} parameter as can be seen in Table~\ref{tab:synbeam} but otherwise vary negligibly during the considered time window. 
Accordingly, a reference model is derived for each receiver band and for each considered  value of the \param{robust} parameter (see Table~\ref{tab:synbeam}), i.e. 2\,$\times$\,9 in total. 
Please note that setting \param{robust} to zero activates the default value of 0.5 so that the cases of \param{robust} = 0.0 and 0.5 are identical. 
For this study, the case with the highest spatial resolution (i.e. \param{robust=-2}) is chosen for the reference models. This choice is motivated by the notion that high spatial and temporal resolution are of fundamental importance for most solar science cases and are thus the ideal outcome.  
After applying the synthesised beam, the reference models are interpolated to the exact same pixel grid that is produced in the imaging process with \tool{SoAP}.

Please note only interferometric observations without any additional Total Power (TP) observation are simulated in this study as the treatment of TP data and the combination with interferometric data is in itself an aspect that potentially needs further development in the future. For that reason, the mean brightness temperature value of the reference model is subtracted. This mean value, which would correspond to a TP offset, is determined across all time steps for the same pixels that are within the mask resulting in the imaging process with \tool{SoAP} (due to the selected primary beam response threshold). 
Examples for reference model maps are shown in Fig.~\ref{fig:tbmaps_ref}.

\subsubsection{Reference models for the ultra-high-cadence case}
\label{sec:reference}

As for the 1\,s cadence series, the original \tool{ART} maps at 0.1\,s cadence, i.e. the data used as input for constructing the uncorrupted measurement sets with  \tool{SASim}, are averaged over frequency and then convolved with a  synthesised beam. This results in 600 reference maps covering a duration of 60\,s at 0.1\,s s cadence. 
The same synthesised beams as for all \tool{SoAP} calculations used in this part of the study is chosen. The chosen beams for the parameter \verb|robust|=0.5  (see Sect.~\ref{sec:imaging}) have the following properties: 
\begin{center}
    \begin{tabular}{|c|c|c|c|}
\hline
 &Major axis [arcsec]&Minor axis [arcsec]&Angle [degree]\\
\hline
Band 3 & 1.865  &     1.334  &     80.99 \\
\hline
Band 6 & 0.805  &     0.571  &     81.63 \\
\hline 
\end{tabular}
\end{center}
Accordingly, a reference model is derived for each receiver band each time frame. 
The resulting reference maps for Band~3 and Band~6 for the first time step in the series are shown in Fig.~\ref{fig:Tbmaps_100ms_allcases}a-b. 

\subsection{Test case selection based on solar observations with ALMA}
\label{sec:obstestcase} 
\subsubsection{Phase variations in the observed ALMA data} 

\begin{figure}[t!]
    \centering
    \hspace*{-3mm}
    \includegraphics[width=15cm]{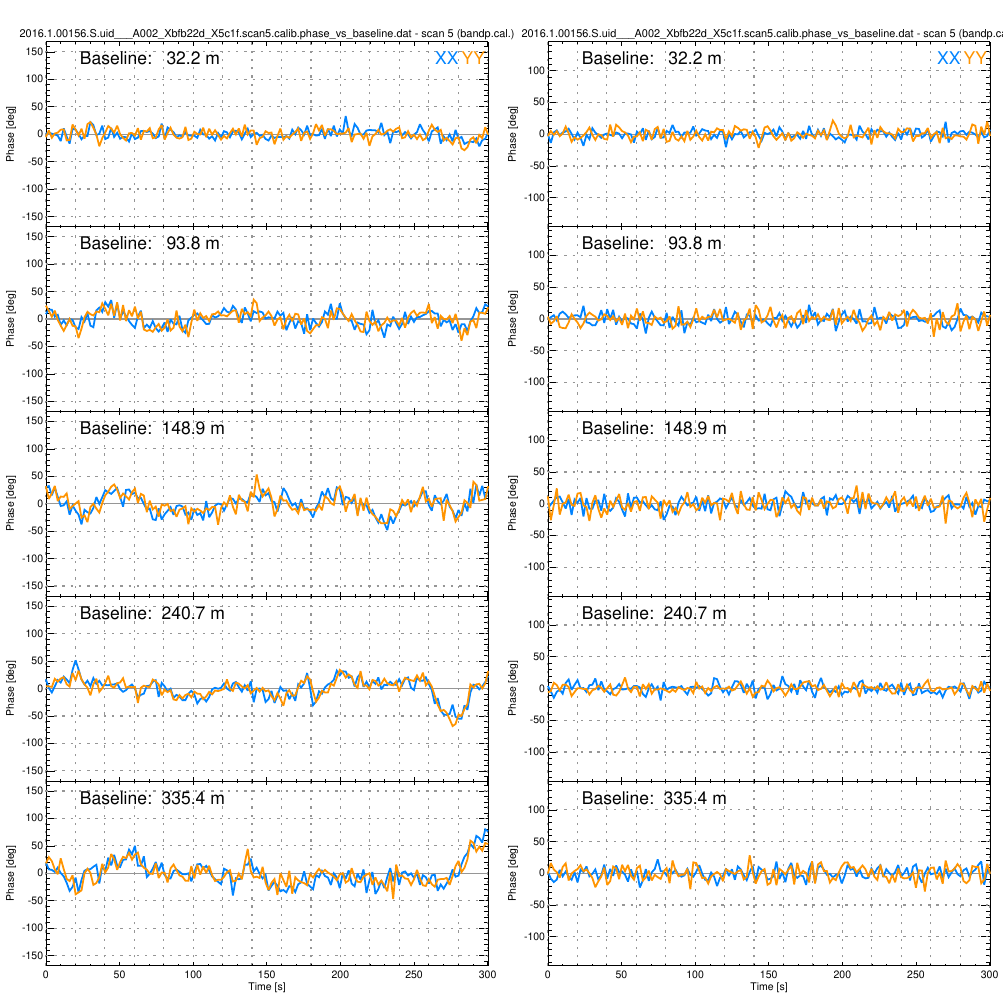}
    \caption{XX and YY phase as function of time for five selected baselines from short to long (top to bottom row) before (left column) and after calibration (right column). The data was taken from the bandpass calibration scan of a real measurement set as part of a solar Band ~6 observation with ALMA (project 2016.1.00156.S). Please note the smaller phase axis range for the post-calibration panels to the right. }
    \label{fig:obs_tseries}
\end{figure}
\begin{figure}[t!]
    \centering
    \includegraphics[width=15cm]{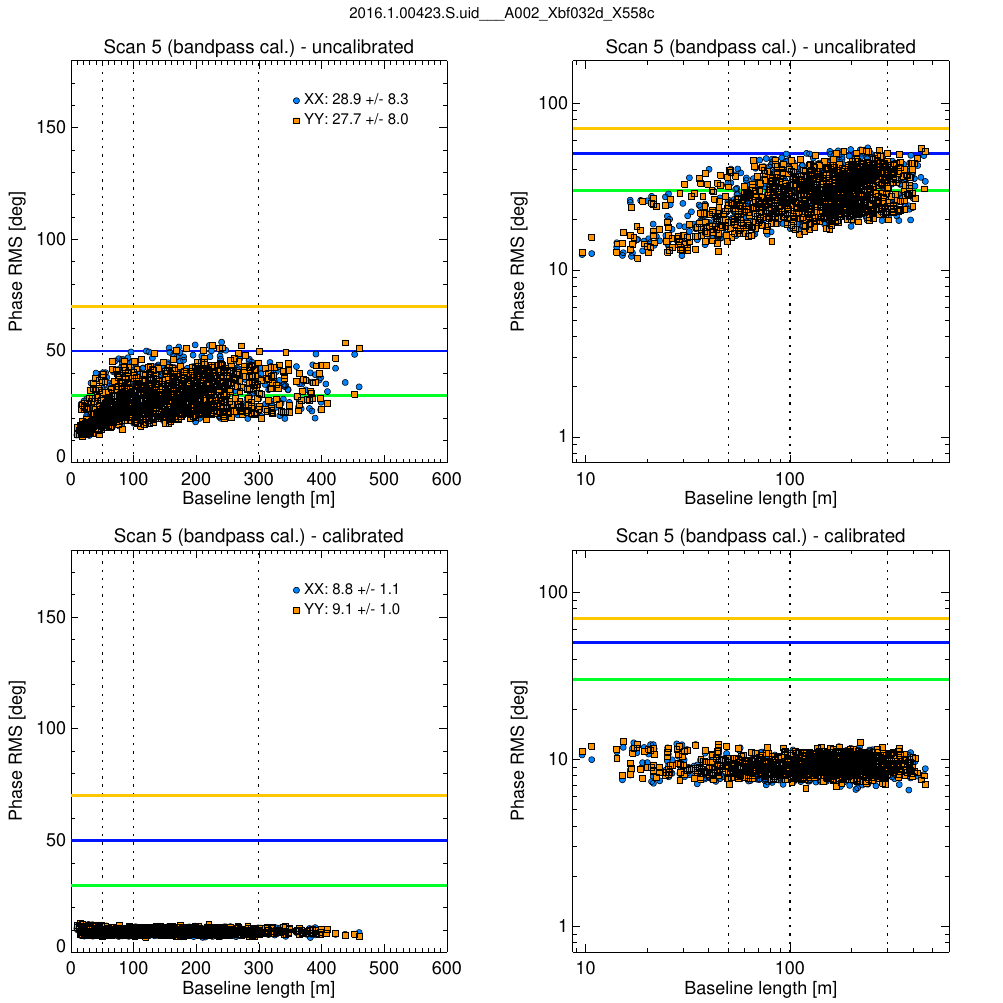}
    \caption{Spatial structure function (SSF) for a bandpass calibration scan from project 2016.1.00423 before (top row) and after calibration (bottom row). The left and right column show the data on a linear and a log-log scale, respectively. Both XX and YY phases are included with different colors. The horizontal lines mark phase RMS limits as used in ALMA's quality assessment. }
    \label{fig:obs_ssf}
\end{figure}

\begin{figure}[pt!]
    \centering
    \hspace*{-2mm}
    \includegraphics[width=16.5cm]{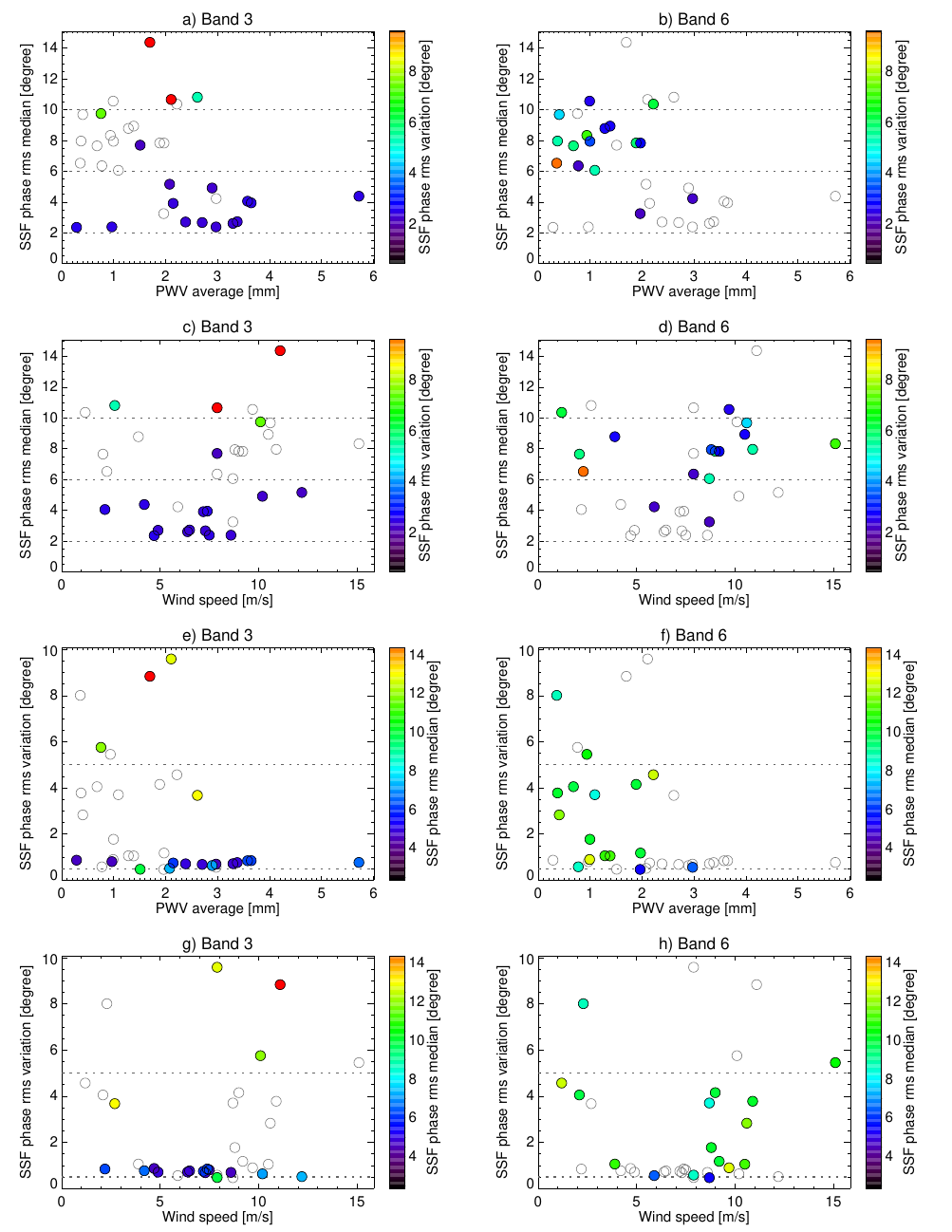}
    \caption{Phase variations in real solar observations with ALMA for Band~3 (left) and Band~6 (right) as shown in Fig.~\ref{fig:obsphases1}. The rows show different combinations of the median and the rms variation of the  post-calibration SSF phase variation values and the PWV and wind speed. 
    For reference, empty circles represent the data for the other receiver band.   
    }
    \label{fig:obsphases2}
\end{figure}

In order to ensure that the simulated measurement sets are sufficiently realistic, a large number of measurement sets from real ALMA observations are analysed. In this report, bandpass calibrator scans are used as they typically cover a longer time window and thus provide a statistically more robust view on the seeing conditions encountered at the ALMA site during daytime observations. For the following analysis, 19 Band~3 and 16 Band~6  bandpass calibrator scans are used. As an example, the  scan 
\verb|2016.1.00423.S.uid___A002_Xbf032d_X558c| 
is illustrated in Fig.~\ref{fig:obs_tseries}, which was obtained in Band~6 in array configuration C-3 with a maximum baseline of $\sim 500$\,m. The figure contains five selected baselines from short to long before and after calibration. 
The significant improvement of the data quality as a result of successful calibration is clearly seen from the reduction of the amplitude of the phase variation and also the removal of large differences between the XX and YY phases, which is most notable for the longest shown baseline (bottom row). 
Please note that the calibration was done with the script that is provided together with each MS from the ALMA Science Archive, thus following the ALMA's suggested calibration for solar MSs.

Degradation of the phase data and the (partial) correction during the calibration process can be visualised and quantified with the Spatial (Phase) Structure Function (SSF), which is here calculated for all considered bandpass calibrator scans. The SSF is given by the root-mean-square (rms) of the phase fluctuation as a function of baseline length  \citep[see the \techdoc\ and, e.g.,][]{alma-memo-529}. 
The SSF for the aforementioned bandpass calibration scan is shown in Fig.~\ref{fig:obs_ssf}. The SSF of the uncalibrated data (top row) exhibits and increase of the phase rms as function of baseline length from $\sim 10$\degree\ for the shortest (ACA) baselines to about $50\degree$ for the longest baselines. As clearly visible in the bottom row of Fig.~\ref{fig:obs_ssf}, the calibration step results in a removal of the slope and thus the dependence on baseline length and in an overall reduction of the phase rms values. In the selected example, the post-calibration values are around 10\degree\ and somewhat below as compared to the up to 50\degree\ before calibration. 
The resulting flat post-calibration SSF is in the following quantified by the median and the standard deviation  of the phase rms values.

\begin{figure}[t!]
    \centering
    \vspace{-7mm}
    \includegraphics[width=15cm]{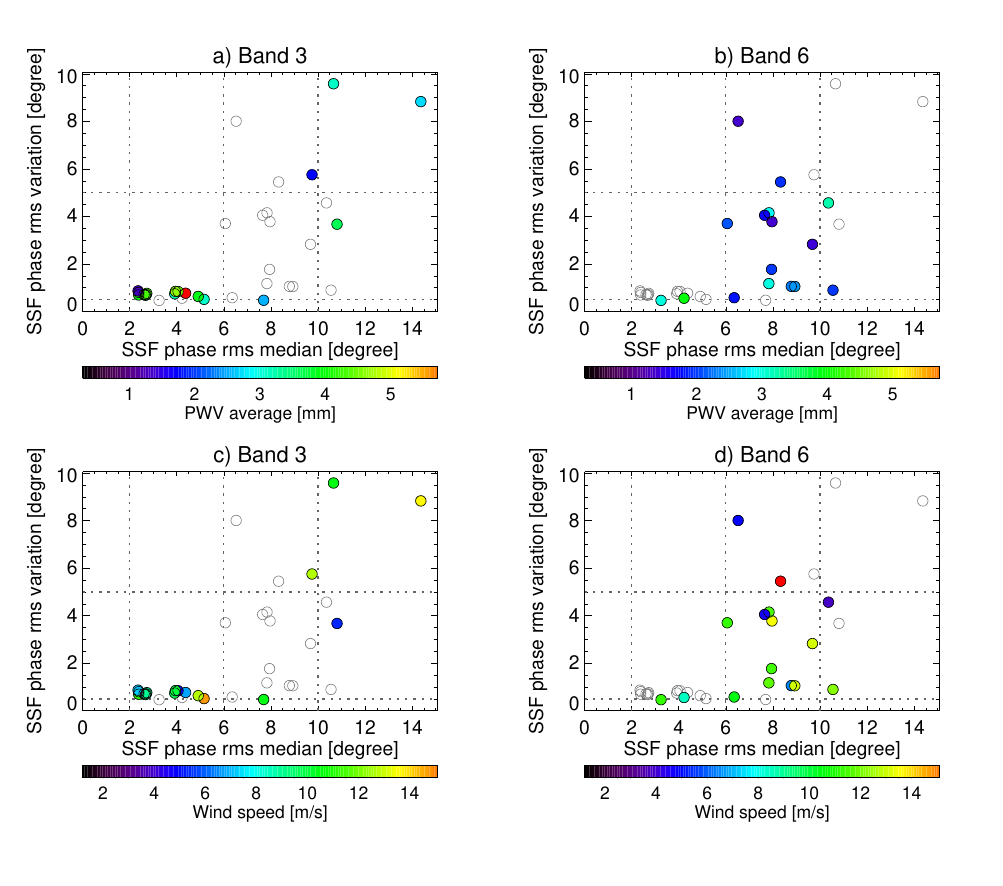}
    \vspace{-9mm}
    \caption{Phase variations in real solar observations with ALMA for Band~3 (left column) and Band~6 (right column). Both rows show the same data with the difference that the symbol color indicates the average PWV in the top row but the wind speed in the bottom row for each set. For reference, empty circles represent the data for the other receiver band.     
    The median and rms variation values are derived from the respective post-calibration SSFs for each considered measurement  set.}
    \label{fig:obsphases1}
\end{figure}

The resulting median and standard deviation values for all considered bandpass calibration scans are presented in Fig.~\ref{fig:obsphases2} compared to the average PWV level and wind speed  at the time of observation.  In addition, the median values are compared to the standard deviation values while at the same being color-coded for to the PWV levels (top row)  and wind speed  (bottom row) in Fig.~\ref{fig:obsphases2}. The plots do not show any clear dependence of the SSF median or variation on PWV level or wind speed. This is particularly important to note as the PWV value is commonly used as a quality indicator for observations and as such even provided in the ALMA Science Archive. Even more important for this study is the fact that PWV is a primary parameter used for the corruption of simulated measurement sets. It is therefore meaningless to create test cases solely based on PWV even for values that occur in actual observations. Instead, an alternative approach was pursued for this study, which establishes test cases directly based on the properties of the corresponding SSF. However, the analysis of real measurement sets suggests that the seeing conditions at the ALMA site are too complex as that they could be satisfactorily characterised by only one or a few parameters. This might be in particular apply to high cadence observing during daytime.  

The SSF parameters displayed in Fig.~\ref{fig:obsphases2} also show that the standard deviation of the post-calibration phase rms values remains consistently small for small to intermediate SSF median values but spreads over a much larger range and much larger values beyond a certain median value. This transition occurs roughly at SSF phase rms median value of 9-10\degree\ for Band~3 and at about 6\degree\ for Band~6, respectively. This finding implies that it is possible to define meaningful and representative test cases for SSFs with smaller median values but the SSFs can differ significantly if these critical median levels are surpassed. This behaviour is most likely connected to increasingly problematic cases with difficult and possibly variable seeing conditions and resulting problems for the calibration stage. At this point, we would like to emphasise that turbulence in Earth's atmosphere with its impact on seeing conditions is a complex phenomenon that affects astronomical observations in general but is certainly particularly challenging at daytime.

\subsubsection{Selected scenarios for the 1s-cadence case}

As the PWV level is not sufficient as a primary parameter for the selection of different seeing and thus phase corruption scenarios, the test cases are selected based on the properties of their SSFs instead. 
The median and the standard variation of the phase rms as shown in the SSF plots in Figs.~\ref{fig:obsphases1}-\ref{fig:obsphases2} are used to define in total seven scenarios from excellent to extreme conditions (see Table~\ref{tab:scenarios_1s}).

\noindent\begin{table}[t!]
\begin{tabular}{|ll|c|c|l|}
\hline
\multicolumn{2}{|l|}{Parameter}&Default& $n$&Values for the 1s-cadence case\\
\hline
\param{pwv} & Total precipitable & 3.0             & 14& 0.0, 0.25, 0.50, 0.75, 1.00, 1.50, 2.00,\\
     &water vapour in mm (PWV)&&& 2.50, 3.00, 4.00, 5.00, 6.00, 7.00, 8.00\\
\hline
\param{deltapwv} &RMS PWV fluctuations as a fraction  & 0.15            & 9&0.0, 0.02, 0.05, 0.10, 0.15, 0.20, 0.30,\\
     &of the PWV parameter (dPWV) &&&0.40, 0.50  \\
\hline
\param{beta} & Exponent of fractional brownian motion &1.1&1&1.1\\
\hline
\param{windspeed} & Wind speed for screen-type corruption (m/s) & 7.0    & 8& 0.0, 1.0, 2.0, 5.0, 7.00, 10.0, 15.0, 20.0\\
\hline
\param{noiselevel} &Additional phase noise as set &0.0 
     & 18 &0.0, 0.05, 0.10, 0.15, 0.20, 0.25, 0.30, 0.40, 0.50,\\
(simple)     &in the simplenoise mode of setnoise.&&& 0.60, 0.70, 0.80, 0.90, 1.00, 1.10, 1.20, 1.30, 1.40\\
\hline
\end{tabular}
\caption{Controllable parameters for the phase corruption with \tool{SASim}. Please note that this parameters are passed on to the \tool{CASA} functions \function{settrop} and \function{setnoise}. The values for the considered phase corruption parameter grid for the 1s-cadence case are listed in the rightmost column. The units for PWV and dPWV are mm, while wind speed and simple noise are given in m/s and Jy, respectively. 
}
    \label{tab:settrop_values}
    \label{tab:settrop}
\end{table}

Reproducing artificial observations with matching SSF properties  makes it necessary to compute corrupted measurement sets for a sufficiently large parameter grid that covers the above defined scenarios commonly encountered during real observations. For this purpose, the \tool{SASim} parameters \param{PWV}, \param{dPWV}, \param{wspd} (wind speed), and \param{smpn} (simple noise) are varied, while the parameter \param{beta} is kept at its default value of 1.1. 
The values for the resulting 4-dimensional parameter grid are listed in Table~\ref{tab:settrop_values}.

Accordingly, phase corruption for all 18140~parameter combinations was applied to the initial uncorrupted calibrator measurement set for each receiver band with \tool{SASim}, and the median and standard deviation of the corresponding SSFs was calculated. In the next step, it was determined for each receiver band and each corruption scenario which parameter combination  produces a SSF with a median and a standard variation of the phase rms closely matching the prescribed values of the selected scenario. This process required the calculation of 
about 200~additional parameter combinations for each band in order to better resolve the parameter space  close to the target parameters for each scenario. The resulting parameter combinations for Band~3 and 6 for all scenarios are listed in Table~\ref{tab:scenarios_1s}. 

\begin{table}[t!]
\centering
\begin{tabular}{|l|l|l|l|l|l|l|l|l|}
\hline
\multicolumn{2}{|c}{}&\multicolumn{2}{|c}{Target}&\multicolumn{5}{|c|}{Matched SASim parameters}\\ 
\hline
\# & Description & SSF-M & SSF-V & PWVa & dPWV & beta & wspd & smpn \\ 
\hline
\multicolumn{9}{|c|}{Band 3}\\ 
\hline
0  & Uncorrupted   &  -    & -     & -    &-     & -    & -    & -    \\
\hline
1&      Excellent& 2.00& 0.20& 0.50& 0.30& 1.10&10.00& 0.40\\
\hline
2&      Very good& 4.00& 0.50& 1.00& 0.40& 1.10& 5.00& 0.70\\
\hline
3&           Good& 6.00& 0.80& 1.50& 0.45& 1.10& 7.00& 0.75\\
\hline
4&       Moderate& 8.00& 1.30& 2.20& 0.45& 1.10& 7.00& 0.80\\
\hline
5&   Challenging&12.00& 3.00& 4.00& 0.45& 1.10& 1.00& 0.90\\
\hline
6&    Problematic&16.00& 4.00& 5.50& 0.45& 1.10& 1.00& 1.20\\
\hline
7&        Extreme&30.00& 4.00& 8.00& 0.45& 1.10&20.00& 2.00\\
\hline
\multicolumn{9}{|c|}{Band 6}\\ 
\hline
0  & Uncorrupted   &  -    & -     & -    &-     & -    & -    & -    \\
\hline
1&      Excellent& 2.00& 0.20& 0.50& 0.02& 1.10& 7.00& 0.40\\
\hline
2&      Very good& 4.00& 0.50& 1.00& 0.03& 1.10& 7.00& 0.50\\
\hline
3&           Good& 6.00& 0.80& 1.50& 0.03& 1.10&10.00& 0.65\\
\hline
4&       Moderate& 8.00& 1.30& 2.00& 0.03& 1.10& 1.00& 1.40\\
\hline
5&   Challenging&12.00& 3.00& 4.00& 0.03& 1.10& 1.00& 1.40\\
\hline
6&    Problematic&16.00& 4.00& 5.50& 0.03& 1.10& 1.00& 1.00\\
\hline
7&        Extreme&30.00& 4.00& 8.00& 0.03& 1.10&20.00& 2.00\\
\hline
\end{tabular}

\caption{Selected scenarios for the Band~3 and Band~6 1s-cadence cases as defined by the median and variation of the Spatial Structure Function (SSF) based on measurement sets obtained with ALMA. Each scenario is matched with a simulated measurement set that produces a SSF with similar  median and variation values. The corresponding SASim corruption parameters are listed for each scenario.  } 
    \label{tab:scenarios_1s}
\end{table}

\subsubsection{Selected scenarios for the ultra-high-cadence case}
\label{sec:weatherscenarios}

As for the 1s-cadence case, the uncorrupted measurement sets for the 100ms-cadence case (see Sect.~\ref{sec:uncorruptms}) is further processed with \tool{SASim} in order to simulate the impact of Earth's atmosphere (and instrumental effects) on the measured visibilities. The resulting phase corruption is controlled with the parameters shown in Table~\ref{tab:settrop}. 
Please note that the additional parameter \verb|simint = 0.1| has to be set for the processing of the time series used here\footnote{D.~Petry (ESO) kindly provided an updated version of CASA that allows to set shorter time steps for settrop() via the new additional parameter. Please see the \techdoc\ for details.}.  As mentioned above, the parameter \verb|pwv| for the perceptible water vapour (PWV) level alone does not produce measurement sets that are representative of real daytime solar observing conditions, even when matching the PWV that is reported for such observations. 
Instead realistic \tool{SASim} parameters are determined by matching the median and variation of the Spatial Structure Function (SSF) of the resulting simulated calibrator measurement set (as part of large parameter grids) to the SSFs of selected observational ALMA calibration measurement sets.  
While these scenarios were developed based on 1\,s cadence data, both for the simulated and most of the real measurement sets, they are applicable to the data at 0.1\,s cadence, too.  

\begin{figure}[t!]
    \centering
    \includegraphics[width=15cm]{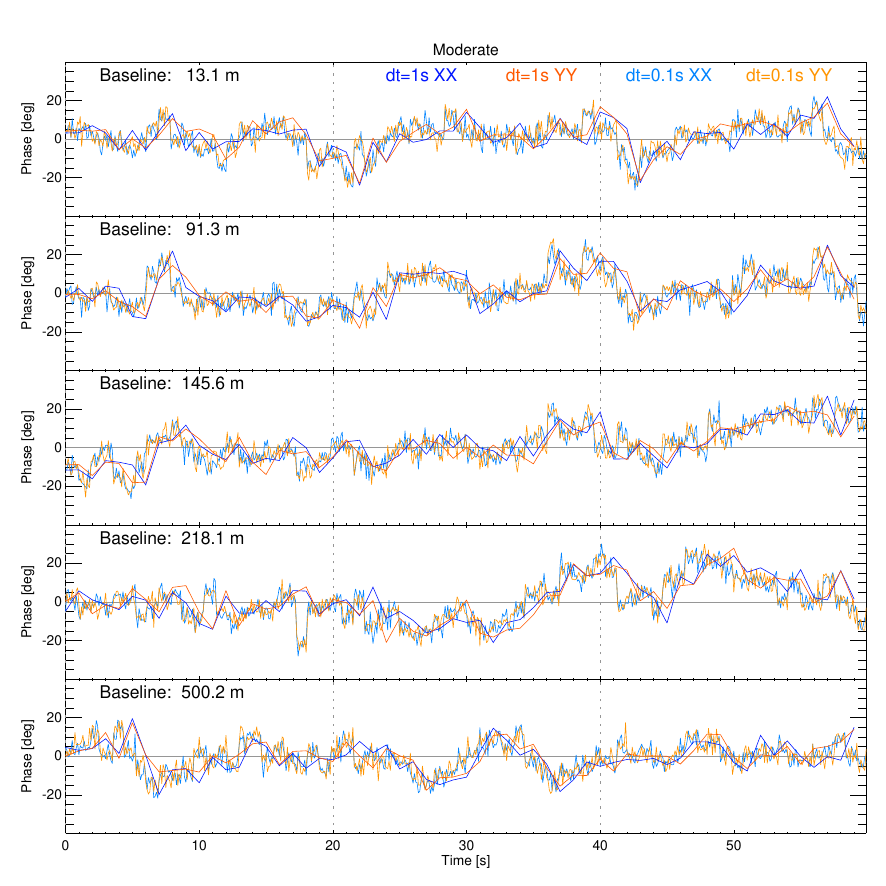}
    \vspace*{-3mm}
    \caption{Phase as function of time for the Band 3 calibrator measurement set for the scenario moderate. The XX and YY phases for the corrupted measurement sets at a cadence of dt=0.1s     (light blue and light orange) are compared to the corresponding simulations at 1.0\,s cadence 
    (dark blue and dark orange lines). 
    A color legend is provided in  the topmost panel.}
    \label{fig:phasevstime_calb3_moderate}
\end{figure}

\begin{figure}[t!]
    \centering
    \includegraphics[width=15cm]{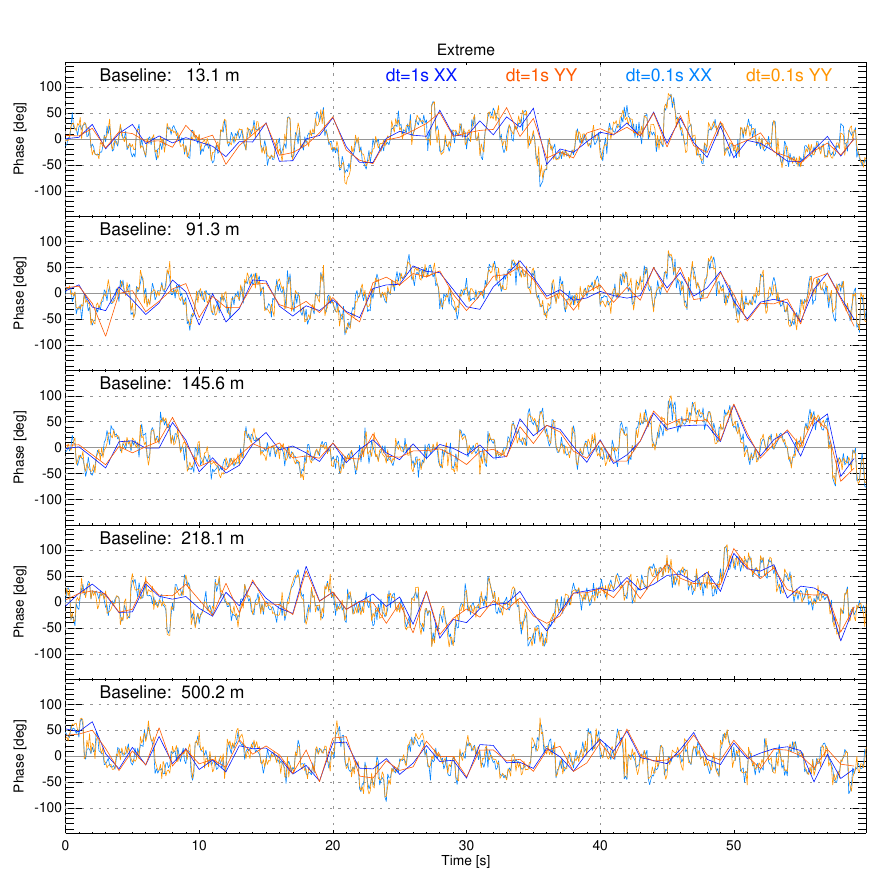}
    \vspace*{-3mm}
    \caption{Phase as function of time for the Band 3 calibrator measurement set for the scenario extreme. The XX and YY phases for the corrupted measurement sets at a cadence of dt=0.1s     (light blue and light orange) are compared to the corresponding simulations at 1.0\,s cadence 
    (dark blue and dark orange lines). 
    A color legend is provided in  the topmost panel.} \label{fig:phasevstime_calb3_extreme}
\end{figure}

\begin{figure}[t!]
    \centering
    \includegraphics[width=15cm]{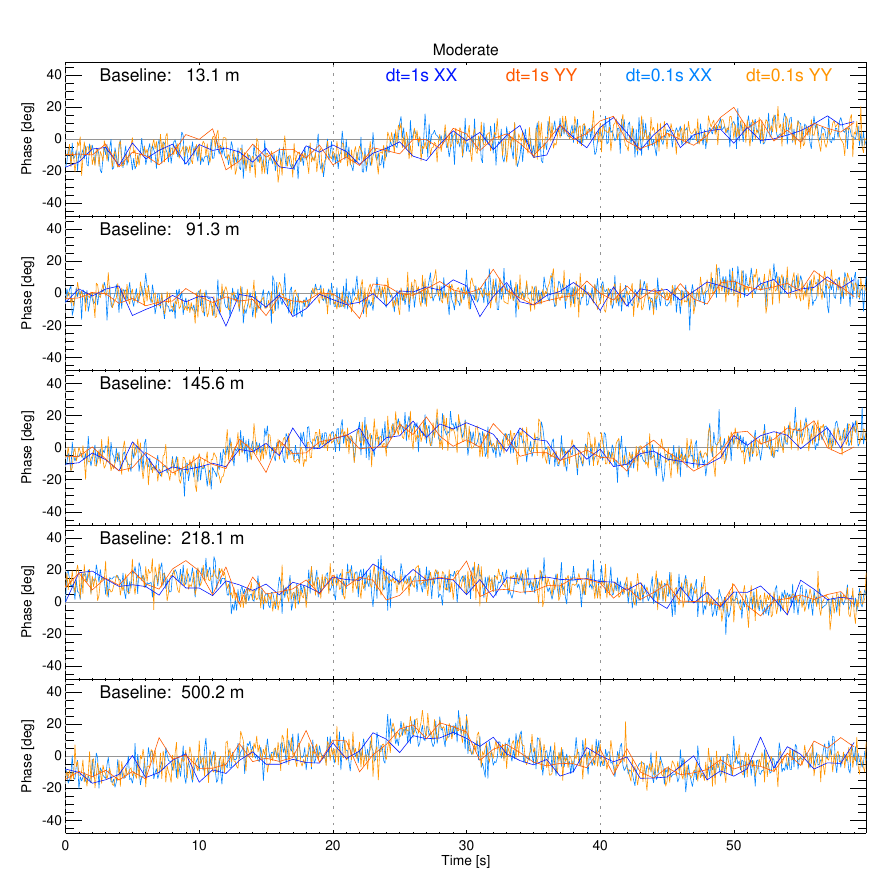}
    \vspace*{-3mm}
    \caption{Phase as function of time for the Band 6 calibrator measurement set for the scenario moderate. The XX and YY phases for the corrupted measurement sets at a cadence of dt=0.1s     (light blue and light orange) are compared to the corresponding simulations at 1.0\,s cadence 
    (dark blue and dark orange lines).  
    A color legend is provided in  the topmost panel.}
    \label{fig:phasevstime_calb6_moderate}
\end{figure}

\begin{figure}[t!]
    \centering
    \includegraphics[width=15cm]{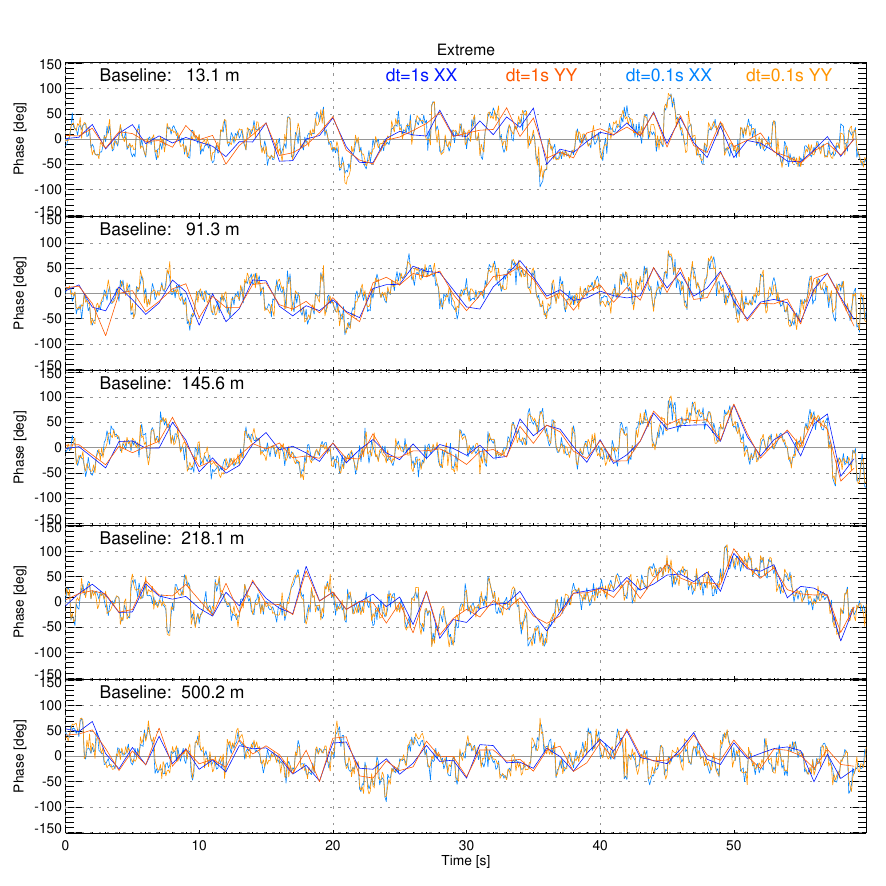}
    \vspace*{-3mm}
    \caption{Phase as function of time for the Band 6 calibrator measurement set for the scenario extreme. The XX and YY phases for the corrupted measurement sets at a cadence of dt=0.1s     (light blue and light orange) are compared to the corresponding simulations at 1.0\,s cadence 
    (dark blue and dark orange lines).  
    A color legend is provided in  the topmost panel.} \label{fig:phasevstime_calb6_extreme}
\end{figure}

\begin{figure}[t]
    \centering
    \includegraphics[width=15cm]{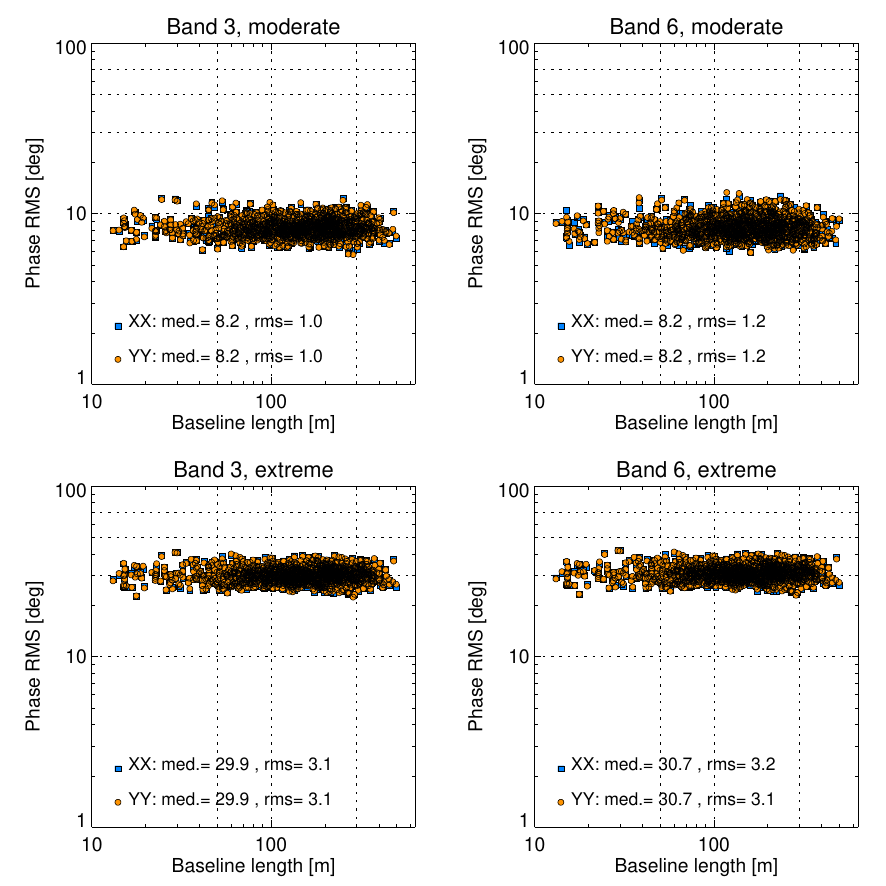}
    \caption{Spatial Structure Function (SSF) for Band~3 (left column) and Band~6 (right column) for the corrupted measurement sets for the scenarios moderate (top) and extreme (bottom).}    
    \label{fig:ssf_100ms}
\end{figure}

For exploring the potential of solar observing at higher cadence, the scenarios \textit{moderate} and \textit{extreme} are selected. The corresponding \tool{SASim} parameters are provided in Table~\ref{tab:scenarios_uh}. 
The resulting phases of the corrupted calibration measurement sets are illustrated in Figs.~\ref{fig:phasevstime_calb3_moderate}-\ref{fig:phasevstime_calb6_extreme} for selected baselines. For comparison, the corresponding results for the measurement sets at 1\,s cadence are plotted, too. Clearly, the phase variations are already significant for the moderate scenario and even much more pronounced for the extreme scenario, thus posing challenges for the imaging step as intended. This effect can also be seen in the SSF plots in Fig.~\ref{fig:ssf_100ms} and the resulting values for the median (SSF-M) and variation (SSF-V) of the SSFs. Please note that the XX and YY phases cover the same value ranges in the SSF plots so that mostly the YY data is visible.

\begin{table}[t!]
\centering
\begin{tabular}{|l|l|l|l|l|l|l|l|l|}
\hline
\multicolumn{2}{|c}{}&\multicolumn{5}{|c|}{SASim parameters}&\multicolumn{2}{c|}{Corrupted MS }\\ 
\hline
\# & Description & PWVa & dPWV & beta & wspd & smpn & SSF-M & SSF-V \\ 
\hline
\multicolumn{9}{|c|}{Band 3}\\ 
\hline
4&       Moderate&  2.20& 0.45& 1.10& 7.00& 0.80&8.17&1.02\\
\hline
7&        Extreme& 8.00& 0.45& 1.10&20.00& 2.00&29.91&3.06\\
\hline
\multicolumn{9}{|c|}{Band 6}\\ 
\hline
4&       Moderate& 2.00& 0.03& 1.10& 1.00& 1.40&8.18&1.18\\
\hline
7&        Extreme& 8.00& 0.03& 1.10&20.00& 2.00&30.70&3.15\\
\hline
\end{tabular}
\caption{Selected scenarios for Band~3 and Band~6 for the 100ms-cadence case as defined by the median and variation of the Spatial Structure Function (SSF) based on measurement sets obtained with ALMA. Each scenario is matched with a simulated measurement set that produces a SSF with similar  median and variation values. The corresponding SASim corruption parameters are listed for each scenario.  } 
    \label{tab:scenarios_uh}
\end{table}

\pagebreak
\subsection{Imaging with \tool{SoAP}}
\label{sec:imaging_soap}

\begin{figure}[pt!]
    \centering
    \vspace{-6mm}
    \hspace{10mm}
    \includegraphics[width=13.5cm]{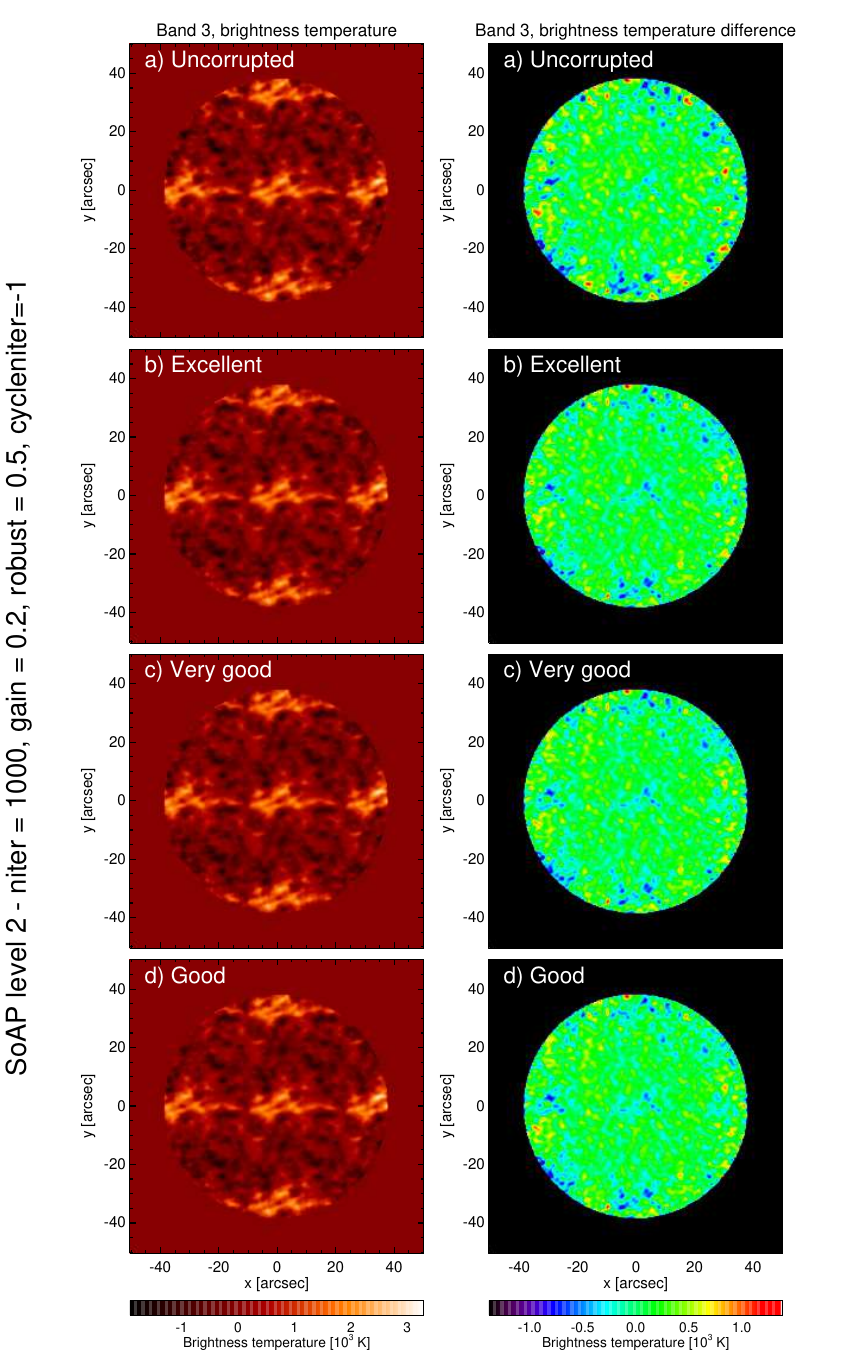}
    \vspace{-2mm}
    \caption{Imaging results for Band~3 for the uncorrupted measurement set at 1\,s cadence (top row) and the scenarios excellent, very good, and good (rows from top to bottom). 
    The brightness temperature maps for the first time step are shown in the left column and the corresponding difference to the reference model (see Fig.~\ref{fig:tbmaps_ref}a) in the right column. The same colour scale is used for all maps in their respective column.}
    \label{fig:tbmaps_b3_1}
\end{figure}

\begin{figure}[pt!]
    \centering
    \vspace{-6mm}
    \hspace{10mm}
    \includegraphics[width=13.5cm]{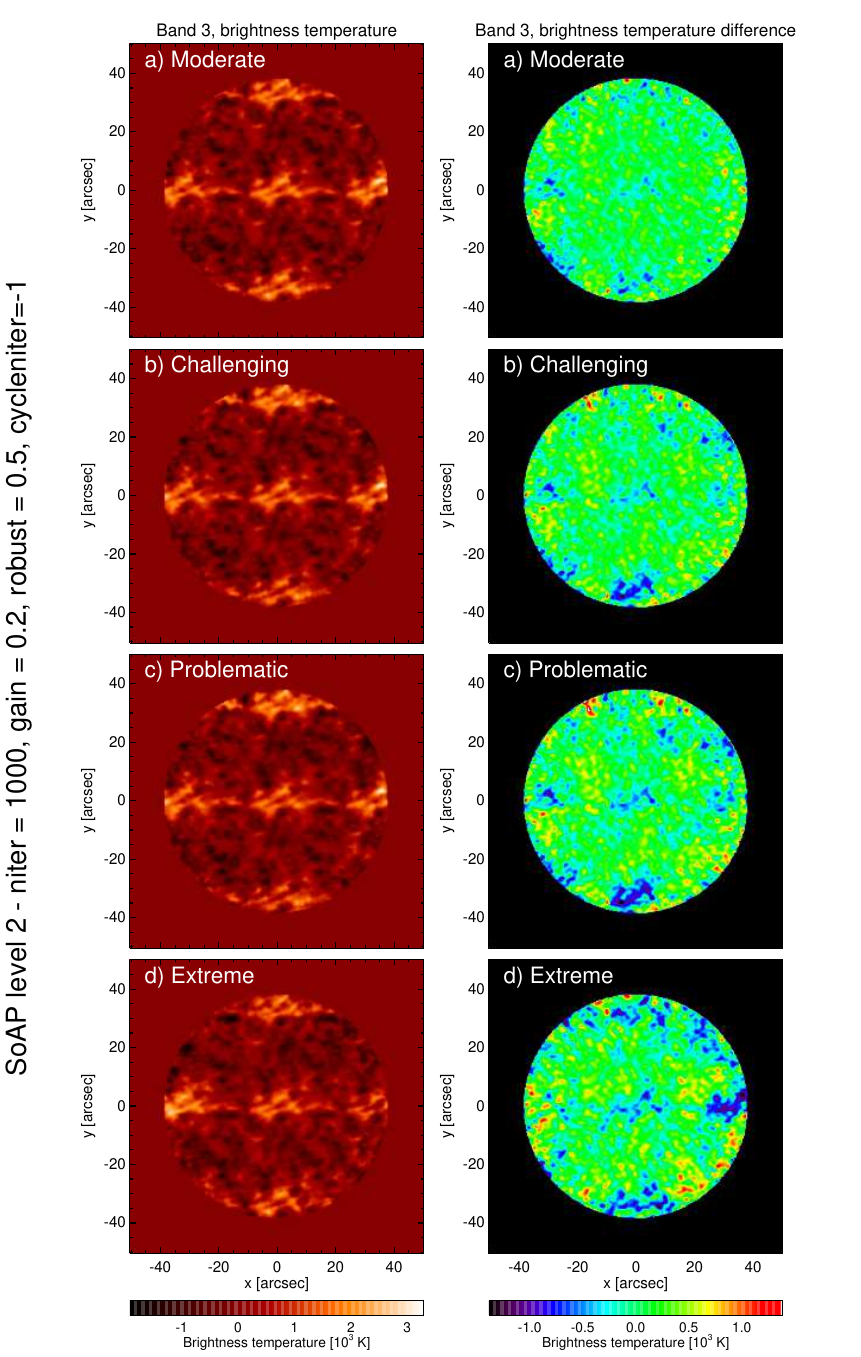}
    \vspace{-2mm}
    \caption{Imaging results for Band~3  at 1\,s cadence for the scenarios moderate, challenging, problematic and extreme (rows from top to bottom). 
    The brightness temperature maps for the first time step are shown in the left column and the corresponding difference to the reference model (see Fig.~\ref{fig:tbmaps_ref}a) in the right column. The same colour scale is used for all maps in their respective column.}
    \label{fig:tbmaps_b3_2}
\end{figure}

\begin{figure}[pt!]
    \centering
    \vspace{-6mm}
    \hspace{10mm}
    \includegraphics[width=13.5cm]{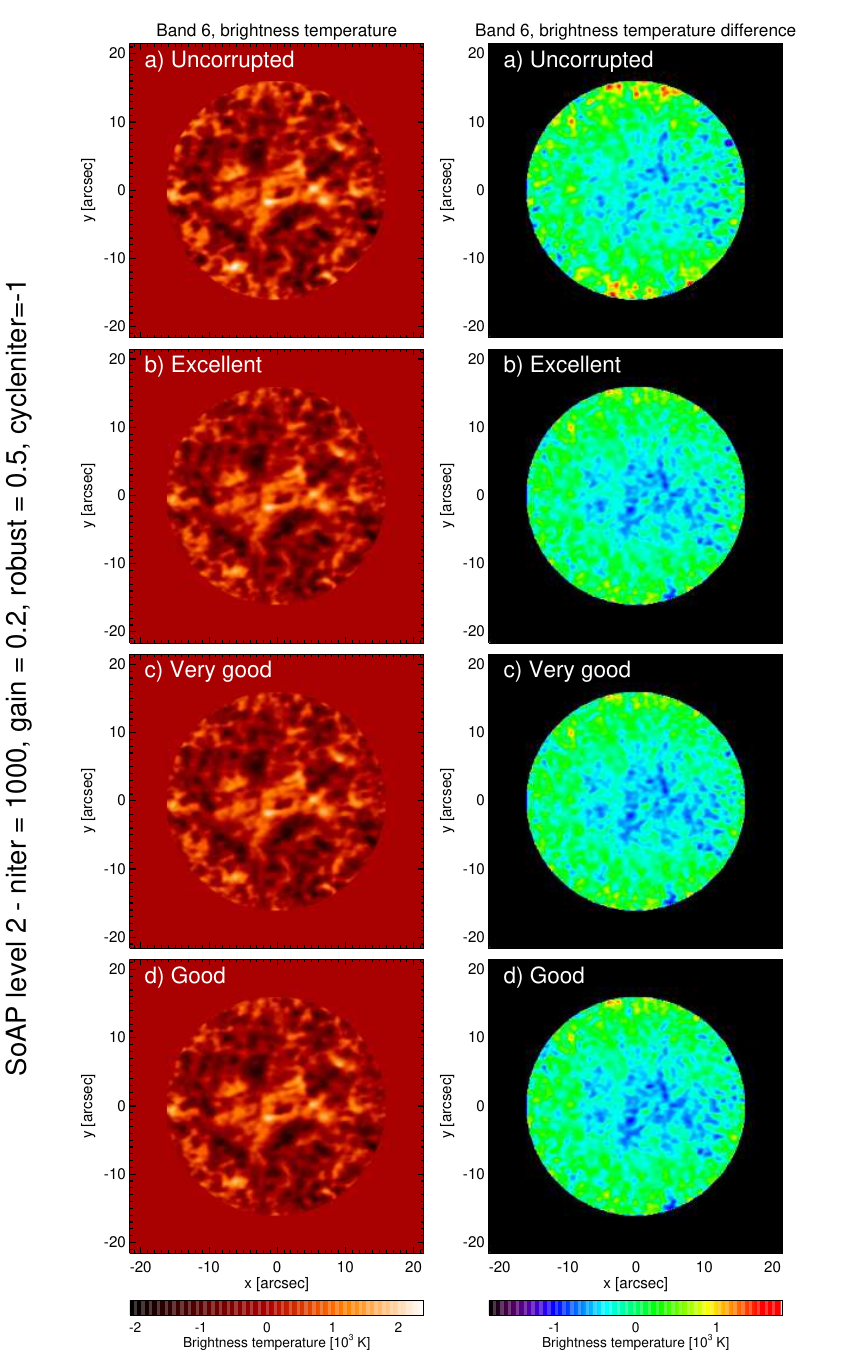}
    \vspace{-2mm}
    \caption{Imaging results for Band~6 for the uncorrupted measurement set  at 1\,s cadence 
 (top row) and the scenarios excellent, very good, and good (rows from top to bottom). 
    The brightness temperature maps for the first time step are shown in the left column and the corresponding difference to the reference model (see Fig.~\ref{fig:tbmaps_ref}b) in the right column. The same colour scale is used for all maps in their respective column.    }
    \label{fig:tbmaps_b6_1}
\end{figure}
\begin{figure}[pt!]
    \centering
    \vspace{-4mm}
    \hspace{10mm}
    \includegraphics[width=13.5cm]{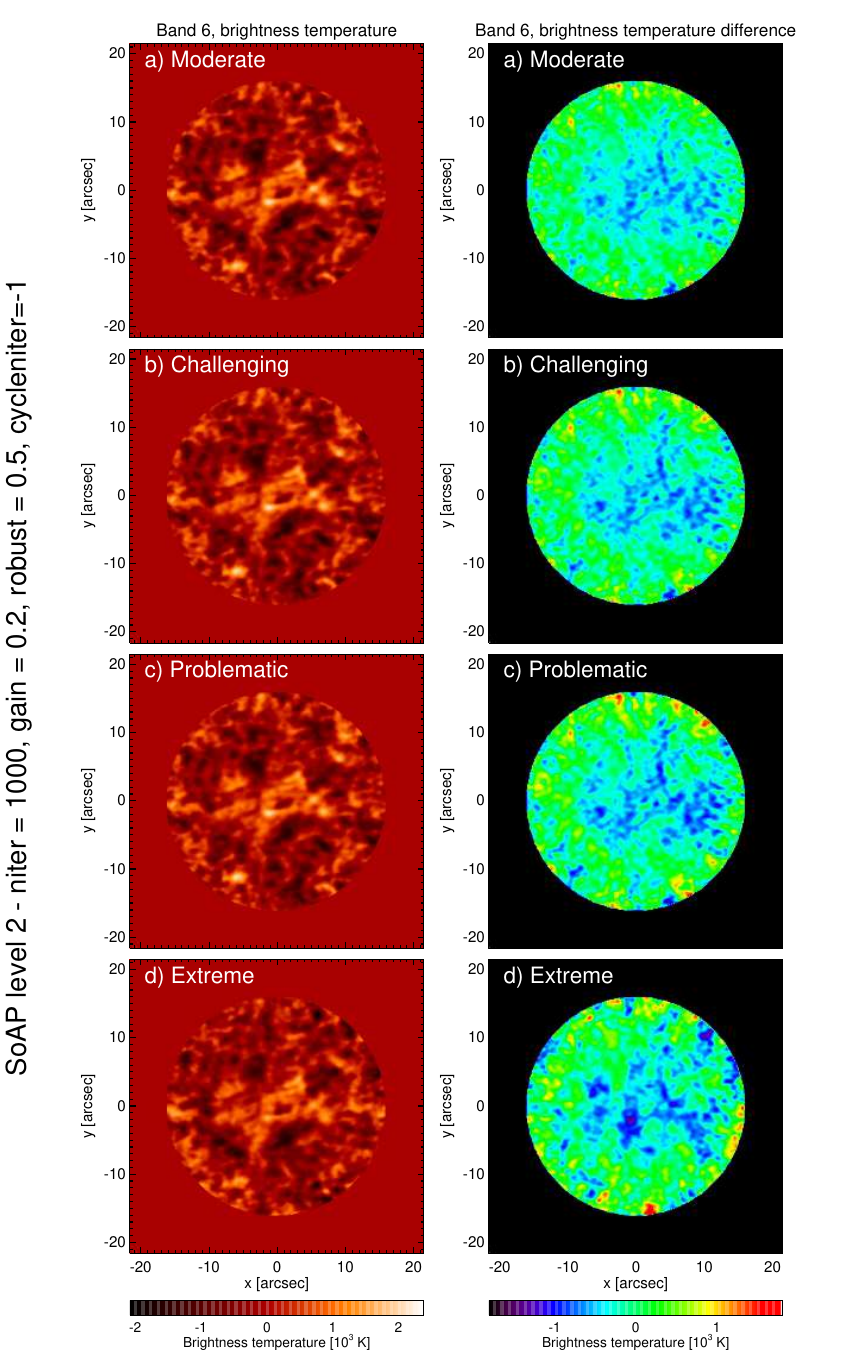}
    \vspace{-2mm}
    \caption{Imaging results for Band~6  at 1\,s cadence for the scenarios moderate, challenging, problematic and extreme (rows from top to bottom). 
    The brightness temperature maps for the first time step are shown in the left column and the corresponding difference to the reference model (see Fig.~\ref{fig:tbmaps_ref}b) in the right column. The same colour scale is used for all maps in their respective column.    }
    \label{fig:tbmaps_b6_2}
\end{figure}

\begin{figure}[t]
    \centering
    \includegraphics{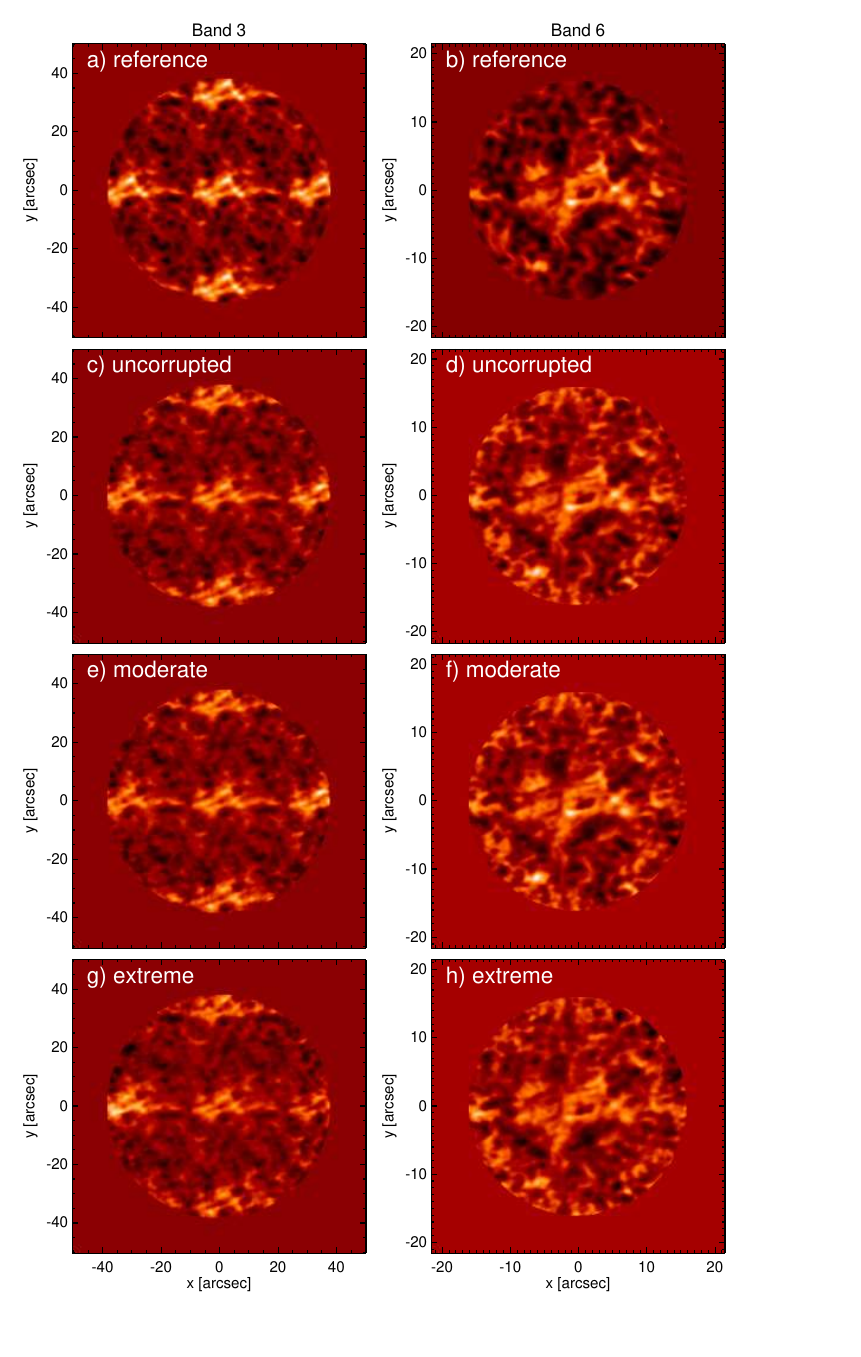}
    \vspace{-12mm}
    \caption{Brightness temperature maps for the first time step in the 0.1\,s cadence series for Band~3 (left) and Band~6 (right).  From top to bottom: The reference sky model (a-b), the imaging results for the uncorrupted case (c-d), and for the corrupted MS for the cases moderate (e-f) and extreme (g-h). 
    }
    \label{fig:Tbmaps_100ms_allcases}    
\end{figure}

The Solar ALMA Pipeline (\tool{SoAP}) uses the \tool{CASA} task \task{tclean}. It is based on the CLEAN algorithm \citep[see, e.g.,][and references therein]{hogbom,2019AJ....158....3R}, which is commonly used for imaging of interferometric data, although in various forms and implementations.  The aim is to deconvolve the dirty map that results from an interferometric measurement set and to deal with the sparse uv sampling. 
The processing thus also aims at removing any imprint of the instrument's non-ideal contributions to its point spread function (such as sidelobes).
The CLEAN algorithm does this iteratively (see the technical document for details) with an outer loop of major cycles and an inner loop of minor cycles, which can be controlled with user-specified parameters. 

In its current implementation, \tool{SoAP} is using preset parameters for \task{tclean} that are based on the experience of the SolarALMA group in Oslo and their collaborators with Cycle~4-5 solar data. It is possible though to override the preset parameters with the values described in Sect.~\ref{sec:clean_parameters}. 
\tool{SoAP} detects the different time stamps in the input measurement set (here the artificial MSs produced with \tool{SASim}, see Sect.~\ref{sec:sasim_desc}) and produces time series of brightness temperature maps.  Examples for the resulting  maps   are shown in Figs.~\ref{fig:tbmaps_b3_1}-\ref{fig:tbmaps_b6_2}. Please note that \code{CASA}~5.7 was used for this step.


\subsection{Systematic imaging parameter grid}
\label{sec:def_paramgrid}

Given the large number of possible parameter combinations and the resulting computation times, the number of tested cases has been limited in a meaningful way. The focus is on the most commonly used setup in the first cycles with regular solar observations.  

\begin{itemize}
    \item \emphbf{Interferometric observations only.} At this stage, no combination with TP contributions is performed within \tool{SASim}.  
    \item \emphbf{Antenna configuration.} While \tool{SASim} allows for simulating different antenna configurations, we focus here on the configuration chosen for the majority of solar ALMA observations so far, namely configuration 3 (more precisely, e.g., in Cycle 5: C43-3). The configuration files alma.cycle5.3.cfg and aca.cycle5.cfg are used for this purpose. According to the ALMA Cycle 5 Technical Handbook\footnote{ALMA Cycle 5 Technical Handbook: Doc 5.3, ver. 1.0, March 21, 2017}, the baselines in this configuration of the 12-m Array range from 15m to 500m, whereas the ACA covers baselines from 9 m to 30 m, resulting in the following angular resolutions (AR) and maximum recoverable scales (MRS):\\[2mm] 
    \begin{tabular}{|l|c|c|}
\hline
 & \textbf{Band 3} & \textbf{Band 6}\\
\hline
Angular resolution 12-m Array& 1.42” & 0.62” \\
\hline
Angular resolution ACA &12.5” & 5.45” \\
\hline
Maximum recoverable scale 12-m Array & 16.2” & 7.02”\\
\hline
Maximum recoverable scale ACA  & 66.7” & 29.0”\\
\hline
    \end{tabular}
    \ \\[2mm]
Due to problems with handling such heterogeneous arrays, the 12\,m and the ACA antennas are combined into the same configuration file and the antenna diameter of the ACA antennas is set to 12\,m.
    \item \emphbf{Single-pointing time series.} As the overall aim of this study is to investigate and optimize high-cadence imaging of the Sun, the focus is on short time series at a cadence of 1s as it was offered since Cycle 5. Mosaics are not considered as they result in only one combined image and thus do not offer any time series. 
    \item \emphbf{Receiver bands:} The focus is on Band~3 and Band~6 as they have been offered since the first regular solar observations in Cycle 4 and are the most used receiver bands so far. 
    \item \emphbf{Full-band maps.} The standard is currently to use all 4x128 spectral channels for the reconstruction of one map, sometimes referred to as ``full-band map'' of ``continuum map''. In principle, solar ALMA observations can be split into sub-bands and even individual spectral channels but this is especially challenging due to the then small number of available visibilities in snapshot mode. The study is therefore limited to the current standard (continuum maps).
    \item \emphbf{Direction / time of observation:} As the Sun moves over the sky in the course of the day, the shape of the synthesised beam will change. It will be the least elongated and thus the closest to a circular shape when the Sun stands highest on the sky at local noon. This effect is small during a typical solar scan of 10min duration but can be notably different for observations during different times of the day. Here, we choose a case of observing close to local noon with only slightly elongated beams (see Table~\ref{tab:synbeam}). This accounts for non-circular beams while still being representative for the majority of solar ALMA observations.  
\end{itemize}

It should be noted that it is straightforward to extend the study to other receiver bands and antenna configurations, even those that are not available in reality yet. Such an extension, however, needs adequate computational and human resources.

\subsubsection{CLEAN parameters} 
\label{sec:clean_parameters}
 
 In this study, the \task{tclean} parameters \param{niter} and \param{gain}, which control  the iteration cycles, and the visibility weighting (sub)-parameter \param{robust} are considered (see Sect.~\ref{sec:imaging_soap} and the  \techdoc{} for more details).

\textbf{niter:}~Controls the (maximum) number of iterations, i.e. major cycles. During one iteration, one  flux component is selected and partially removed from the residual image. 
When  \param{niter}=0, only the initial major cycle is executed (which produces the dirty map etc.). 

\vspace{-3mm}
{\footnotesize{\begin{tabular}{|p{1.5cm}|p{1.5cm}|p{1.5cm}|p{1.5cm}|p{1.5cm}|p{1.5cm}|p{1.5cm}|}
\hline
\textbf{niter}&  100  &  1000   & 2500  &  5000&   10000&   25000\\
\hline
\end{tabular}
}}

\ \\
\textbf{{gain:}}~The (loop) gain controls the minor (or inner) cycles during which a fraction of the flux is removed from the residual image.  

{\footnotesize{\begin{tabular}{|p{1.5cm}|p{1.5cm}|p{1.5cm}|p{1.5cm}|p{1.5cm}|p{1.5cm}|p{1.5cm}|p{1.5cm}|}
\hline
\textbf{gain}& 0.02& 0.05& 0.10& 0.20& 0.50\\
\hline
\end{tabular}
}}

\ \\
\textbf{{robust:}}~The (sub-)parameter \param{robust} controls the weighting scheme (in the Briggs mode) that is used when mapping the visibilities onto a regular grid in uv-space. A value of -2.0 results in essentially \textit{uniform} weighting, whereas a value of 2.0 essentially leads to \textit{natural} weighting. 
Uniform weighting gears the solution to maximum angular resolution (with a small synthesised beam), whereas natural weighting is a better choice in terms of sensitivity but produces a larger synthesised beam. 
A value of \param{robust}=0 is defined as compromise between both weighting schemes. 

{\footnotesize{\begin{tabular}{|p{2cm}|p{2cm}|p{2cm}|p{2cm}|p{2cm}|p{2cm}|}
\hline
\textbf{robust}& -2.0&-0.5& 0.0& +0.5& +2.0\\
\hline
\end{tabular}
}}

\pagebreak
\textbf{{Other parameters:}}~The various CLEAN implementations, including the \tool{CASA} task \task{tclean} used here, have a large number of user-controllable parameters beyond those chosen for this study. For instance, the parameter \param{threshold} is not considered here as it has in practice no relevance due to other stop criteria connected to above mentioned parameters prevailing.

\subsubsection{Algorithm}

As a default for this study, the processing of (artificial) ALMA data has been performed with the Solar ALMA Pipeline level~2, which builds up on the \tool{CASA} task \task{tclean} \citep[see, e.g.][]{casa}. More specifically, the  multi-scale (multi-frequency) CLEAN algorithm \citep{2011A&A...532A..71R} as implemented in \tool{CASA} is used. 
Here, we investigate alternative choices for the deconvolver provided in \tool{CASA} \task{tclean}\footnote{See the \tool{CASA} documentation on deconvolution algorithms:   
\url{https://casa.nrao.edu/casadocs-devel/stable/imaging/synthesis-imaging/deconvolution-algorithms}} (see Table~\ref{tab:tclean_algorithms}).

\begin{table}
\begin{tabular}{|l|l|c|c|}
 \hline
\textbf{test case}& \textbf{Algorithm name} & \task{tclean} \textbf{argument} & \textbf{Reference} \\
 \hline
\tool{SoAP} lvl2& MultiScale Clean & multiscale &\citet{multiscale} \\
 \hline
hogbom& Hogbom Clean & hogbom & \citet{hogbom} \\
 \hline
clark& Clark Clean & clark &\citet{clark} \\
 \hline
mtmfs& Multi-Term (MultiScale) Multi-Frequency  & mtmfs & \cite{mtmfs}\\
 &  Synthesis &   & \\ 
\hline
\multicolumn{1}{l}{}&\multicolumn{3}{|l|}{\textbf{Additional case}}\\
\hline
\tool{SoAP} lvl3& MultiScale Clean + self-calibration & multiscale &\citet{multiscale} \\
\hline
\end{tabular}
\caption{CASA deconvolvers used in experiments.} %
\label{tab:tclean_algorithms}
\end{table}

For each algorithm, the 60\,s long sequences without phase corruption  for Band~3 and Band~6 are  processed. To enable a fair comparison, the following parameters are chosen in all cases: 
\param{niter} = 15000,\quad \param{gain} = 0.02, \quad  \param{robust} = 0.5.\\
\tool{SoAP} level~3 is considered as an additional test case. 
It uses the same deconvolver as \tool{SoAP} level~2 but includes self-calibration, which involves  a multilevel averaging for phase correction. Please refer to the \techdoc{} for more details on \tool{SoAP}. 
 
Please note that experiments with the Maximum Entropy Method 
\citep[MEM, ][]{mem} 
as implemented in \tool{CASA} have been performed. Apart from low computational performance of the current implementation, which would pose a significant problem for the processing of solar data, other technical aspects prevented the inclusion in this report. It is recommended, however, that the usability of MEM for the processing of solar ALMA data is investigated in the future. 

\subsubsection{Imaging pixel grid}
\label{sec:setup_pixelgrid}

During the reconstruction of  interferometric images, a discrete pixel grid has to be chosen. Typically, the pixel size of the brightness temperature maps is set as a fraction of the synthesised (CLEAN) beam width such that the map is slightly over-sampled. 
The original pixel size of the \tool{ART} input maps (see Sect.~\ref{sec:artcalc} and Table~\ref{tab:artdata}) for both Band~3 and Band~6 is 0.066'' and thus much smaller than the default values of  0.30'' for Band~3 and 0.13'' for Band~6 for the images reconstructed with \tool{SoAP}. 
Here, finer pixel grids are used in order to investigate if higher pixel grid resolution leads to a more accurate reproduction of the source images at small (sub-resolution) spatial scales. 
The following pixel sizes are considered: 

\noindent\begin{tabular}{|p{2cm}|p{3cm}|p{2cm}|p{2cm}|}
\hline
\textbf{Band 3}& 0.30" (default) & 0.22" & 0.15" \\
\hline
\textbf{Band 6}& 0.13" (default) & 0.10" & 0.07" \\
\hline
\end{tabular}
\ \\

The \tool{SASim} measurement sets for PWV\,=\,0\,mm are used for this test. 
\tool{SoAP} is then executed for each of the different pixel sizes as given above for the two bands. The CLEAN parameters are set to \param{niter}=100000, \param{gain}=0.05, \param{robust}=0.5.

The impact of the pixel size on the resulting \tool{SoAP} image sequences is  inspected by comparing the corresponding spatial power spectra. 
To enable one-to-one comparisons, adequate reference models that correspond to 
the reference models for the default pixel size (see Sect.~\ref{sec:refmodel}) but for the chosen pixel sizes. For this purpose, the \tool{ART} maps are resized to the pixel sizes used in \tool{SoAP} and then convolved with the respective beams in each band. In addition, the spatial power spectrum procedure was also applied to the original \tool{ART} maps at the simulation grid size (see Sect.~\ref{sec:artcalc}). The latter can be understood as a case with superior angular resolution, for which the synthesised beam size is comparable to the pixel size and which would correspond to maximum ALMA baselines of $\sim 11$\,km for Band~3 and  $\sim 5$\,km for Band~6, respectively.

\subsection{Imaging of ultra-high-cadence sequences}
\label{sec:imaging}

In order to get a first idea about the potential prospects of exploiting a higher cadence for increasing the data quality for solar observations with ALMA, the approach described above for the 1s-cadence sequence is applied to a small set of experiments for the 100ms-cadence sequence. 
The Band~3 and Band~6 sequences of artificial observations with a cadence of 0.1\,s (thus 600 frames for the covered 60\,s) are produced and phase corruption with the Solar ALMA Simulator (\code{SASim}) for different weather scenarios is applied (see Sect.~\ref{sec:weatherscenarios}).  
Here we restrict the analysis to the moderate and extreme scenarios as those are representative of situations for which advanced data processing may significantly improve the data quality. 
Measurement sets are produced with \code{SASim}) for Band~3 and 6 for the uncorrupted case and for the selected scenarios.  

\begin{figure}[pt!]
    \centering
    \vspace{-4mm}
    \includegraphics[width=14cm]{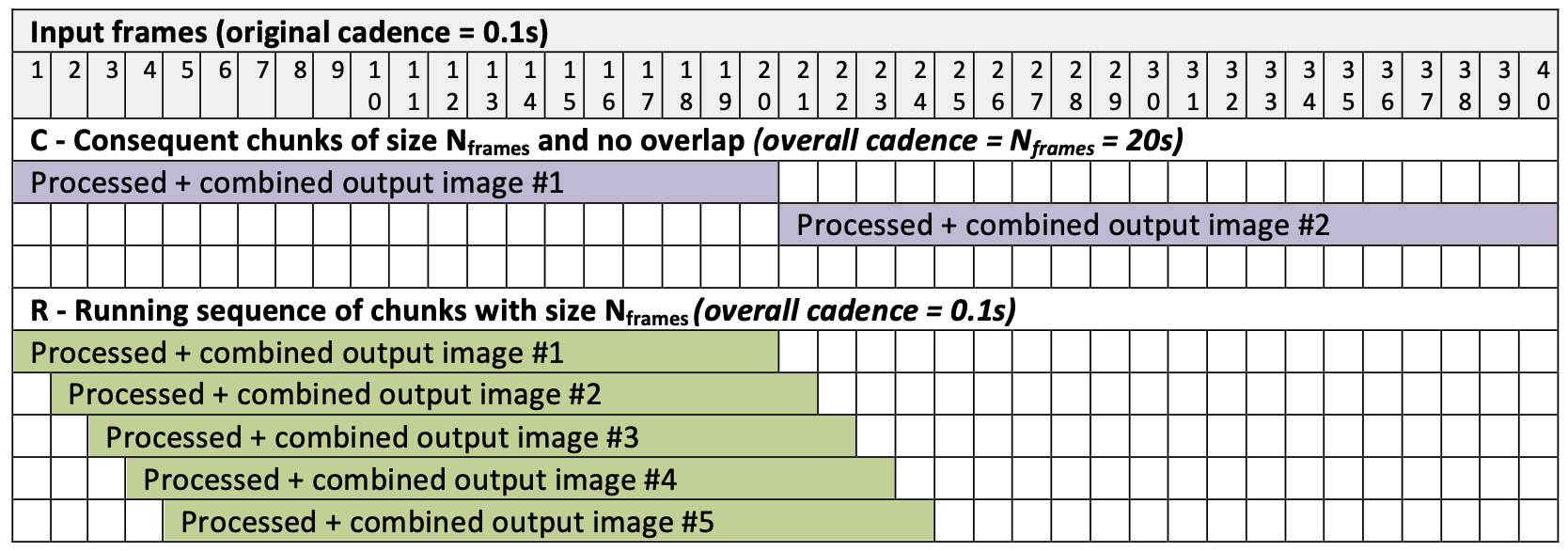}
    \vspace*{-3mm}
    \caption{Illustration of the time-binning (top) and sliding time window (bottom) techniques for an example of 20~consecutive time steps at 0.1\,s cadence.}
    \label{fig:mode_setup}
\end{figure}

Standard data reduction and imaging is carried out for all frames of these measurement sets with the Solar ALMA Pipeline (\code{SoAP}), resulting in time series of images (in FITS format) at a cadence of 0.1\,s (see Sect.~\ref{sec:basicimaging_100ms}).  
These data correspond to those at 1\,s cadence as described above. 
In addition, a sliding time window (see Fig.~\ref{fig:mode_setup} for an illustration) and self-calibration are considered (see Sects.~\ref{sec:slidingtimewindow}-\ref{sec:selfcalibration}).  
For the latter, first, the corrupted measurement sets are processed with \tool{SoAP}-level 2 without sliding time windows and then for a range of values for the sliding window size (\verb|twsize|). Then, self-calibration is applied to the corrupted measurements sets. Finally, the resulting self-calibrated measurements sets are used for imaging with tool{SoAP}-level 2 for the same values of \verb|twsize| as above. 
The considered cases are summarised in Table~\ref{tab:imagingcases}.

\subsubsection{Standard (basic) imaging}
\label{sec:basicimaging_100ms}

Imaging is performed for the uncorrupted and corrupted science target measurement sets for each frame independently. Applying \tool{SoAP} (see Sect.~\ref{sec:simseq_100ms}) then results in time series of 600 frames for each measurement set. 
The resulting data is illustrated in Fig.~\ref{fig:Tbmaps_100ms_allcases}.
The following default \tool{SoAP}  (CLEAN) parameters were used:\\[-8mm] 
\begin{verbatim}    
    niter: 1000    robust: 0.5    gain: 0.025    cycleniter: -1
\end{verbatim}
Please note that neither the brightness temperature maps for the spectral channels   nor the reference model maps  are  multiplied by the primary beam prior to being used as input for \code{SASim}. Application of the primary beam occurs as part of \task{simobserve()} when it is called during the execution of \code{SASim}. Please note that 
a mask (based on the primary beam response) is applied within \code{SoAP} and also to the reference model maps.
Please refer to the \techdoc\ for more information regarding \tool{SoAP} and the parameters. 

Already by comparing the maps for the same single time step in Fig.~\ref{fig:Tbmaps_100ms_allcases}, the impact of phase corruption becomes clearly visible. Stronger corruption as simulated in the extreme scenario indeed adds noise to the final maps as intended. 
These results and the corresponding results for the measurement sets for the original cadence of  1\,s will be used as reference for the imaging strategies that exploit the time domain for improving the overall image quality.

\begin{table}[!b]
    \centering
    \begin{tabular}{|c|c|c|}
        \hline
         &  self-calibration&twsize [s]\\
        \hline
        X & ---& --- \\
                \hline
        W1 & ---& 1.0\\
        W2 & ---& 2.0\\
        W5 & ---& 5.0\\
        W10 & ---& 10.0\\
                 \hline
        S & yes& ---\\
                \hline
        SW1 & yes& 1.0\\
        SW2 & yes& 2.0\\
        SW5 & yes& 5.0\\
        SW10 & yes& 10.0\\
        \hline
    \end{tabular}
    \caption{Considered imaging cases. }
    \label{tab:imagingcases}
\end{table}

\subsubsection{Sliding Time Window/Time-binning}
\label{sec:slidingtimewindow}

The sliding time results in an overlap of the time ranges that are used to construct the final images while maintaining the cadence of the input data (see Fig.~\ref{fig:mode_setup}). An alternative approach that avoids such overlap is to only construct images for consecutive time windows. While this time-binning approach will save computation time, the cadence of the final image series will be reduced to the user-selected time window size. 
This mode has originally been referred to as ``burst mode'' in analogy to the observing techniques used for solar observations at visible wavelengths. During the development of the sliding time window technique, it became obvious that it also covers the planned burst mode. The latter would produce combined frames that are a subset of the sequences produced with the sliding time window technique. The burst mode is therefore no longer pursued separately as it is covered by the sliding time window mode. 

A new processing mode of \code{SoAP} combines the visibility data across a user-specified time window that combines the data from neighbouring time steps (at 0.1\,s cadence) for producing one combined image each. 
As the central time step is increasing, the original cadence of 0.1\,s can be preserved.    
This option is activated and controlled with the \tool{SASim}-\tool{SoAP} parameter \verb|twsize| specified in units of seconds. Setting this parameter to \verb|null| deactivates this option. This approach can be used with \tool{SoAP} level as set with parameter \verb|lvl|. The input measurement set can also be the result of a prior self-calibration step.  

\subsubsection{Self-calibration}
\label{sec:selfcalibration}

In general, self-calibration algorithms estimate and correct for the errors in the complex visibility measurements including those that are introduced by the instrumentation. It involves measuring the instrument's response to known signals, and then using that information to correct for errors in the measured data. This is typically done by applying a mathematical model that describes the instrument's response and the errors it introduces. 
It should be noted that, for radio/(sub)-mm interferometric observations,  self-calibration can be applied on phases and amplitudes and that sufficient information on the source model is required, which will be complex for a time varying source like the Sun.

The estimate of the errors can be done by minimizing the difference between the observed and modeled visibilities. Once the errors are estimated, the visibility measurements are corrected by dividing the observed visibilities by the estimated errors. 
The corrected data is then used to reconstruct the final image. 
Self-calibration usually results in more accurate imaging. For solar observations, however, the pronounced time variability of the source poses particular challenges. While an adequately long time window is needed to ensure the construction of a accurate sky model, temporal variations of the source can make this fundamental part of the self-calibration process difficult. 
 
Self-calibration of a measurement set with  \tool{SoAP} is carried out by setting the \tool{SASim}-\tool{SoAP} parameter \verb|lvl| to 3. The procedure then starts with the window size as set with the parameter \verb|max_window_size| and iterates until the window size set by the parameter \verb|min_window_size| is reached. The results are then saved as a new measurement set. For this study,   \verb|max_window_size=30| and  \verb|min_window_size=1| are used, which results in six iterations with self-calibration windows of 30, 15, 8, 4, 2, and finally 1\,s.

\clearpage
\subsection{Imaging quality indicators}
\label{sec:qi_desc}

\subsubsection{Introduction}
\label{sec:qi_intro}

In the following, different metrics for the quality of the imaging results are discussed and developed. 
After reviewing commonly used metrics such as dynamic range in Sect.~\ref{sec:commonmetrics}, alternative quality indicators are introduced in Sects.~\ref{sec:qi_new}-\ref{sec:qiintro_spr}. 
It is important to emphasise that the aim is not to simply evaluate the overall quality of a given image but rather to determine accurately how well the original source image is reproduced after the imaging stage for a given scenario and parameter combination. 
The new quality indicators introduced here are therefore designed as quantitative measures for the quality of the imaging results with respect to the original (reference) images. See Table~\ref{tab:qi} for an overview. 

The brightness temperature maps in this study are strictly brightness temperature differences as a result of the simulated interferometric imaging without application of a TP offset. The values, which for simplicity are referred to as brightness temperatures ($T_\mathrm{b}$), span several 1000\,K for positive and negative values. The maps are filled with emission and do not contain any sky background that would allow for the determination of noise levels. Also, the overall quality of reproducing time series is assessed as this aspect is crucial for optimising the imaging results for ALMA's solar observing mode. 

 
\subsubsection{Common metrics}
\label{sec:commonmetrics}

In the following, we discuss metrics that are commonly used for observational data and explain how they are modified and/or why the TDR/SPR indicators should be used instead.  

\noindent\paragraph{Contrast} 
For an image with values $I$, the contrast $\mathcal{C}$ is defined as 
\begin{equation}
    \mathcal{C} = \frac{I_\mathrm{max} - I_\mathrm{min}}{I_\mathrm{min} + I_\mathrm{max}}    
\end{equation}
This definition is not useful for interferometric brightness temperature maps as the pixel values are both negative and positive with a average value close to zero. Alternatively, the full brightness temperature range can be calculated as 
\begin{equation}
    \mathcal{C}_T = T_\mathrm{b,max} + |T_\mathrm{b,min}|
\end{equation}
While this metric can be used to assess the sharpness of an image, it bears no information on how well the original image is reproduced.

\noindent\paragraph{Dynamic range}
According to \citet{1993A&A...271..697C}, the dynamic range $\mathrm{DR}$ is defined as the ratio of the peak brightness in an image to the off-source error, which then provides a single-number measure of the image contrast. The DR of an image with values $I$ can then be calculated as 
\begin{equation}
   \mathrm{DR} = \log_2 I_\mathrm{max} - \log_2 I_\mathrm{min}   
\end{equation}
with the maximum pixel value $I_\mathrm{max}$ and the minimum value $I_\mathrm{max}$ as set by the off-source error level. 
As mentioned above, the primary beam of interferometric brightness temperature maps in this study  is filled with emission so that the off-source error level is not known. Also, the negative brightness temperature values in $T_\mathrm{b}$, prohibit the use of the definition of the DR given above. Alternatively, one can adopt the definition 
\begin{equation}
   \mathrm{DR'} = \log_2 |T_\mathrm{b,max}|    
\end{equation}
but, like for the contrast discussed above, this metric does not provide a measure for how well the original image has been reproduced. It would be possible to compare the $\mathrm{DR'}$ of the original map to the imaging result but this metric then solely relies on the maximum values in the images. 

\noindent\paragraph{Imaging fidelity} 

\citet{1993A&A...271..697C} define the fidelity as the ratio of a pixel value to the error between the true sky distribution and the reconstructed image. The fidelity has therefore the same dimensions as the input image and in principle provides a signal-to-noise (SNR) estimate for each pixel. In an attempt to condense the information into a single number, the median of the (imaging) fidelity can be defined as 
\begin{equation}
   \mathit{f} = \mathrm{median} \left\lbrace \frac{T_\mathrm{b}}{T_\mathrm{b} - T_\mathrm{b,ref}} \right\rbrace
\end{equation}
or, given the occurrence of negative values, as 
\begin{equation}
   \mathit{f} = \mathrm{median} \left\lbrace \frac{| T_\mathrm{b}|}{|T_\mathrm{b} - T_\mathrm{b,ref}|} \right\rbrace
\end{equation} 
In principle, this metric then captures the typical relative error in the reconstructed brightness temperature map, although this is not an ideal definition for a value distribution that is centred around a small absolute value (close to zero).

\noindent\paragraph{Visibility SNR curve}
The metric as described by  \citet{1993A&A...271..697C} is derived from the difference between the reconstructed image and the (convolved) model image, which is then Fourier-transformed, radially averaged and finally divided into the radially-binned Fourier transform of the model. This metric can only be calculated if the true brightness (temperature) distribution of the source is known. 
 
\subsubsection{New imaging quality indicators}
\label{sec:qi_new}

As described above, the aim of this study is to determine which imaging parameters result in the most accurate reproduction of the original source image. This task requires to quantitatively evaluate the differences between imaging products and the original (reference) maps. Next to differences in brightness temperature values, also the good reproduction of the spatial structure of the source image is wanted.  
The new imaging quality indicators (QI) introduced and used in this study are therefore divided into two groups: 
\begin{itemize}
    \item \textbf{Direct brightness temperature differences} as measures for the accuracy of the reconstructed brightness temperature values of the source (see Sect.~\ref{sec:qiintro_tdr}). The abbreviation \textbf{TDR} stands for the radially averaged brightness temperature difference with respect to the reference model.\\
    
    \item \textbf{Differences in the spatial power spectra} as measures for the reproduction of the spatial structure of the source (see Sect.~\ref{sec:qiintro_spr}). The abbreviation \textbf{SPR} stands for the ratio of the spatial power spectra with respect to the reference model.
\end{itemize}

\begin{table}[bp!]
    \centering
    \begin{tabular}{|c|l|}
    \hline
    Abbreviation&Description\\
    \hline
    \hline
    TDR& Brightness temperature difference with respect to the corresponding \\[-1mm]
    &reference model (in radial bins  or weighted average, across whole time series)\\
    \hline
    \textbf{TDRA}&Average of TDR\\
    \textbf{TDRV}&Standard deviation of TDR \\
    \textbf{TDR+}&TDRA and TDRV combined, weighted average across 
     all spatial bins \\
    \hline
    \hline
    SPR &Spatial power ratio with respect to the corresponding reference \\[-1mm] &model (in spatial bins or weighted average, across whole time series))\\
    \hline
    \textbf{SPRA}&Average of SPR over a bin\\
    \textbf{SPRV}&Standard deviation of SPR \\
    \textbf{SPR+}&SPRA and SPRV ombined, weighted average across 
     all spatial bins \\
    \hline
    \hline
    \textbf{UQI}& Unified Quality Indicator combining TDRA, TDRV, SPRA, SPRV \\
    \hline
    \end{tabular}
    \caption{Quality indicators used in this study. Please note that the TDR and SPR indicators are always calculated within spatial bins   but can then evaluated as corresponding weighted averages. An additional {b} indicates that the quantity refers to individual bins (e.g., TDRAb) but the weighted average otherwise.  See Sects.~\ref{sec:qiintro_tdr}-\ref{sec:qiintro_uqi} for further explanation.} 
    \label{tab:qi}
\end{table}

Please note that these indicators probe similar properties as the metrics described in Sect.~\ref{sec:commonmetrics} but are adjusted for the particular data sets used in this study (see also Sect.~\ref{sec:qi_intro}).

For a given scenario and receiver band, the quality indicators are primarily evaluated as cubes  with the three dimensions being \param{niter}-\param{gain}-\param{robust}. The QI values in these cubes are   mapped to a range of 0-1 with 0 being the best result and 1 the worst result in that cube. 
This process results in scaled cubes for all quality indicators for each receiver band, weather scenario, and each spatial (scale) bin, which are introduced in the following sections. 
We note that, depending on the foreseen scientific goal, the optimal parameter choice can be found via the TDR indicators or via the SPR indicators or via  a combination of both.

As will be detailed in the next sections, these quantities are derived for different spatial bins covering different distance ranges from the axis and different spatial scale bins. This choice is made in view of the dependence of the image quality on the distance from the axis (i.e. the centre of the field-of-view, FOV) and the resulting change in signal-to-noise due to the primary beam response.  
It is important to note that variations within a spatial (scale) bin can coincidentally cancel out so that a low average value, i.e. a brightness temperature difference (TDRA) close to 0\,K or a spatial power ratio (SPRA) close to 1.0 alone would be misleading. 
These potential problems are mitigated through several measures: 
\begin{enumerate}
    \item Only the absolute brightness temperature differences with respect to the reference model are evaluated although this means that this value hardly ever will go to a value close to zero for realistic imaging set-ups as the reference model represents the ultimately ideal case with complete $uv$-coverage, no instrumental noise and no phase corruption. 
    \item Weighted averages across the spatial bins account for the impact of the primary beam response although the exact choice of these weights can be changed depending on the requirements of the scientific application. For this study, the focus the outermost parts of the FOV are weighted lower as there the larges deviations occur.  
    \item Combing the average and the standard deviation of the difference with respect to the reference model provides a more reliable metric for the overall quality. The combined quality indicators TDR and SPR are calculated by adding the scaled average cubes (TDRA or SPRA) and the corresponding standard deviation cubes (TDRV or SPRV) with equal weight (0.5 + 0.5). 
    \item Finally, the deviations in terms of brightness temperature values and reproduction of the source structure (as evaluated with spatial power spectra), which presents a metric of the overall imag reproduction quality, are evaluated together in form of a Unified Quality Indicator (UQI). 
    See Sect.~\ref{sec:qiintro_uqi} for more information regarding the UQI and     Sect.~\ref{sec:res_uqi} for the final results based on the UQI). 
\end{enumerate}

Please note that the following sections provide detailed descriptions of the considered metrics (see Table~\ref{tab:qi}) and also contain approaches that are not used for the final evaluation. These descriptions are provided as contribution of a possible further development of quality indicators in the future that might be motivated by different requirements and priorities of more specialised science cases.

\begin{figure}[t]
    \vspace{-4mm}
    \centering
    \includegraphics{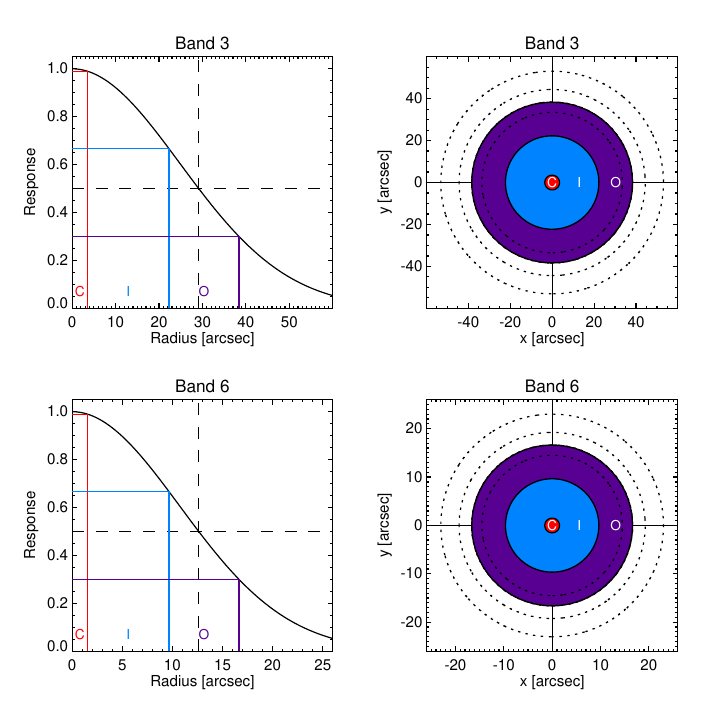}
    \vspace{-4mm}
    \caption{Definition of concentric regions and primary beam levels used for the quality indicators for Band~3 (top) and Band~6 (bottom). Left: Primary Beam Response with the HPHW marked by dashed lines. Right: Concentric regions in the image: C: Centre, I: Inner, O: Outer. }
    \label{fig:concentric}
\end{figure}

\begin{table}[bp!]
\vspace{5mm}
    \centering
    \begin{tabular}{|c|l|c|c|c|c|c|}
\hline
\multicolumn{2}{|c|}{}&
\multicolumn{2}{c|}{inner boundary}&
\multicolumn{2}{c|}{outer boundary}&enclosed area\\
\cline{3-6}
\multicolumn{2}{|c|}{Region}&radius$^{*}$ &  PBR$^{**}$ & radius$^{*}$ &  PBR$^{**}$  & fraction$^{***}$\\
\hline
C & Centre&             0.0 &   100.0\,\%&  12.0\,\%&   99.0\,\%&   1.4\,\%\\
\hline
I & Inner&   12.0\,\%&   99.0\,\%&   76.6\,\%&   66.6\,\%&   57.2\,\%\\
\hline
O & Outer&          76.6\,\%&   66.6\,\%&   131.8\,\%&  30.0\,\%&   115.0\,\%\\
\hline
\multicolumn{1}{l}{$^{*}$} &
\multicolumn{6}{l}{\footnotesize{With respect to the radius at half power (PBR=50\%)}}\\[-1mm]
\multicolumn{1}{l}{$^{**}$}&
\multicolumn{6}{l}{\footnotesize{With respect to Primary Beam Response; PBR=50\% at HWHP (Half Width Half Power)}}\\[-1mm]
\multicolumn{1}{l}{$^{***}$}  &
\multicolumn{6}{l}{\footnotesize{With respect to primary area with a diameter of HWHP (PBR=50\%)}}\\
    \end{tabular}
    \caption{Definition of concentric regions (spatial bins) in terms of radius from the axis (i.e. centre of the brightness temperature map) and the corresponding primary beam response for Band~3 and Band~6. The rightmost column shows the corresponding area of the regions with respect to the primary beam area at a primary beam response of 50\%.}
    \label{tab:def_rbin_rel}
    \vspace{-5mm}
\end{table}

\subsubsection{Brightness temperature differences}
\label{sec:qiintro_tdr}

\noindent\paragraph{Overall definition}

A straightforward way to evaluate the overall deviation for a given reconstructed map ($T_\mathrm{b} (x,y,t)$) from the original (reference) map ($T_\mathrm{b,ref} (x,y,t)$) is to calculate the average and standard deviation of the brightness temperature difference 
$\Delta T_\mathrm{b} (x,y,t) = T_\mathrm{b} (x,y,t) - T_\mathrm{b,ref} (x,y,t)$ as 
but, as noted in Sect.~\ref{sec:qi_new}, these differences can cancel out so that it is preferable to rather use the absolute values of these differences. Accordingly, the following two quantities are defined:  
\begin{equation}
    \mathrm{TDM} = \left< | \Delta T_\mathrm{b}| (x,y,t)  \right>_{x,y,t}
     = \left< | T_\mathrm{b} - T_\mathrm{b,ref}| (x,y,t) \right>_{x,y,t}
    \label{eq:def_tdm}
\end{equation}
and 
\begin{equation}
    \mathrm{TDV} = \frac{1}{N}  \sqrt{ \sum ( | \Delta T_\mathrm{b}| (x,y,t) )^2 }
    = \frac{1}{N}  \sqrt{ \sum ( T_\mathrm{b}  - T_\mathrm{b,ref})^2 (x,y,t)   }
    \label{eq:def_tdv}
\end{equation}
For each image set, i.e. a time series of brightness temperature maps $T_\mathrm{b} (x,y,t)$ for a given scenario, band, and imaging parameter combination, a pair of TDM and TDV values results. 
These two metrics can then be combined into one combined quality indicator by scaling the TDM and TDV values to the full TDM and TDV range for a given scenario $i$ and receiver band $j$:  

\begin{equation}
    \mathrm{TDC}_{ijk} = 
    a_m \ \frac{\mathrm{TDM}_{ijk} - \mathrm{TDM}_{\mathrm{min},ij}} {\mathrm{TDM}_{\mathrm{max},ij} - \mathrm{TDM_{\mathrm{min},ij}}}
    \ + \ 
    a_v \ \frac{\mathrm{TDV}_{ijk} - \mathrm{TDV}_{\mathrm{min},ij}}{\mathrm{TDV}_{\mathrm{max},ij} - \mathrm{TDV}_{\mathrm{min},ij}}
\end{equation}
with the index $k$ representing a specific imaging parameter combination. In this study, the average and variation are weighted equally, i.e. $ a_m = a_v = \frac{1}{2}$.

\begin{table}[!b]
    \centering
    \begin{tabular}{|c|c|c|c|c|c|}
\hline
\multicolumn{2}{|c|}{}&\multicolumn{2}{c|}{\textbf{Band 3}}&\multicolumn{2}{c|}{\textbf{Band 6}}\\
\cline{3-6}
\multicolumn{2}{|c|}{Region} & min. radius & max. radius &min. radius &max. radius \\
\multicolumn{2}{|c|}{ }&[arcsec]&[arcsec]&[arcsec]&[arcsec]\\
\hline
C & Centre&         0.0”&   3.5”&   0.0”&   1.5”\\
\hline
I & Inner&   3.5”&   22.3”&  1.5”&   9.7”\\
\hline
O & Outer&          22.3”&  38.4”&  9.7”&   16.6”\\
\hline
    \end{tabular}
    \caption{Radii for the boundaries of the concentric regions (spatial bins) for Band~3 and Band~6.}
    \label{tab:tdr_binbound}
\end{table}
\begin{figure}[t!]
    \centering
    \vspace{-2mm}
    \includegraphics{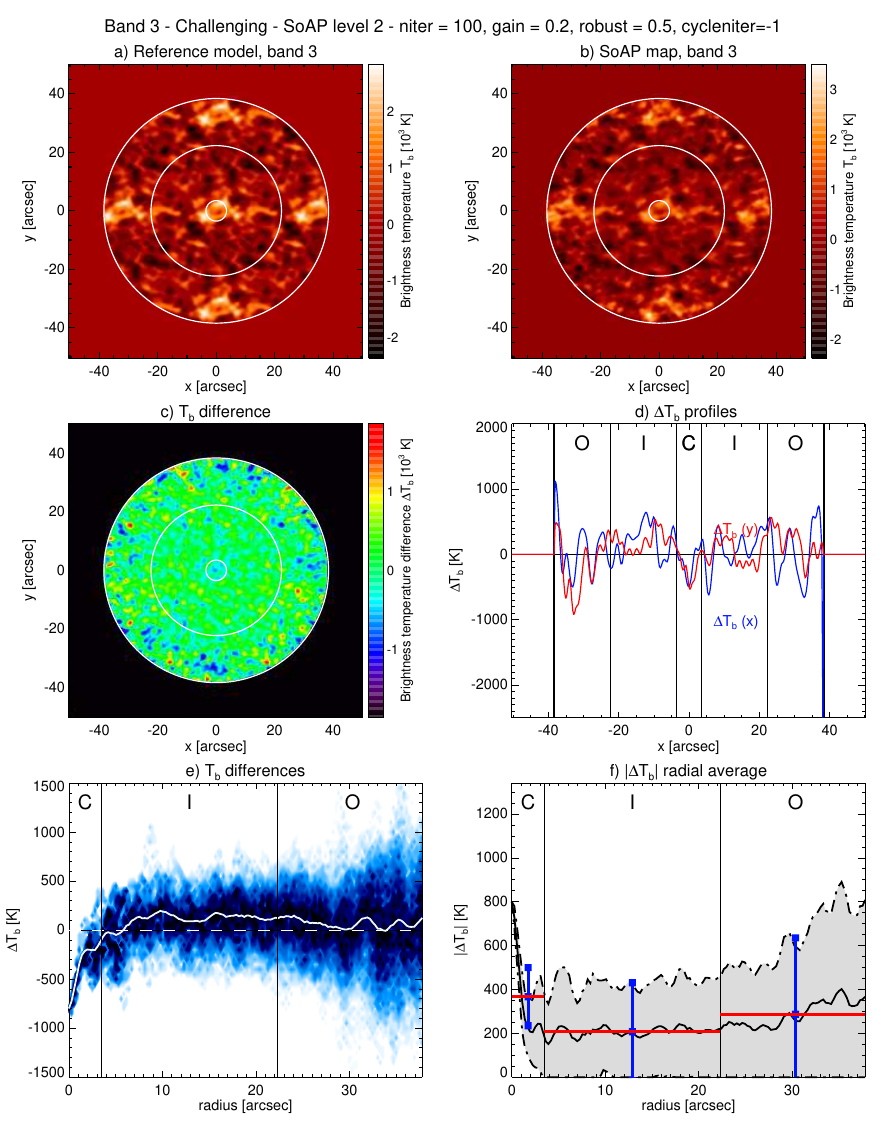}
    \vspace{-2mm}
    \caption{Calculation of the brightness temperature difference $\Delta T_\mathrm{b}$  for Band~3 for a selected time step and exemplary test case (see top for parameters). 
    \textbf{a)}~The reference model $T_\mathrm{b,ref}\,(x,y)$ (\tool{ART} map convolved with synthesised beam). 
    \textbf{b)}~\tool{SoAP} output map $T_\mathrm{b,c}\,(x,y)$. 
    \textbf{c)}~Difference of \tool{SoAP} map and reference model $\Delta T_\mathrm{b} = T_\mathrm{b,c} - T_\mathrm{b,ref}$. 
    \textbf{d)}~Profiles of $\Delta T_\mathrm{b}$ through the centre along the x- and y-axis.  
    \textbf{e)}~Histogram of $\Delta T_\mathrm{b}$ values as function of radius (distance from centre). 
    \textbf{f)}~Average radial profile of the absolute difference $|\Delta T_\mathrm{b}|$ (black solid) and the range enclosed by one standard deviation (grey area) but here over the whole time series. The concentric regions (bins) are marked in all panels. The resulting values of the average (TDRA) and standard deviation (TDRV) quality indicators are shown in panel~f for all bins as red and blue lines, respectively. }
    \label{fig:tdr_illustb3}
\end{figure}
\begin{figure}[t!]
    \centering
    \vspace{-2mm}
    \includegraphics{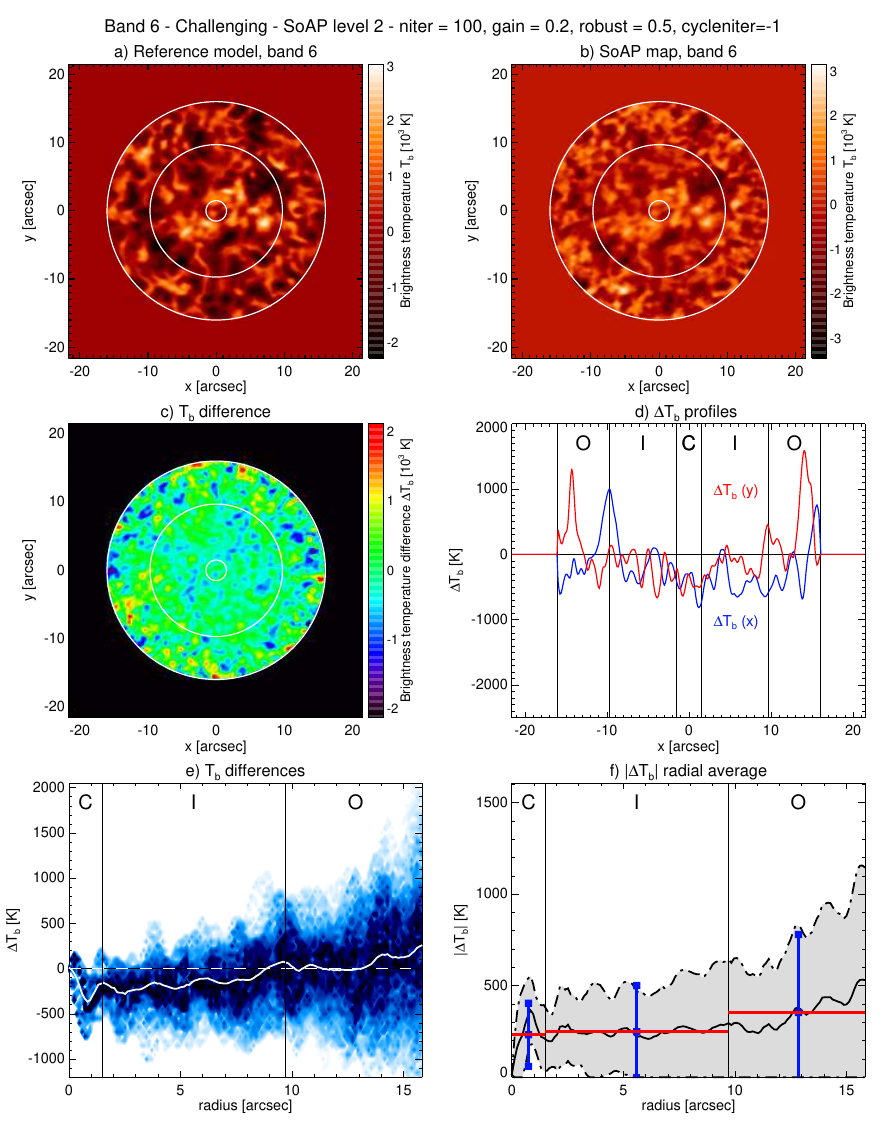}
    \vspace{-2mm}
    \caption{Calculation of the brightness temperature difference $\Delta T_\mathrm{b}$  for Band~6 for a selected time step and exemplary test case (see top for parameters). 
    \textbf{a)}~The reference model $T_\mathrm{b,ref}\,(x,y)$ (\tool{ART} map convolved with synthesised beam). 
    \textbf{b)}~\tool{SoAP} output map $T_\mathrm{b,c}\,(x,y)$. 
    \textbf{c)}~Difference of \tool{SoAP} map and reference model $\Delta T_\mathrm{b} = T_\mathrm{b,c} - T_\mathrm{b,ref}$. 
    \textbf{d)}~Profiles of $\Delta T_\mathrm{b}$ through the centre along the x- and y-axis.  
    \textbf{e)}~Histogram of $\Delta T_\mathrm{b}$ values as function of radius (distance from centre). 
    \textbf{f)}~Average radial profile of the absolute difference $|\Delta T_\mathrm{b}|$ (black solid) and the range enclosed by one standard deviation (grey area) but here over the whole time series. The concentric regions (bins) are marked in all panels. The resulting values of the average (TDRA) and standard deviation (TDRV) quality indicators are shown in panel~f for all bins as red and blue lines, respectively. }
    \label{fig:tdr_illustb6}
\end{figure}

\noindent\paragraph{Definition of concentric regions}
\label{sec:concentric}

In order to capture the variation of the imaging quality as a function of radius, the given brightness temperature map is divided into three concentric regions. The boundaries are defined as levels of the  primary beam response (PBR) of a 12m antenna  and are thus connected to the sensitivity as function of distance from the axis, i.e. from the centre of the map. The boundaries can also be expressed as the corresponding relative radii relative to the Half Width Half Power (HWHP) of the primary beam. The definitions of the boundaries between the Centre (C), Inner (I), and Outer (O) region are summarised in Table~\ref{tab:def_rbin_rel} and illustrated in Fig.~\ref{fig:concentric}.  
The Centre region has a diameter that corresponds to approximately five times the synthesised beam width (depending on the used value for the \param{robust} parameter) and to roughly 12\,\% of the 12m primary beam width (FWHP). The Centre region covers a very small fraction of primary beam area but captures the imaging performance at and close to the beam axis. The Inner region covers a much larger area and reaches out 76.6\,\% of the primary width. This region is in practice essential for any (standard)  scientific analysis of the maps. Finally, the Outer region covers larger distances from the beam axis and includes even areas with a PBR of down to 30.0\,\%. It is the largest region in terms of covered area but also the one with the lowest SNR.

\begin{table}[!t]
    \centering
    \begin{tabular}{|c|c|c|c|c|}
\hline
Spatial&\multicolumn{2}{c|}{\textbf{Band 3}}&\multicolumn{2}{c|}{\textbf{Band 6}}\\
\cline{2-5}
 scale bin&lower limit&upper limit&lower limit&upper limit\\
\hline
SUBRES& 0.30” &1.42”&0.13”&0.62”\\
\hline
SMALL &1.42”& 4.26”&0.62”&1.86”\\
\hline
MEDIUM& 4.26”&12.50”& 1.86”& 5.45”\\
\hline
LARGE& 12.50”& 66.70”& 5.45”& 29.0”\\
\hline
    \end{tabular}
    \caption{Definition of spatial scale bins used for the quality indicators based on spatial power ratios. Please note that the SUBRES bin is only defined for completeness and might be used for future analysis of noise contributions at the these scales. For the calculation of the quality indicators in this study, the SMALL to LARGE bins are used.}
    \label{tab:def_spatialscalebins}
\end{table}

\begin{table}[b!]
    \centering
    \vspace{5mm}
        \begin{tabular}{|c|c|c|c|}
    \hline
    \multicolumn{2}{|c}{\textbf{TDR+}}& 
    \multicolumn{2}{|c|}{\textbf{SPR+}}\\
    \hline
    bin&    weight&     bin&    weight\\
    \hline
    CENTRE&     0.10&   SUBRES& 0.0\\
    \hline
    INNER&      0.60&   SMALL&  0.40\\
    \hline
    OUTER&      0.30&   MEDIUM& 0.40\\
    \hline
    \multicolumn{2}{|c|}{ }&LARGE&    0.20\\
    \hline
    \end{tabular}
    \caption{Weights for the individual regional and spatial scale bins for the calculation of overall parameter combinations. }
    \label{tab:comboweights}
\end{table}

Following the ALMA Cycle 5 Technical Handbook\footnote{ALMA Cycle 5 Technical Handbook: Doc 5.3, ver. 1.0, March 21, 2017}, we set the primary beam width for an actual 12m ALMA antenna as the Half Power Beam Width (HPBW $\sim 1.13 \lambda/D$, due to non-uniform illumination). This results in primary beam widths of 58.3” for Band~3 and 24.4” for Band~6, respectively. Please note that the mask of imaging products can be set as a threshold in the primary beam response, which is usually 
 set to smaller values than the actual primary beam width in \tool{SoAP}.  
The exact width of the mask (i.e. the part of the map with actual data) in the \tool{SoAP} output can therefore in principle vary depending on the PBR threshold chosen during the imaging process. However, for this study, we strictly apply the outer boundaries defined above. 
The resulting boundaries between the three concentric regions in absolute units of arcsec  for Band~3 and Band~6 are given in Table~\ref{tab:tdr_binbound}.

\noindent\paragraph{Average brightness temperature difference as function of radius}

In Figures~\ref{fig:tdr_illustb3} and \ref{fig:tdr_illustb6}, examples of Band~3 and Band~6 brightness temperature maps for a randomly selected time step  are shown. Panel~c of each figure shows the  difference $\Delta T_\mathrm{b}\,(x,y,)$ of those brightness temperature maps that are produced with \tool{SoAP} and the corresponding reference models. The difference is also shown for cuts along the $x$- and $y$-axis through the centre in panel~d, whereas panels e and f show the radial average of absolute the difference ($\langle|\Delta T_\mathrm{b}|\rangle_r (r)$). 
It should be noted that the exact distribution of the
$\langle|\Delta T_\mathrm{b}|\rangle_r (r)$ 
values in Figs.~\ref{fig:tdr_illustb3} and \ref{fig:tdr_illustb6} will vary over time in response to the dynamics in the model atmosphere. The 3D simulations reveal significant variations of  the atmospheric structure and (gas) temperatures, which (at least qualitatively) agree with observations across a large range of wavelengths. That means also that the exact brightness temperatures and their gradients can vary significantly across the simulated field-of-view and time span. 
Another  difference between the maps shown for Band~3 and Band~6 is the periodic repetition of the original model (see Sect.~\ref{sec:sasim_desc} and Table~\ref{tab:artdata}), which results in the bright feature in the middle of the original model appearing once in the Band~6 maps but several times in the Band~3 maps. 
The simulated Band~3 maps thus  have bright features in the outer parts of the simulated field-of-view in addition to the central one, which contribute substantially to the flux across the primary beam. This effect  impacts the radial profiles  of the brightness temperatures both in the CLEANed maps and the reference maps and thus the resulting radially averaged differences. In essence, it could be argued that the simulated scenario for Band~3 is thus slightly different from the scenario for Band~6 in terms of contained bright features as compared to Quiet Sun regions but such differences are also expected for real observations as a direct result of the largely different primary beam sizes for the two bands. 

\noindent\paragraph{Calculation of TDR quality indicators.}
The calculation of the TDR quality indicators is illustrated for Band~3 and Band~6 in Figs.~\ref{fig:tdr_illustb3} and \ref{fig:tdr_illustb6}, respectively. 
For each receiver band, weather scenario, and considered imaging parameter combination (see Table~\ref{tab:scenarios_1s} and  Sect.~\ref{sec:clean_parameters}), the following steps are performed: 
\begin{enumerate}
    \item The brightness temperature differences are calculated time step by time step as the difference of the brightness temperature map for the  tested case ($T_\mathrm{b}^\mathrm{c}\,(x,y,t)$, see Fig.~\ref{fig:tdr_illustb3}b and Fig.~\ref{fig:tdr_illustb6}b)  and the corresponding reference model ($T_\mathrm{b}^\mathrm{ref}\,(x,y,t)$  see Figs.~\ref{fig:tdr_illustb3}a and \ref{fig:tdr_illustb6}a): 
\begin{equation}
 \Delta T_\mathrm{b}^\mathrm{c}\,(x,y,t) \ = \ T_\mathrm{b}^\mathrm{c}\,(x,y,t) - T_\mathrm{b}^\mathrm{ref}\,(x,y,t)
\end{equation}
The resulting brightness temperature difference map is shown for a selected time step in  Figs.~\ref{fig:tdr_illustb3}c and \ref{fig:tdr_illustb6}c. \\

\item The difference maps are radially averaged (with the radius being the distance from the centre of the maps), which results in a radial profile 
$\overline{\Delta T_\mathrm{b}^\mathrm{c}}\,(r,t)$ for each time step. 
See Figs.~\ref{fig:tdr_illustb3}e and \ref{fig:tdr_illustb6}e) for examples for a selected time step.\\   

\item The brightness temperature difference as function of radius is then calculated as time-average of the radial profiles of the absolute difference values: 
\begin{equation}
\mathrm{TDR} \ = \ \left<|\overline{\Delta T_\mathrm{b}^\mathrm{c}}|\,(r,t) \right>_t
 \end{equation}

\item For each of the three regions (C,I,O, see Sect.\ref{sec:concentric}), the average (TDRAb) and the standard deviation (TDRVb) of the brightness temperature is calculated. See Figs.~\ref{fig:tdr_illustb3} and \ref{fig:tdr_illustb6} for illustration.\\ 
\item The TDRAb and TDRVb values are combined into the combined TDRA and TDRA indicators by calculating the weighted averages across all bins. The weights are specified in Table~\ref{tab:comboweights}. 
The Inner region was weighted highest. The centre was given less weight due its lower statistical significance and as the differences can be affected by the normalisation over the whole primary beam area. The Outer region is considered but with a lower weight.The outer regions of brightness temperature maps should always be analysed with caution as larger brightness temperature uncertainties are to be expected. \\
The respective TDRA and TDRV cubes have dimensions set by the considered \param{niter}, \param{gain}, and \param{robust} values. Next, the individual TDRA cubes (i.e. for each receiver band and scenario) are each scaled to a range from 0 to 1 so that the combination [\param{niter}, \param{gain}, \param{robust}] with the lowest value corresponds the best choice for TDRA, respectively.   The same procedure is followed for the TDRV cubes.  Finally, the scaled TDRA and TDRV cubes are combined with equal weight into the TDR+ quality indicator. A TDR+ value is thus calculated for each combination of receiver band and weather scenario. 
\end{enumerate}
 
The resulting quality indicator values are compared and discussed in Sect.~\ref{sec:result_qitdr}.

\subsubsection{Spatial power spectra} 
\label{sec:qiintro_spr}

\noindent\paragraph{Definition of spatial scale bins}
\label{sec:def_spatialscalebins}
For the analysis of the spatial power spectra for the resulting \tool{SoAP} maps, the covered spatial scale range is divided into four bins, namely SUBRES, SMALL, MEDIUM, and LARGE (see Table~\ref{tab:def_spatialscalebins}), although the SUBRES bin is not considered for the calculation of the quality indicators. 
The boundaries between these bins are chosen with respect to ALMA's instrumental properties  (see also Sect.~\ref{sec:def_paramgrid}). 
The lowest spatial scale boundary is set by the grid pixel size of the images produced by \tool{SoAP}, which is here  0.3” for Band~3 and 0.13” for Band~6. The pixel grid is oversampling with respect to the nominal interferometric angular resolution (i.e. the width of the synthesised beam). Spatial scales below the latter are therefore grouped into the sub-resolution (SUBRES) bin. An initial investigation of the resulting power spectra revealed that it is useful to define the next bin boundary as three times the nominal interferometric angular resolution (the SMALL bin), whereas larger scales are again divided into the MEDIUM and LARGE bin. The division between MEDIUM and LARGE is here set by the angular resolution of the ACA so that the largest scales sampled by the 12-m Array fall into the LARGE bin. The latter therefore covers the complete range of scales that is sampled by the ACA. The largest considered spatial scale and thus the limit for the LARGE bin is set by the maximum recoverable scale (MRS) of the ACA. 

Even larger scales are not accounted for as they are larger than the typical image sizes. 

\begin{figure}[t!]
    \centering
    \vspace{-1mm}
    \includegraphics[width=\textwidth]{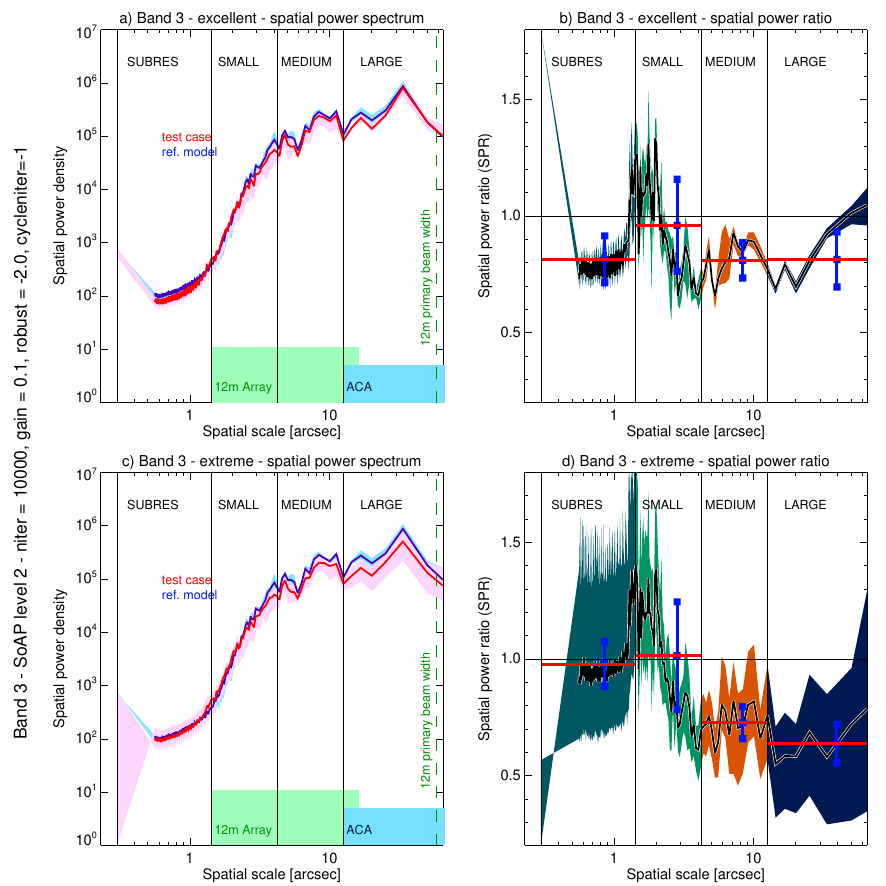}
    \caption{Examples of spatial power spectra for Band~3 for the excellent (top row) 
    and extreme (bottom row) scenario. The selected \tool{SoAP} parameters are shown to the left. In the left column, the power spectrum for the selected data set is plotted in red with the value range of the time series represented by light red shades and the time-average by a solid red line. The corresponding reference model is plotted in blue.  At the bottom, the spatial scale ranges are marked that are defined by the angular resolution and maximum resolvable scales for the 12-m Array and the ACA, respectively. The beam width of a 12-m antenna is presented as vertical dashed line. At the top, the different spatial scale bins for labelled.
    The corresponding spatial power ratios (SPR = $P_\mathrm{test case} / P_\mathrm{ref.model}$ ) are shown in the right column.  The shaded areas mark the value ranges in the different bins for the full time series. The horizontal red lines represent the time-averaged value (SPRA) for each bin, whereas the blue vertical lines represent the standard deviations (SPRV) for each bin. }
    \label{fig:spr_illustrat_b3}
\end{figure}
\begin{figure}[t!]
    \centering
        \vspace{-1mm}
    \includegraphics[width=\textwidth]{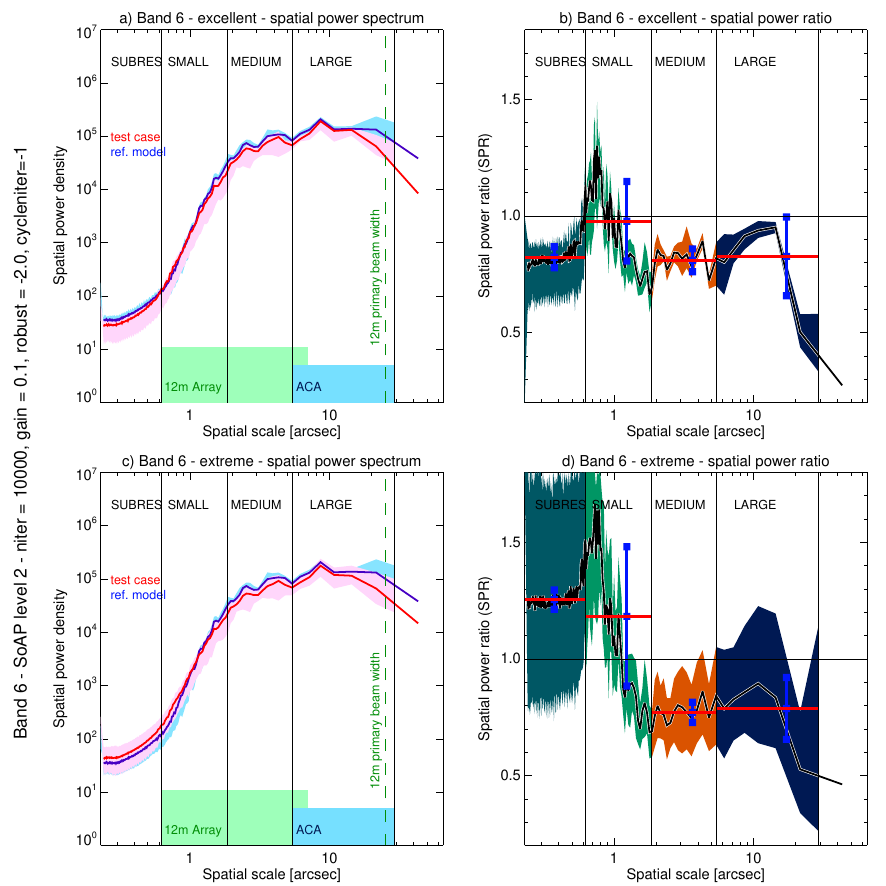}
    \caption{Examples of spatial power spectra for Band~6 for the excellent (top row) 
    and extreme (bottom row) scenario. The selected \tool{SoAP} parameters are shown to the left. In the left column, the power spectrum for the selected data set is plotted in red with the value range of the time series represented by light red shades and the time-average by a solid red line. The corresponding reference model is plotted in blue.  At the bottom, the spatial scale ranges are marked that are defined by the angular resolution and maximum resolvable scales for the 12-m Array and the ACA, respectively. The beam width of a 12-m antenna is presented as vertical dashed line. At the top, the different spatial scale bins for labelled.
    The corresponding spatial power ratios (SPR = $P_\mathrm{test case} / P_\mathrm{ref.model}$ ) are shown in the right column.  The shaded areas mark the value ranges in the different bins for the full time series. The horizontal red lines represent the time-averaged value (SPRA) for each bin, whereas the blue vertical lines represent the standard deviations (SPRV) for each bin. }
    \label{fig:spr_illustrat_b6}
\end{figure}

\noindent\paragraph{Calculation of SPR quality indicators}
Examples for the spatial power spectra for selected parameter combination are shown in Figs.~\ref{fig:spr_illustrat_b3} and \ref{fig:spr_illustrat_b6} 
for Band~3 and Band~6, respectively. The power spectra for the corresponding reference models and the variation of the power spectra during the time series are also shown. 
The differences between the power spectra of the \tool{SoAP} maps and the reference model maps are difficult to discern in the typical log-log presentation but become more evident when plotting the ratio of the spatial power spectrum for a \tool{SoAP} map and the corresponding reference model map (see panels b and d in the figures).  The examples clearly show that the power spectrum for the tested cases deviate from the reference spectrum but that this deviation depends on the spatial scale and differs for the receiver bands, scenario and (not displayed) the imaging parameter combination. The degree to which the power spectrum of the reference case is reproduced is used here as a key quality indicator for the chosen imaging parameter combination. 
For each receiver band, weather scenario, and tested imaging parameter combination, the calculation of the SPR quality indicators   is carried out as follows: 
\begin{enumerate}
    \item The spatial power spectrum is calculated time step by time step for each \tool{SoAP} map and the corresponding reference model map. The power spectra are radially averaged, resulting in spatial power spectra for the tested case (see red shades in Fig.~\ref{fig:spr_illustrat_b3}a,c and Fig.~\ref{fig:spr_illustrat_b6}a,c) and the reference model  (see blue shades in the same figure panels) for each time step. 
    \item For each time step, the spatial power spectrum for the tested case ($P_\mathrm{c}\,(k,t)$) is divided by the power spectrum for the reference model ($P_\mathrm{ref}\,(k,t)$) 
    \begin{equation}
        \mathcal{S}\,(k,t) \ = \ \frac{P_\mathrm{c}\,(k,t)}{P_\mathrm{ref}\,(k,t)}
    \end{equation}
    as function of the spatial wave number $k$.  
    The resulting spatial power ratios are shown in 
    Fig.~\ref{fig:spr_illustrat_b3}b,d and Fig.~\ref{fig:spr_illustrat_b6}b,d as coloured shades as function of spatial scale $\overline{x} = \frac{k}{2\,\pi}$.
    \item The SPR quality indicator as function of spatial scale is then calculated as time-average of the spatial power ratio: 
    \begin{equation}
    \mathrm{SPR}\,(\overline{x}) \ = \ \left< \overline{\mathcal{S}_\mathrm{c}}\,(\overline{x},t) \right>_t
     \end{equation}
    The result is shown as a black line in Fig.~\ref{fig:spr_illustrat_b3}b,d and Fig.~\ref{fig:spr_illustrat_b6}b,d.
    \item For each of the four spatial scale bins (SUBRES, SMALL, MEDIUM, LARGE, see Sect.~\ref{sec:def_spatialscalebins}), the average (SPRA) and the standard deviation (SPRV) of the spatial power spectrum is calculated. Examples for the resulting values shown in Fig.~\ref{fig:spr_illustrat_b3}b,d and Fig.~\ref{fig:spr_illustrat_b6}b,d as red lines (SPRA) and blue lines (SPRV) for all four bins.
    \item The SPRA values are calculated as weighted averages across the spatial bins from the respective values in the bins (SPRAb). The same procedure is followed for the SPRV values.  The weights are specified in Table~\ref{tab:comboweights}. 
    The largest weight was given to the SMALL and MEDIUM scales since they are best sampled by ALMA. The LARGE scales enter accordingly only with small weight, while the SUBRES bin is not considered for the quality indicator. \\  
    The resulting SPRA and SPRV cubes have dimensions set by the considered \param{niter}, \param{gain}, and \param{robust} values. 
    The individual SPRA cubes (i.e. for each receiver band and scenario) are then  scaled each to a range from 0 to 1 so that the combination [\param{niter}, \param{gain}, \param{robust}] with the lowest value corresponds the best choice for SPRA, respectively.   The same procedure is followed for the SPRV cubes.  Finally, the scaled SPRA and SPRV cubes are combined with equal weight into the SPR+ quality indicator. A SPR+ value is thus calculated for each combination of receiver band and weather scenario. 
\end{enumerate}

The resulting SPRC  quality indicator values are compared and discussed in Sect.~\ref{sec:result_qispr}.

\subsubsection{Unified Quality Indicator} 
\label{sec:qiintro_uqi}

The TDR quality indicators discussed above prioritise the accurate reproduction of brightness temperature values, while the  SPR indicator assess how accurately the spatial distribution, i.e. brightness temperature pattern, is reproduced. 
For most scientific applications, both aspects are equally important. Consequently, the combination of the TDR and SPR indicators into an Unified Quality Indicator (UQI) is chosen for the final evaluation of the imaging process in this study. 
The minimum UQI value in a resulting combined cube for a given receiver band and weather scenario then indicates the parameter combination \param{niter}-\param{gain}-\param{robust}  that produces the  best overall results. 
The determined optimum parameter combination is thus a compromise of a minimal brightness temperature differences with respect to the reference model across most of the field-of-view (with the error-prone outer areas being weighted less) and resulting spatial power spectrum that closely resembles that of the corresponding reference model.

\clearpage
\section{Results regarding optimal processing of solar ALMA data} 

\subsection{CLEAN parameter grid} 

For each weather scenario and each receiver band, the quality indicators described in Sect.~\ref{sec:qi_desc} are available for all considered parameter combinations \param{niter}-\param{gain}-\param{robust} for all concentric regions and all spatial scale bins. 
The aim of this section is now to determine the parameter combinations that produce the best imaging results. 
The results for the TDR  quality indicators and the SPR   quality indicators are discussed separately in Sects.~\ref{sec:result_qitdr} and \ref{sec:result_qispr}, respectively. 
The final evaluation based on the Unified Quality Indicator is presented in Sect.~\ref
{sec:res_uqi}.

\subsubsection{Brightness temperature differences}
\label{sec:result_qitdr}

\begin{figure}[tp!]
    \centering
    \includegraphics{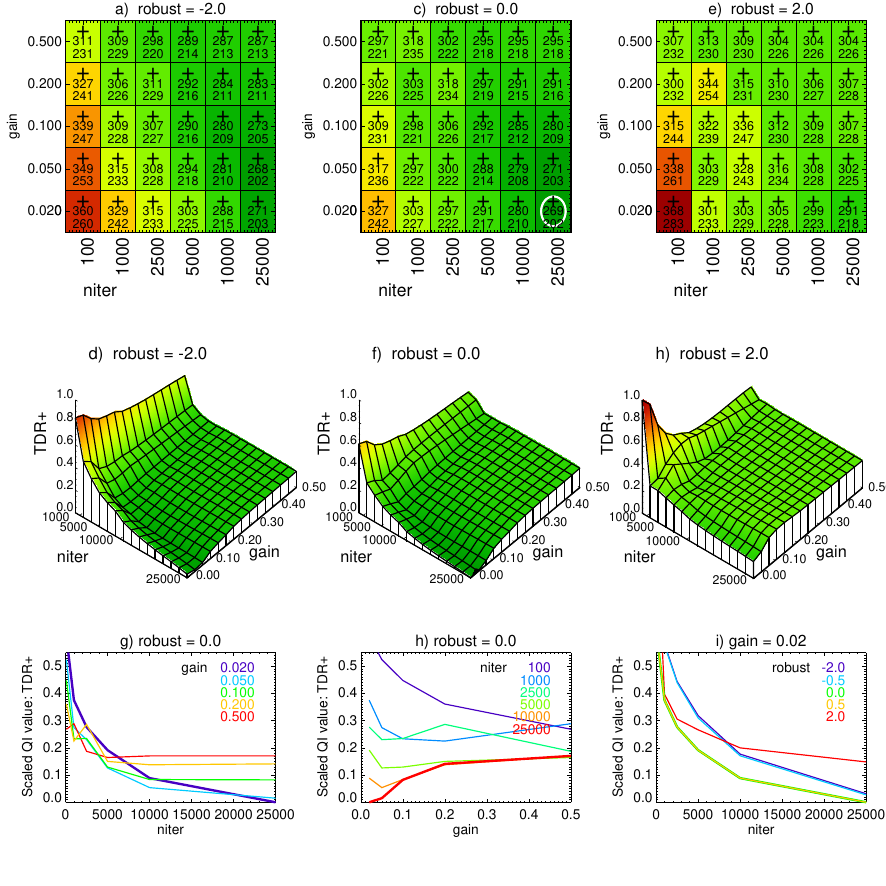}
    \vspace*{-8mm}
    \caption{Determination of the optimal imaging parameter combination based on the combined TDR+ quality indicator for Band~6 for the problematic scenario. 
    \textbf{a-c)}~\param{niter}-\param{gain} diagrams for the three out of the five \param{robust} values (left to right). 
    In each  panel, the quality of the \tool{SoAP} images is presented as color depending on the \param{niter} and \param{gain} parameters, resulting in 6\,$\times$\,5 coloured squares. Dark green stands for the best choice, while dark red represents the poorest choice. Cases with \param{niter}=0 are not considered for finding the optimal choice as they refer to dirty maps.    
    A plus or minus in each square indicates if the brightness temperature is on average (TDRA) lower or higher than in the reference model. 
    The numbers in each panel specify the average (top) and standard deviation of the brightness temperature difference with respect to the reference model. 
    The best choice in each row, i.e. the best combination of \param{niter}-\param{gain}-\param{robust} is marked with a white circle.    
    \textbf{d-f)}~3D surface plots for the TDR+ indicator corresponding to the respective parameter subspaces (i.e. \param{robust} values) in the row above. Please note that the data has been interpolated to a finer grid for a better visualisation of trends. 
    \textbf{g-i)}~TDR+ as function of two parameters with the third dimension fixed to the best choice within the cube. From left to right, the panels show on the main axis \param{niter}, \param{gain}, and \param{niter}, whereas the different colored lines represent the \param{gain}, \param{niter}, or \param{robust} dimension as stated in the panel legend, respectively. The respective cases with the best parameter choices are marked with thick lines.
    }
    \label{fig:qi_illus}
\end{figure}

\begin{figure}[hbt!]
    \centering
    \includegraphics[width=12.8cm]{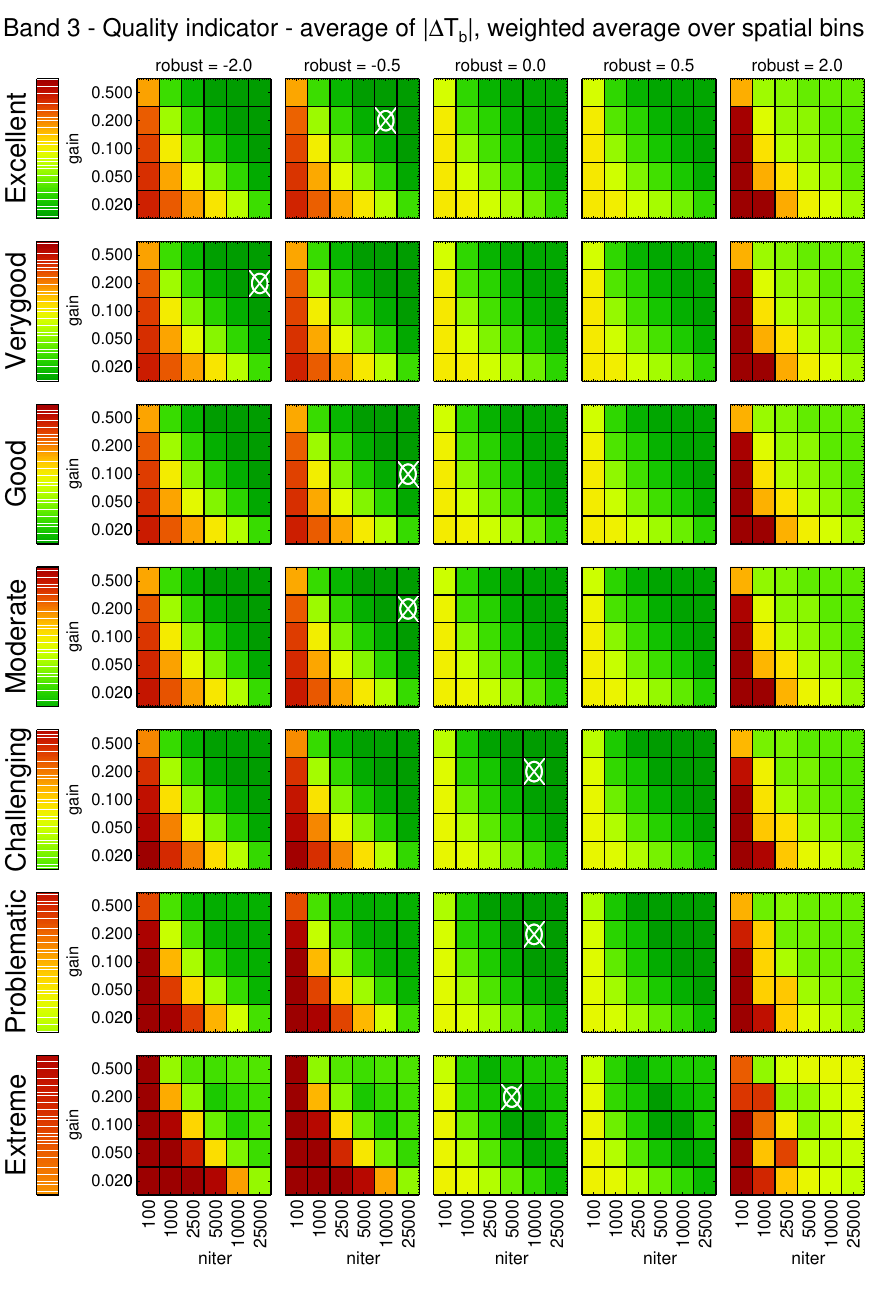}
    \vspace*{-4mm}
    \caption{Comparison of the TDRA quality indicator for Band~3. The individual rows show the \param{niter}-\param{gain} plane for the five considered \param{robust}  values (in each column). The colours represent the value of the quality indicator scaled across the whole row with dark green representing the best result and dark red the worst result, respectively. In each row (i.e. for each scenario) the best parameter combination is marked with a white cross in a circle. 
    As the values for the quality indicator differ across the different scenarios, the color bar to the left next to the scenario label show the covered value range with respect to all scenarios, again from best (dark green) to worst (dark red). 
    }
    \label{fig:qiresults_tdpa_b3}
\end{figure}
\begin{figure}[hbt!]
    \centering
    \includegraphics[width=12.8cm]{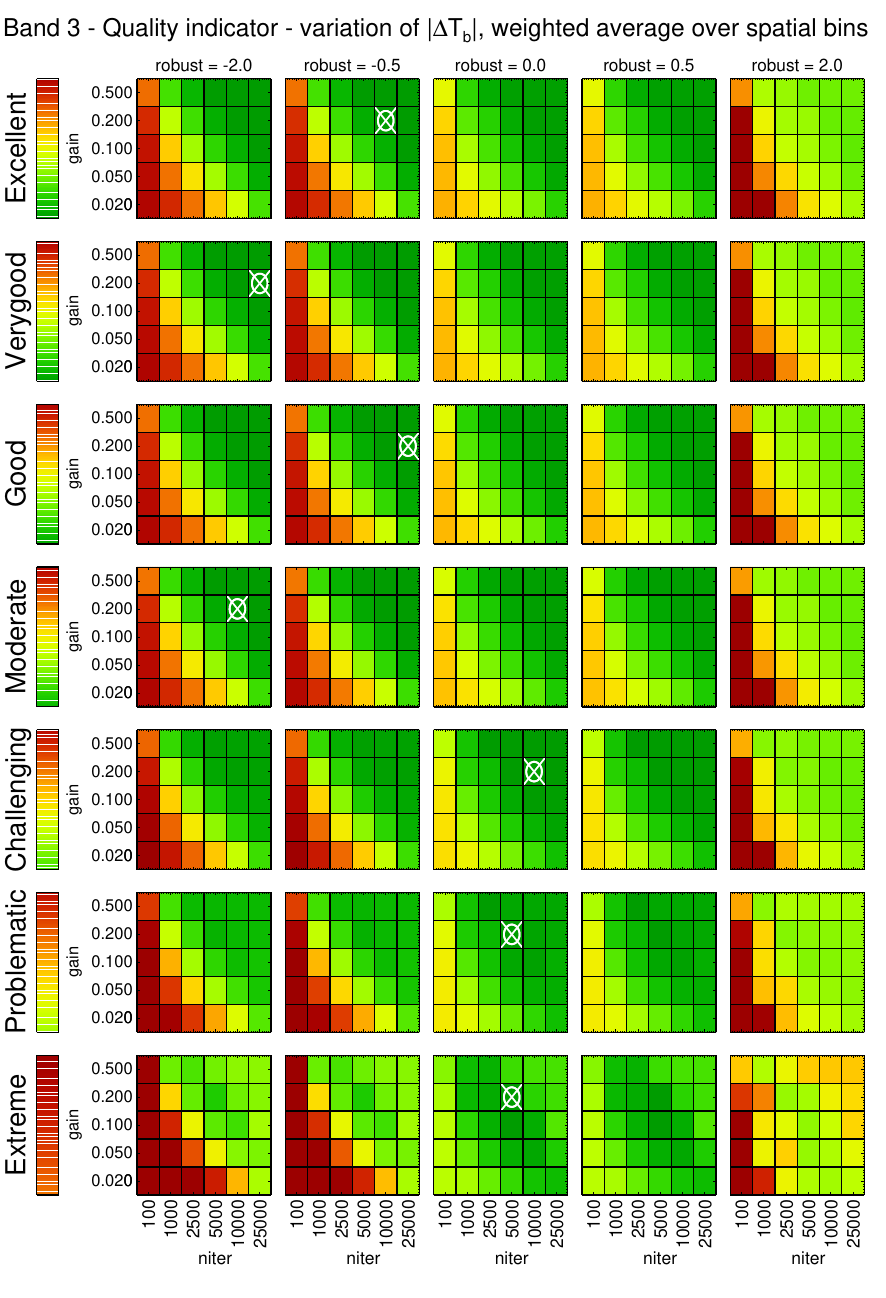}
    \vspace{-6mm}
    \caption{Comparison of the TDRV quality indicator for Band~3. See the caption of Fig.~\ref{fig:qiresults_tdpa_b3} for an explanation.} 
    \label{fig:qiresults_tdpv_b3}
    \vspace{-4mm}
\end{figure}
\begin{figure}[t!]
    \centering
    \includegraphics[width=12.8cm]{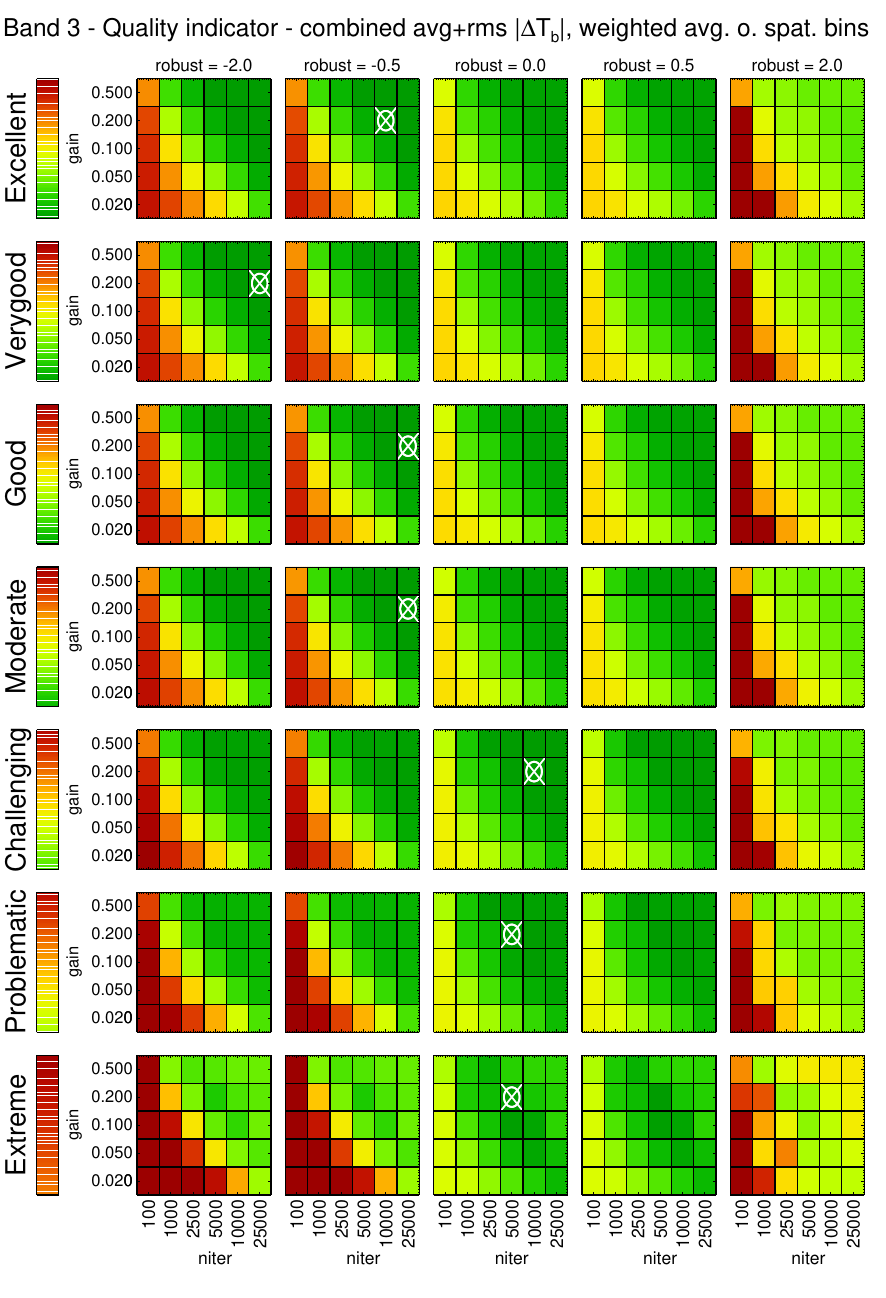}
    \vspace{-6mm}
    \caption{Comparison of the TDR+ quality indicator for Band~3. See the caption of Fig.~\ref{fig:qiresults_tdpa_b3} for an explanation. This quality indicator combines the average and variation of the deviations from the reference models in terms brightness temperature values and spatial power spectra in produce image time series.} 
    \label{fig:qiresults_tdpc_b3}
\end{figure}

\begin{figure}[hbt!]
    \centering
    \includegraphics[width=12.8cm]{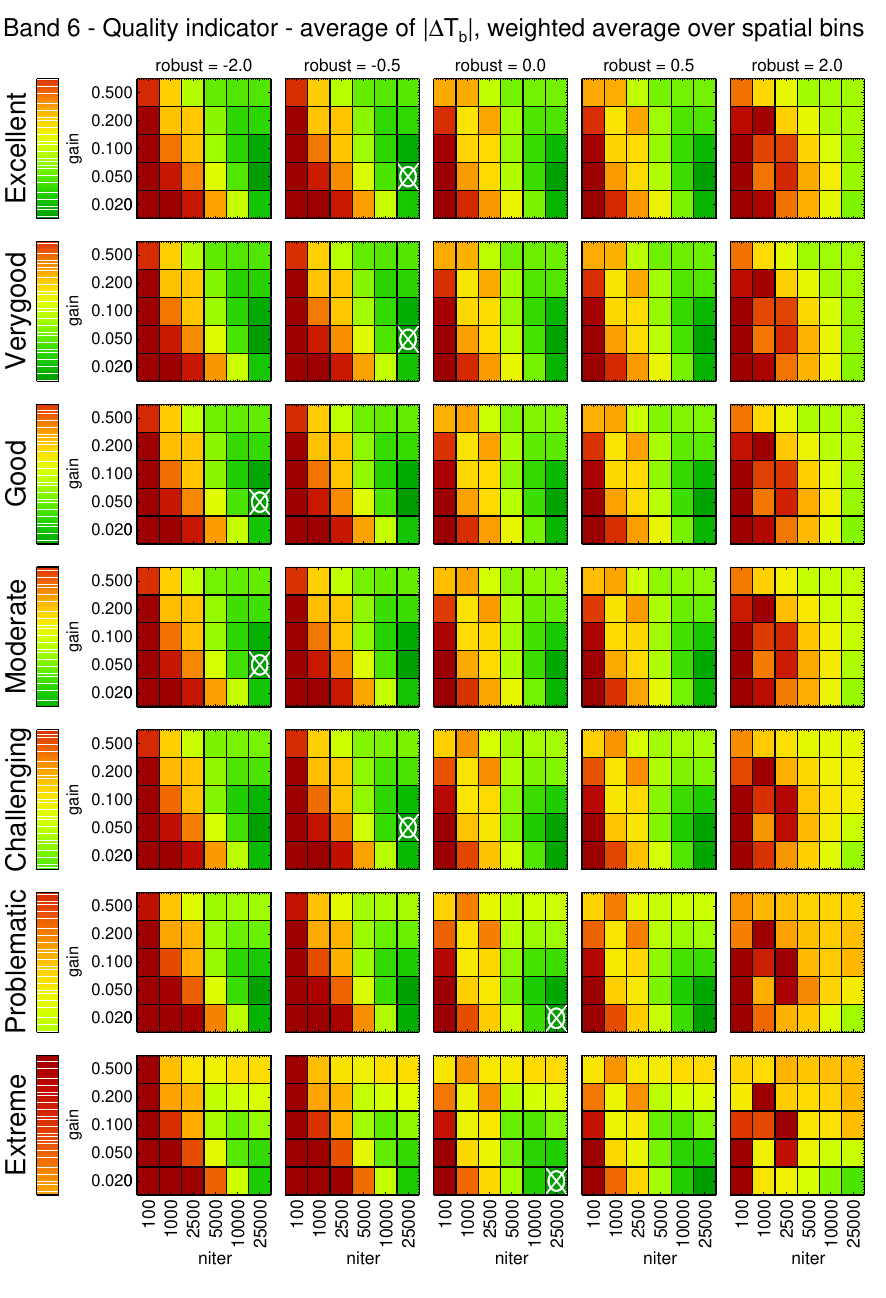}
    \vspace{-6mm}
    \caption{Comparison of the TDRA quality indicator for Band~6. See the caption of Fig.~\ref{fig:qiresults_tdpa_b3} for an explanation.} 
    \label{fig:qiresults_tdpa_b6}
\end{figure}
\begin{figure}[t!]
    \centering
    \includegraphics[width=12.8cm]{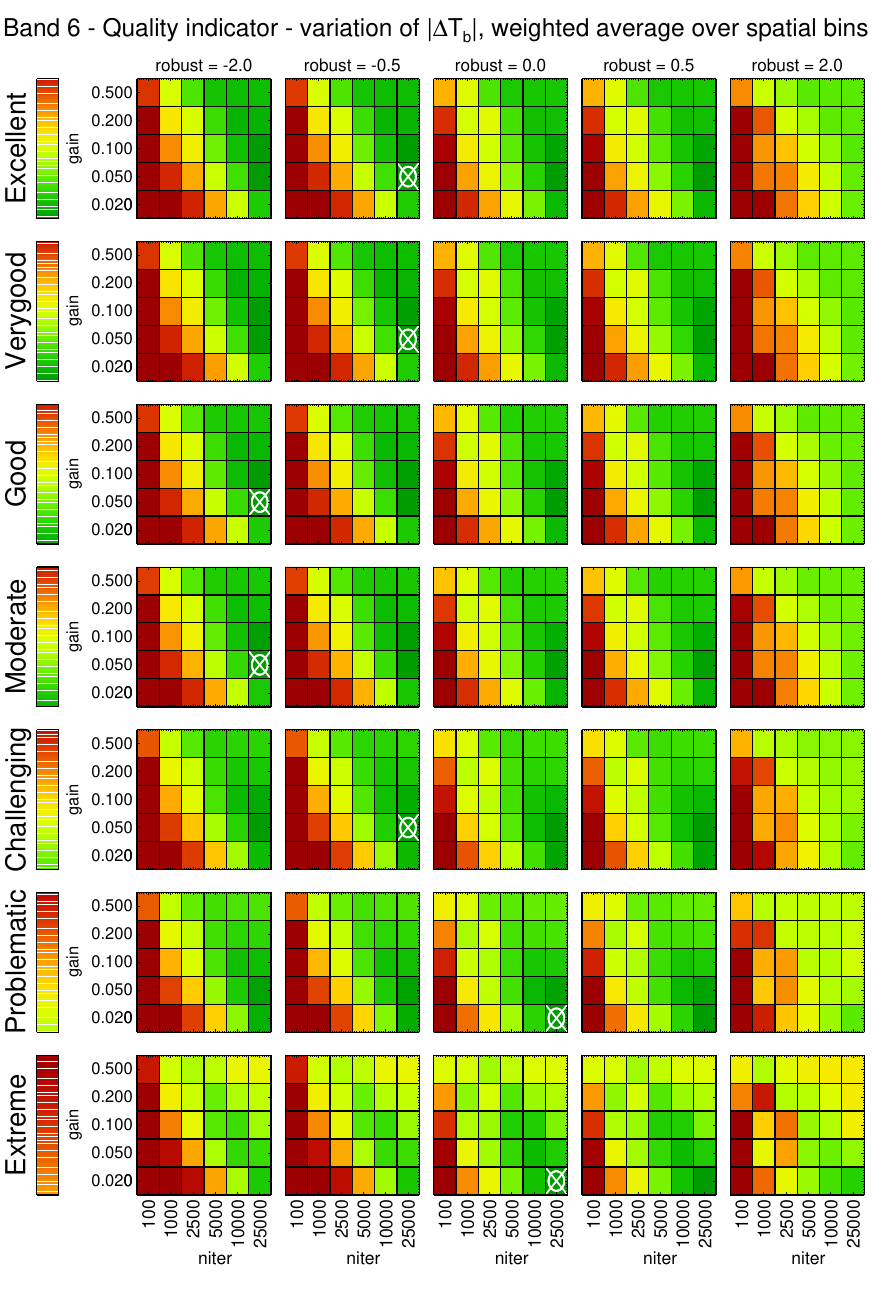}
    \vspace{-6mm}
    \caption{Comparison of the TDRV quality indicator for Band~6. See the caption of Fig.~\ref{fig:qiresults_tdpa_b3} for an explanation.} 
    \label{fig:qiresults_tdpv_b6}
\end{figure}
\begin{figure}[t!]
    \centering
    \includegraphics[width=12.8cm]{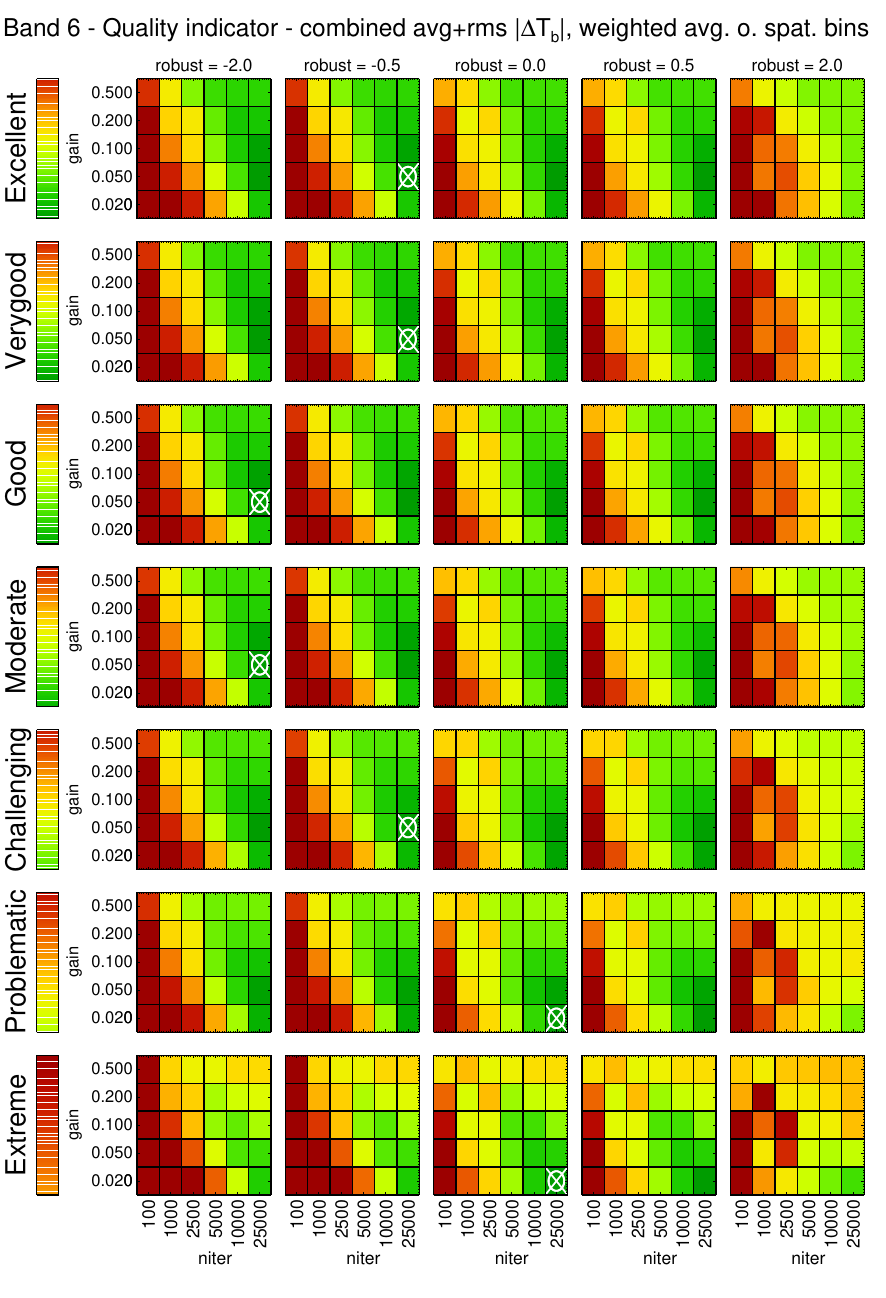}
    \vspace{-6mm}
    \caption{Comparison of the TDR+ quality indicator for Band~6. See the caption of Fig.~\ref{fig:qiresults_tdpa_b3} for an explanation. This quality indicator combines the average and variation of the deviations from the reference models in terms brightness temperature values and spatial power spectra in produce image time series.} 
    \label{fig:qiresults_tdpc_b6}
\end{figure}

The results for the brightness temperature difference indicators are presented in Figs.~Fig.~\ref{fig:qi_illus}-\ref{fig:qiresults_tdpc_b6}.
Please see Fig.~\ref{fig:qi_illus} for an illustration of the combined quality indicator TDR+ in the \param{niter}-\param{gain}-\param{robust} parameter space and how the optimal imaging parameter combination is determined. 
The shown quality indicator is calculated from the weighted average of  the mean and standard deviation of the  brightness temperature differences with respect to the reference model. Please see Table~\ref{tab:comboweights} for the weight for the  different spatial bins.  
For this illustration, the Band~6 data for the problematic scenario are selected. 
The top row shows the quality indicator in the \param{niter}-\param{gain} plane for three selected \param{robust} values. The colour-coding is chosen across the whole three-dimensional cube (\param{niter}-\param{gain}-\param{robust}), which reveals trends and the best parameter combination for the selected receiver band and scenario.
In this case, it gets clear that the best choice for producing the  brightness temperature is achieved with combination \param{niter}=25000, \param{gain}=0.02, and \param{robust}=0.0 (see white circle in the top row) although other combinations (as visible from similar dark green colours) would almost as good results. 
Please note cases for \param{niter}=0 are not considered for finding the optimal choice as they refer to the dirty maps.  
The same type of diagrams is shown for Band~3 and 6 for the TDRA, TDRV, and TDR+ quality indicators in Figs.~\ref{fig:qiresults_tdpa_b3}-\ref{fig:qiresults_tdpc_b6}. 

The other two rows in    Fig.~\ref{fig:qi_illus} provide alternative visualisations of the TDR+ indicator in the \param{niter}-\param{gain}-\param{robust} space. For this example, an increasing value of \param{niter} often leads to improved results but this clearly depends on the chosen \param{gain} value. Not unexpectedly, it turns out that the choices of \param{niter} and \param{gain} are not fully independent and that different combinations the two parameters can lead to comparable results, as far as measured with the chosen quality indicator. In this particular example, a small value for \param{gain} in combination with larger value for \param{niter} and 
\param{robust=0.0} produces the best results. 
As will be discussed below, the exact behaviour and implications for the best parameter choice depend on the selected receiver band, scenario, spatial bin, and also quality indicator. 

\noindent\paragraph{Band 3}
The results for the quality indicators for the average absolute brightness temperature difference (TDRA), the variation of this difference (TDRV), and the combination of both (TDR+) are displayed in Figs.~\ref{fig:qiresults_tdpa_b3}-\ref{fig:qiresults_tdpc_b3}, respectively.  
The TDRA indicator shows good results for combination of higher \param{niter} (10\,000 - 25\,000), moderate \param{gain} (typically 0.2), and medium \param{robust} (mostly -0.5 - 0.0). 
The TDRV indicator shows the same trend and implies the same choices for optimal imaging parameters. Interestingly, both TDRA and TDRV do not show further improvement in the extreme scenario for higher values of \param{niter} and \param{gain}, indicating the limitations in terms of quality that can be achieved for highly corrupted measurement sets. 
Please note that quality that can be achieved for the extreme scenario is much lower than for the less corrupted scenarios as can be seen from comparing the color bars to the very left of each figure with each other. 
Since the TDRA and TDRV results are similar, also the combined TDR+ leads to the same conclusions for Band~3. 
 
\noindent\paragraph{Band 6}
The corresponding results for Band~6 for the TDRA, TDRV, and TDR+ indicators exhibit much stronger restrictions for the optimal choice of imaging parameters. While a larger  number for \param{niter} generally improves the quality, its deteriorates with increasing \param{gain}. This effect becomes more pronounced for the more difficult the weather scenario is. As a result, TDRA points at a combination of high \param{niter} and low \param{gain}. The same conclusion is derived from the TDRV and TDR+ results for Band~6. As for Band~3, a low to moderate \param{robust} value is preferable, while values larger than 0.0 are not recommended.

\begin{figure}[t!]
    \centering
    \includegraphics[width=12.8cm]{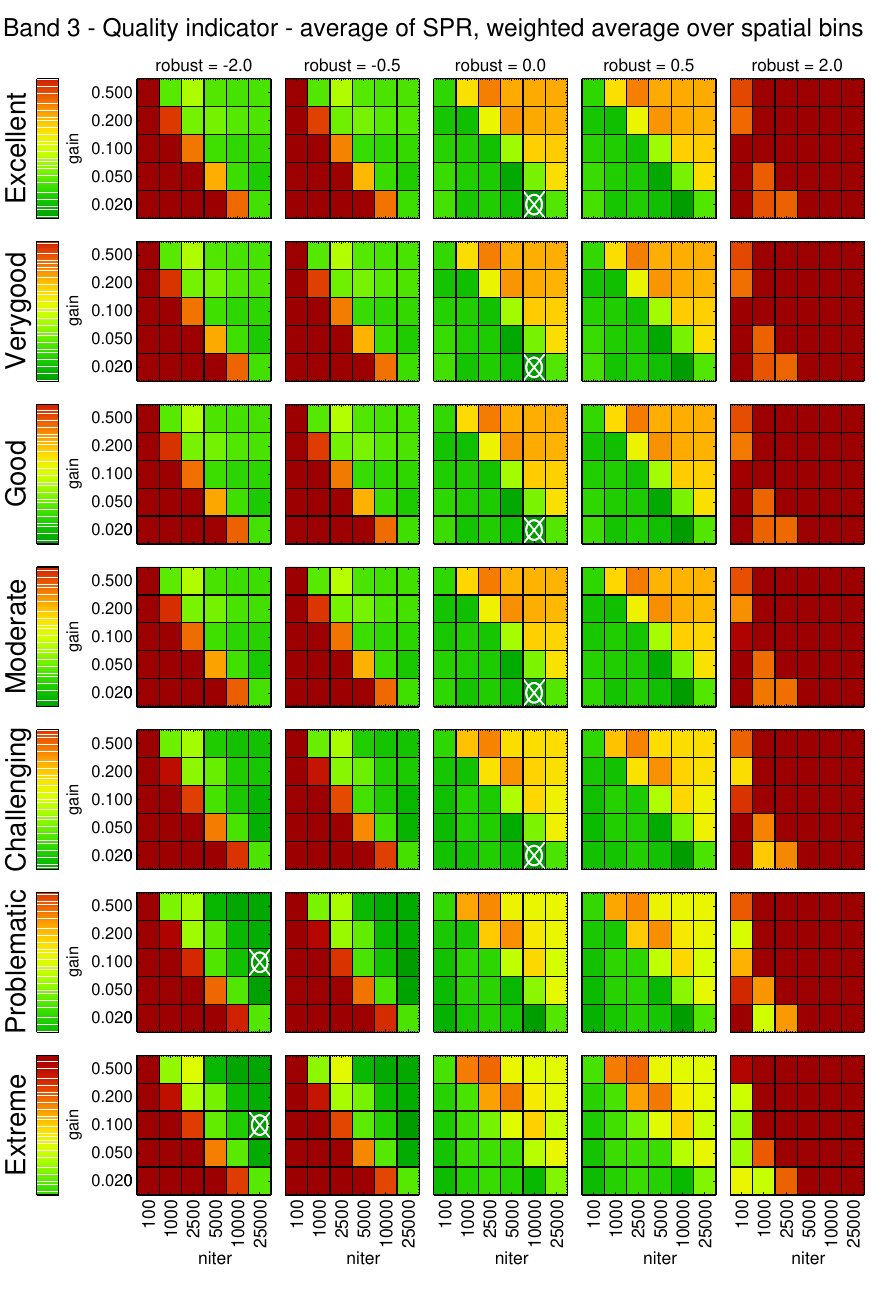}
    \vspace{-6mm}
    \caption{Comparison of the SPRA quality indicator for Band~3, which is based on the ratio of the spatial power spectra for the produced image series and the reference models. See the caption of Fig.~\ref{fig:qiresults_tdpa_b3} for an explanation.}    
    \label{fig:qiresults_spra_b3}
\end{figure}
\begin{figure}[t!]
    \centering
    \includegraphics[width=12.8cm]{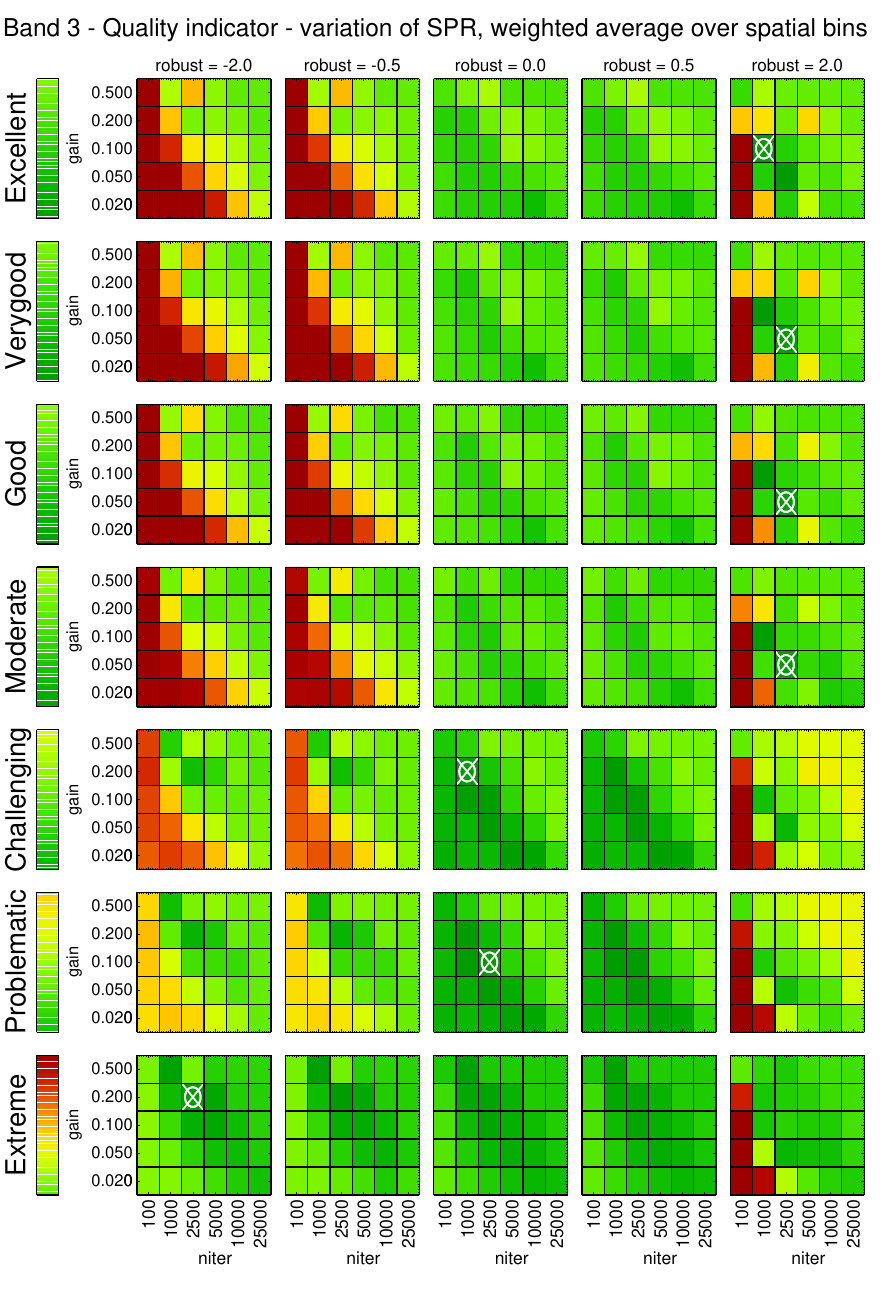}
    \vspace{-6mm}
    \caption{Comparison of the SPRV quality indicator for Band~3, which is based on the ratio of the spatial power spectra for the produced image series and the reference models. See the caption of Fig.~\ref{fig:qiresults_tdpa_b3} for an explanation.}    
    \label{fig:qiresults_sprv_b3}
\end{figure}
\begin{figure}[t!]
    \centering
    \includegraphics[width=12.8cm]{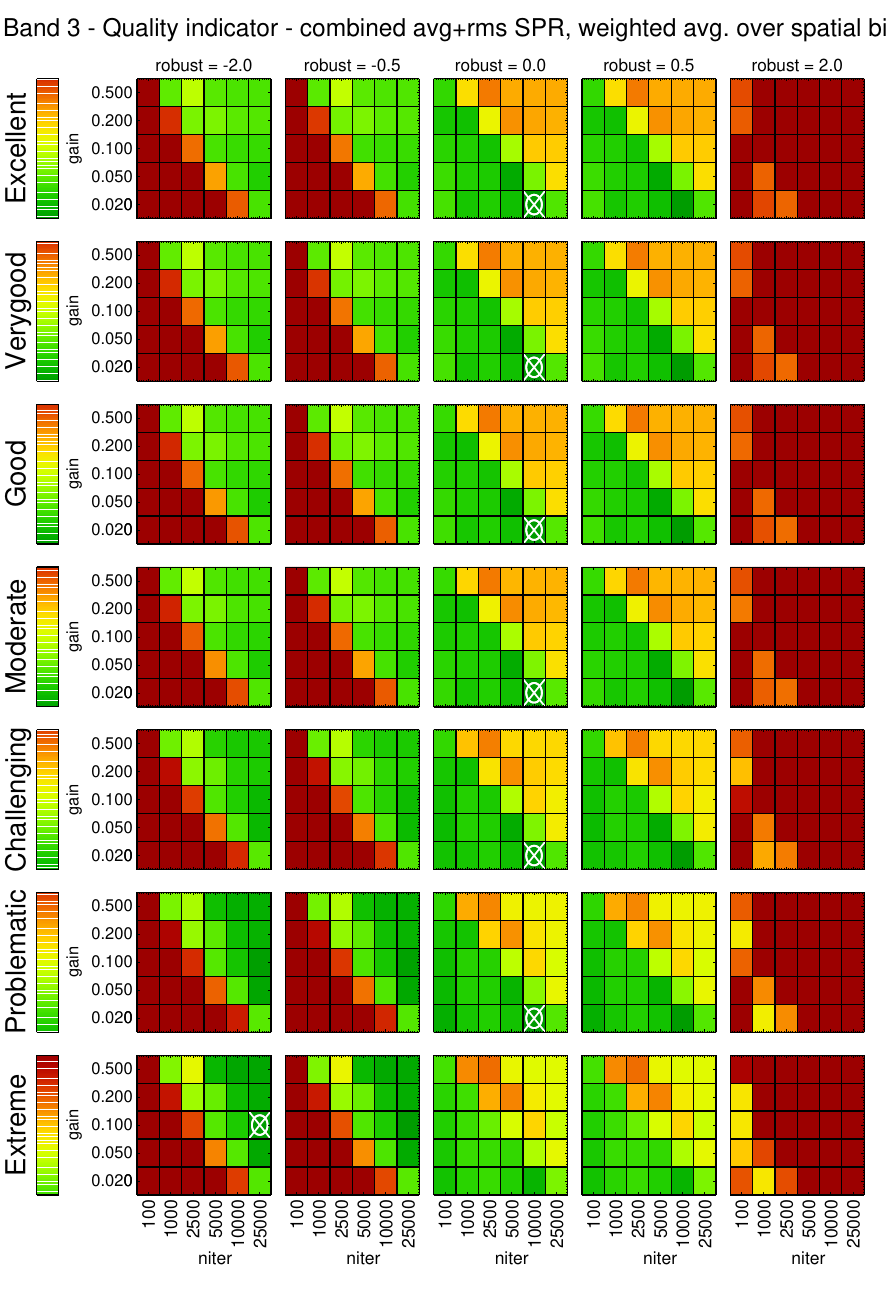}
    \vspace{-6mm}
    \caption{Comparison of the SPR+ quality indicator for Band~3, which is based on the ratio of the spatial power spectra for the produced image series and the reference models. This quality indicator combines the average and variation of the deviations from the reference models in terms brightness temperature values and spatial power spectra in produce image time series. See the caption of Fig.~\ref{fig:qiresults_tdpa_b3} for an explanation. }    
    \label{fig:qiresults_sprc_b3}
\end{figure}
\begin{figure}[t!]
    \centering
    \includegraphics[width=12.8cm]{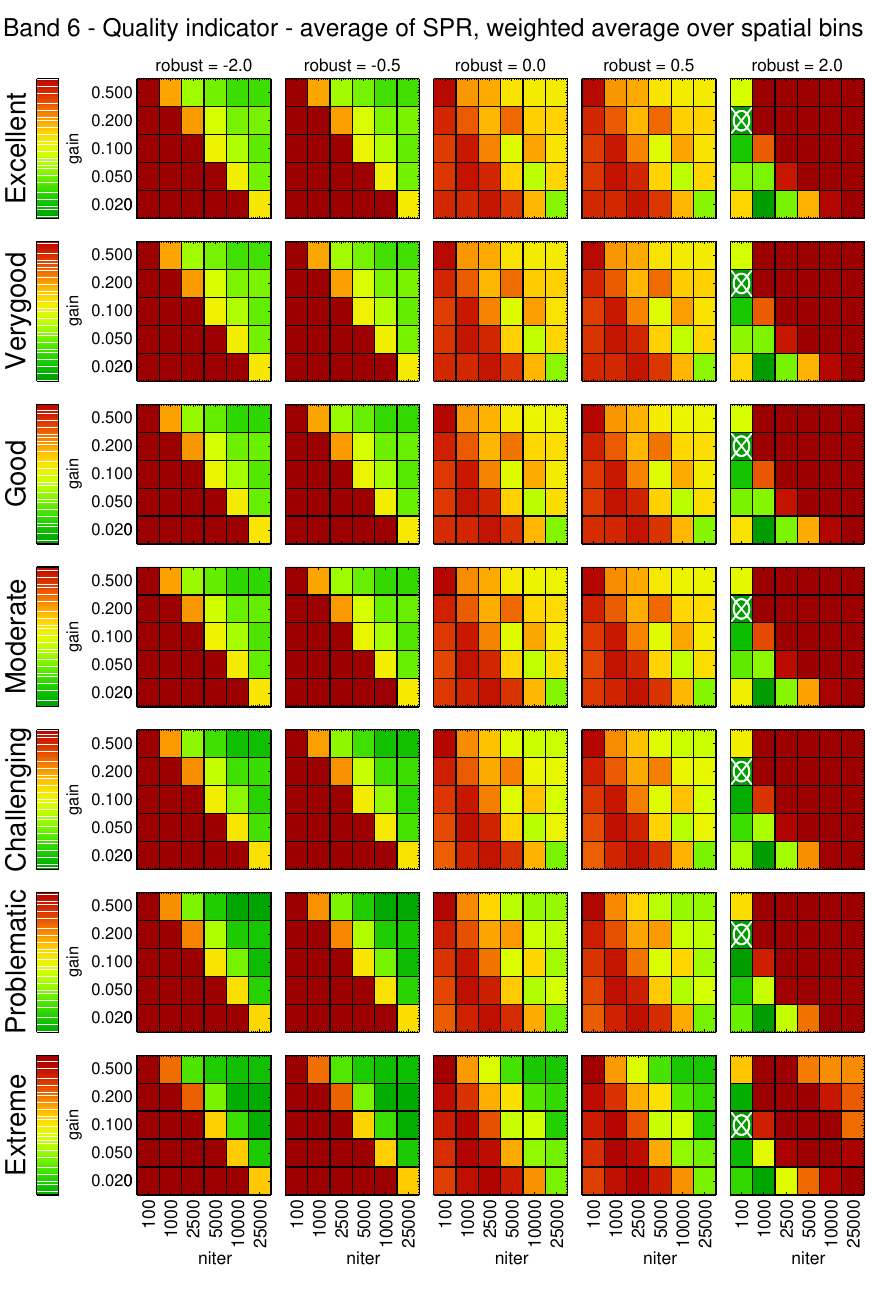}
    \vspace{-6mm}
    \caption{Comparison of the SPRA quality indicator for Band~6, which is based on the ratio of the spatial power spectra for the produced image series and the reference models. See the caption of Fig.~\ref{fig:qiresults_tdpa_b6} for an explanation.}    
    \label{fig:qiresults_spra_b6}
\end{figure}
\begin{figure}[t!]
    \centering
    \includegraphics[width=12.8cm]{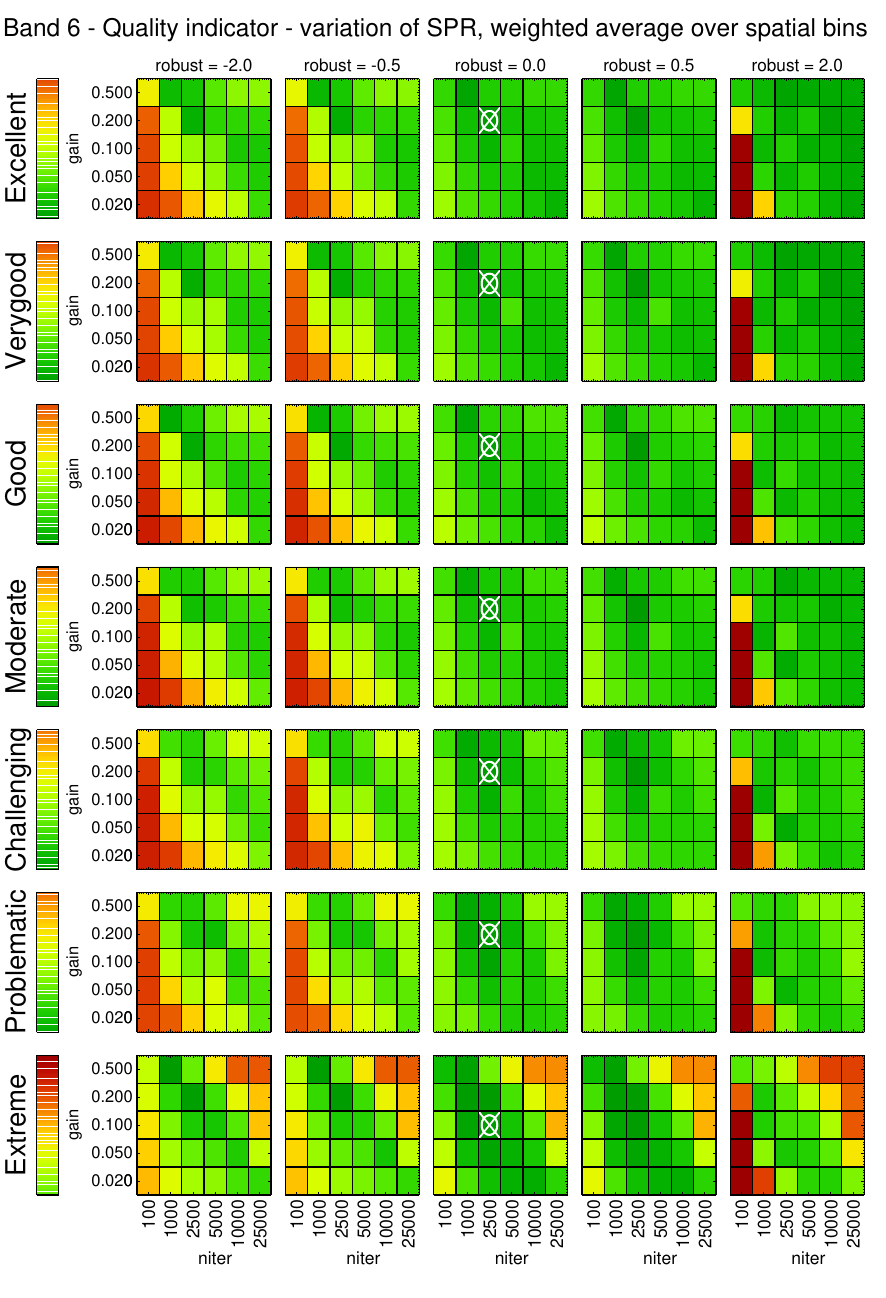}
    \vspace{-6mm}
    \caption{Comparison of the SPRV quality indicator for Band~6, which is based on the ratio of the spatial power spectra for the produced image series and the reference models. See the caption of Fig.~\ref{fig:qiresults_tdpa_b6} for an explanation.}    
    \label{fig:qiresults_sprv_b6}
\end{figure}
\begin{figure}[t!]
    \centering
    \includegraphics[width=12.8cm]{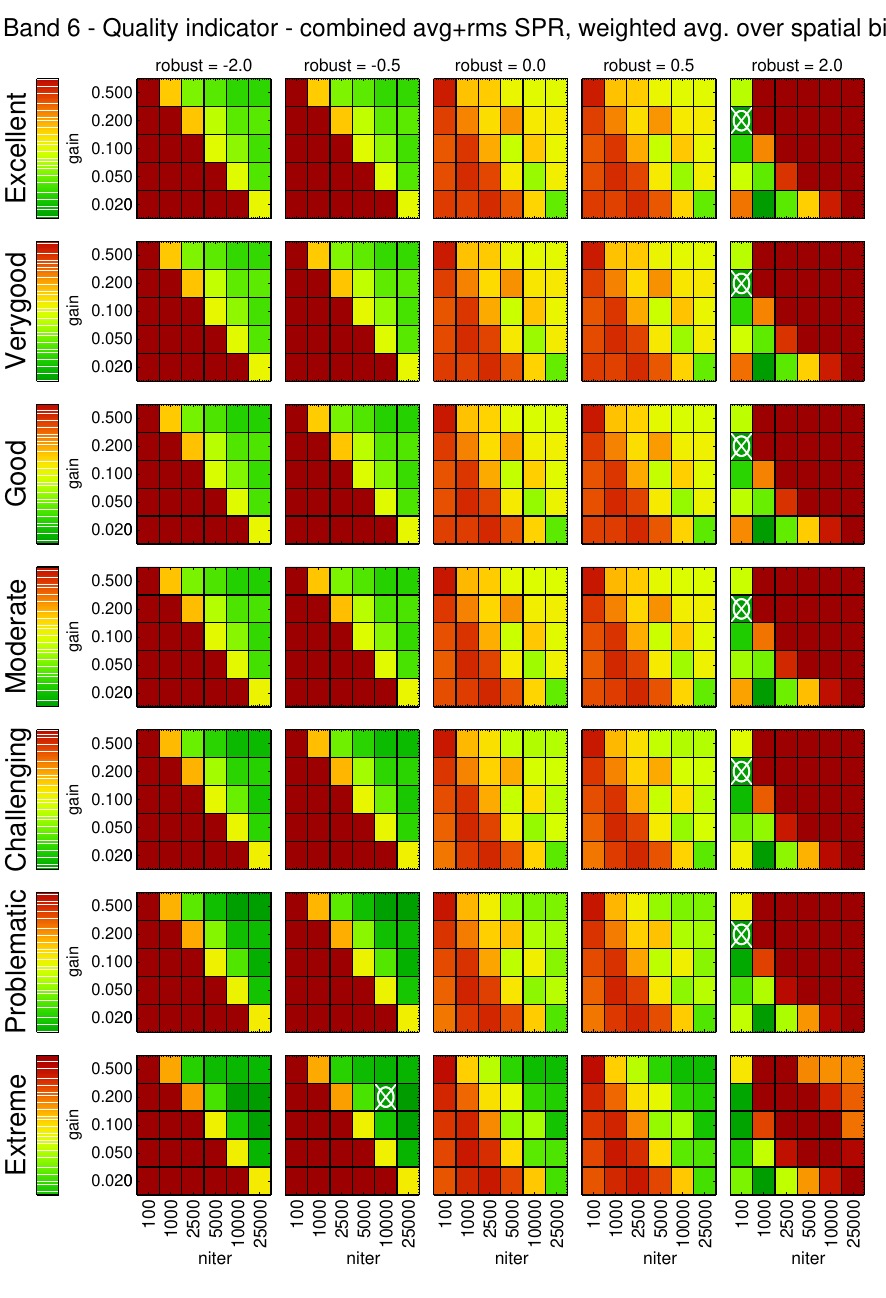}
    \vspace{-6mm}
    \caption{Comparison of the SPR+ quality indicator for Band~6, which is based on the ratio of the spatial power spectra for the produced image series and the reference models. This quality indicator combines the average and variation of the deviations from the reference models in terms brightness temperature values and spatial power spectra in produce image time series. See the caption of Fig.~\ref{fig:qiresults_tdpa_b3} for an explanation.}    
    \label{fig:qiresults_sprc_b6}
\end{figure}

\pagebreak
\subsubsection{Spatial power ratios}
\label{sec:result_qispr}

The resulting SPR  quality indicator values are compared for the different cases in   Figs.~\ref{fig:qiresults_spra_b3}-\ref{fig:qiresults_sprc_b6}. 
The optimum parameter combinations \param{niter}-\param{gain}-\param{robust} based on the SPRC quality indicator are determined in the same way as illustrated for the TDRC indicator in Fig.~\ref{fig:qi_illus} and described in the previous subsection. 

\noindent\paragraph{Band 3}
As for the TDR results, the SPR indicator turns out worst for combinations of low \param{niter} and \param{gain}. 
Please refer to Figs.~\ref{fig:qiresults_spra_b3}-\ref{fig:qiresults_sprc_b3} for the results of the SPR indicators for Band~3.  
This is particularly true for  low or high values for  \param{robust} whereas a medium values generally fairs better. A strong difference to the TDR indicators is, however, that the range of the SPR indicators is much larger across all scenarios, resulting in stronger restrictions for the optimal parameter choices. In particular the SPRA indicator favours a combination of large \param{niter}, low \param{gain}, and \param{robust}=0. An exception are the problematic and extreme scenarios for which a low \param{robust} value and moderate \param{gain} leads to a better reproduction of the average spatial power spectrum. 
In contrast, the SPRV indicator for the less corrupted Band~3 cases implies good results even for high \param{robust} values and not much difference within the \param{niter}-\param{gain} plane as long as the \param{robust} value is not low (i.e. for \param{robust}~$\geq 0$). The problematic and extreme scenario again tend to produce lower quality and are on that level less susceptible to the exact choice of imaging parameters, maybe except for a weak preference for medium \param{niter}-\param{gain} combinations. 
The SPR+ indicator is strongly impacted by the behaviour of the SPRA indicator. Consequently, a combination of moderately high \param{niter}, low \param{gain}, and \param{robust}=0 is recommended. The only notable exception is the generally worse scenario extreme, for which a combination of high \param{niter}, medium \param{gain}, and \param{robust}=-2 might help to recover the spatial power spectrum to some extent.

\noindent\paragraph{Band 6}
The SPRA indicator shows a similar behaviour for Band~6 than for Band~3 for \param{robust}~$< 0$ but implies poor quality even for higher \param{robust} values. Surprisingly, a very low \param{niter} with moderate to high \param{gain} and \param{robust}=2 is indicated as best choice for all scenarios but a possible almost as good alternative could a combination of high \param{niter}, high \param{gain}, and \param{robust}=2. 
The SPRV indicator shows similar results for Band~6 as for Band~3, suggestion a  combination of medium \param{niter}, medium \param{gain}, and \param{robust}=0. 
When combined into the SPR+ indicator, a combination of moderate to high \param{niter}, low \param{gain}, and \param{robust}=0 is recommended for all scenarios with exception for the extreme scenario. In the latter case, a combination of high \param{niter}, \mbox{medium \param{gain}, and \param{robust}=-2 produces better results.}

\begin{figure}[t!]
    \centering
    \includegraphics[width=12.8cm]{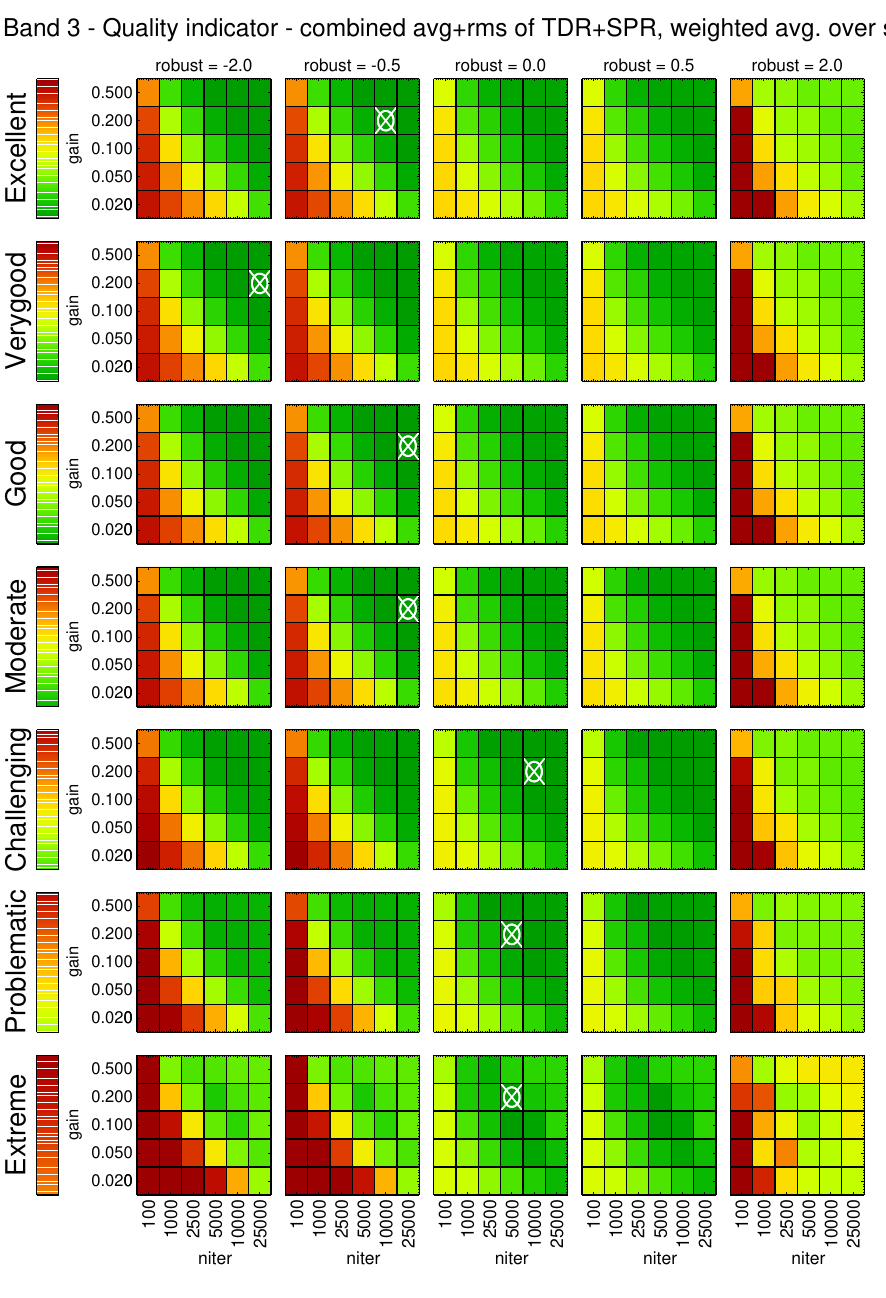}
    \vspace{-6mm}
    \caption{Comparison of the UQI quality indicator for Band~3, which combines the averages and variations weighted across all spatial bins for deviations of the brightness temperature values and the spatial power spectra  for the produced image series with respect to the reference models. See the caption of Fig.~\ref{fig:qiresults_tdpa_b3} for an explanation.}
    \label{fig:qiresults_UQI_b3}
\end{figure}

\begin{figure}[t!]
    \centering
    \includegraphics[width=12.8cm]{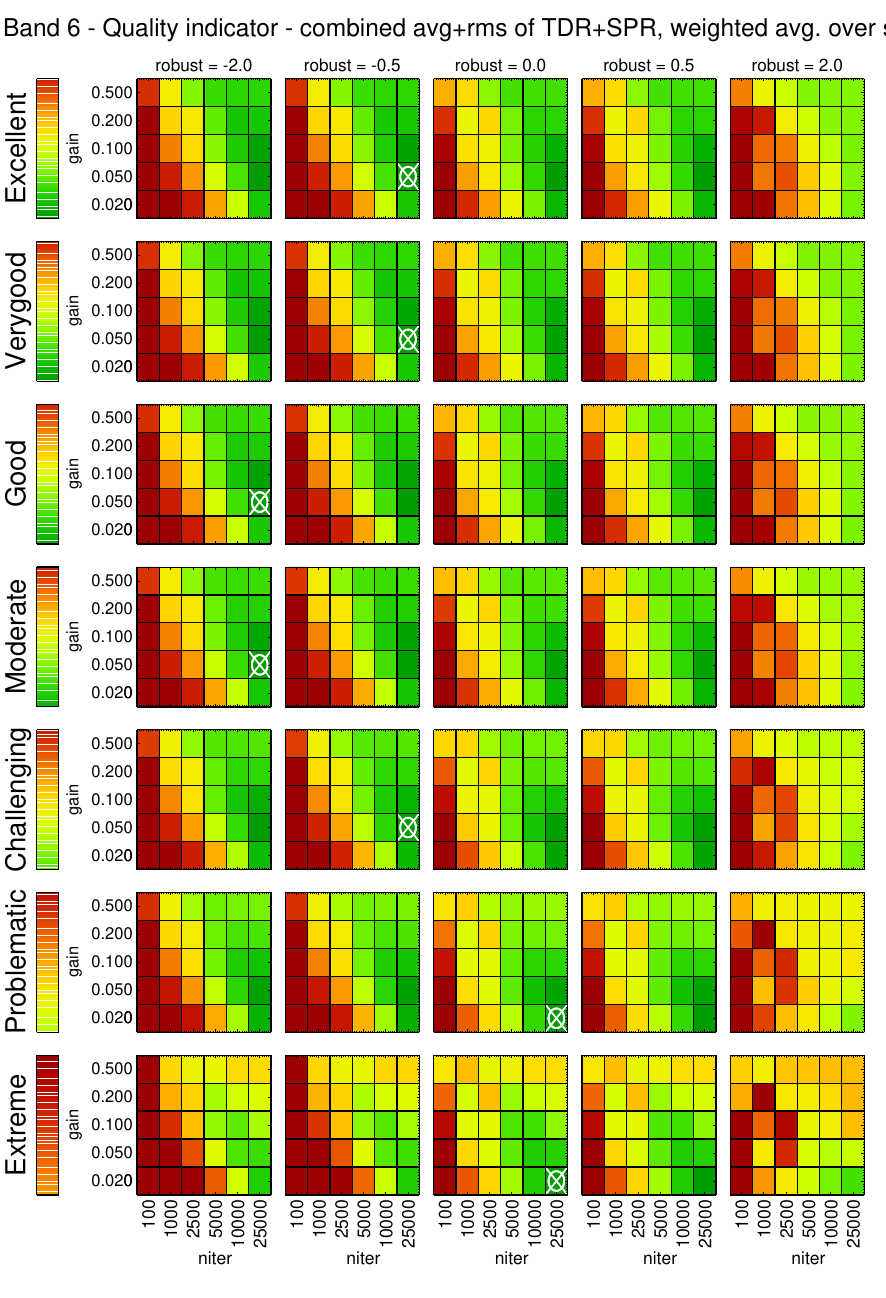}
    \vspace{-6mm}
    \caption{Comparison of the UQI quality indicator for Band~6, which combines the averages and variations weighted across all spatial bins for deviations of the brightness temperature values and the spatial power spectra  for the produced image series with respect to the reference models. See the caption of Fig.~\ref{fig:qiresults_tdpa_b3} for an explanation.}
    \label{fig:qiresults_UQI_b6}
\end{figure}

\clearpage

\begin{table}[b!]
    \centering
    \begin{tabular}{|l|c|c|c|c|c|c|}
    \hline
    \multicolumn{7}{|c|}{\textbf{Band 3}}\\
    \cline{2-7}
    &\multicolumn{2}{c}{Centre}&\multicolumn{2}{|c}{Inner}&\multicolumn{2}{|c|}{Outer}\\
    \cline{2-7}
 &TDRA & TDRV & TDRA & TDRV & TDRA & TDRV\\
\hline
\tool{SoAP} lvl 2    &    -188  &     43  &    -100  &     90  &     228  &     95\\
\tool{SoAP} lvl 3    &    -265  &     70  &    -161  &     87  &     141  &     67\\
Clark         &    -181  &     39  &    -128  &     51  &     250  &    158\\
Hogbom        &    -181  &     39  &    -129  &     51  &     250  &    158\\
mtmfs         &    -177  &     37  &     -85  &     60  &     174  &     85\\
\hline
\multicolumn{7}{|c|}{\textbf{Band 6}}\\
\cline{2-7}
&\multicolumn{2}{c}{Centre}&\multicolumn{2}{|c}{Inner}&\multicolumn{2}{|c|}{Outer}\\
\cline{2-7}
&TDRA & TDRV & TDRA & TDRV & TDRA & TDRV \\
\hline
\tool{SoAP} lvl 2    &       4  &     49  &      13  &     40  &      68  &     34\\
\tool{SoAP} lvl 3    &     215  &     39  &     178  &     46  &     214  &     57\\
Clark         &      97  &     57  &      46  &     57  &      69  &     70\\
Hogbom        &      98  &     57  &      47  &     57  &      70  &     70\\
mtmfs         &     151  &     62  &     115  &     42  &     158  &     63\\
\hline
\multicolumn{7}{c}{}\\[-4mm]
\hline
\multicolumn{7}{|c|}{\textbf{Band 3}}\\
\cline{2-7}
&\multicolumn{2}{c}{SMALL}&\multicolumn{2}{|c}{MEDIUM}&\multicolumn{2}{|c|}{LARGE}\\
\cline{2-7}
& SPRA & SPRV & SPRA & SPRV & SPRA & SPRV \\
\hline
\tool{SoAP} lvl 2    &    
1.28  &   0.36  &    1.07  &   0.22  &    1.05  &   0.12\\
\tool{SoAP} lvl 3    & 
1.02  &   0.14  &    1.05  &   0.20  &    0.99  &   0.07\\
Clark         & 
1.27  &   0.30  &    1.10  &   0.21  &    1.17  &   0.21\\
Hogbom        &   
1.27  &   0.30  &    1.10  &   0.21  &    1.17  &   0.21\\
mtmfs         & 
1.19  &   0.34  &    0.97  &   0.18  &    1.02  &   0.06\\
\hline
\multicolumn{7}{|c|}{\textbf{Band 6}}\\
\cline{2-7}
&\multicolumn{2}{c}{SMALL}&\multicolumn{2}{|c}{MEDIUM}&\multicolumn{2}{|c|}{LARGE}\\
\cline{2-7}
& SPRA & SPRV & SPRA & SPRV & SPRA & SPRV \\
\hline
\tool{SoAP} lvl 2    &   
1.21  &   0.25  &    1.11  &   0.25  &    1.06  &   0.22\\
\tool{SoAP} lvl 3    &  
0.96  &   0.12  &    1.11  &   0.25  &    1.09  &   0.21\\
Clark         &   
1.29  &   0.22  &    1.22  &   0.28  &    1.24  &   0.35\\
Hogbom        &    
1.30  &   0.22  &    1.22  &   0.28  &    1.24  &   0.35\\
mtmfs         &   
1.09  &   0.22  &    1.08  &   0.24  &    1.08  &   0.22\\
 \hline
   \end{tabular}
    \caption{Quality indicators for the tested convolver algorithms with the brightness temperatures differences (TDRA, TDRV) for Band~3 and Band~6 in the upper half and the spatial power ratios (SPRA, SPRV) for Band~3 and Band~6 in the lower half.}
    \label{tab:qi_algo}
\end{table}

\subsubsection{Comprehensive quality assessment based on the Unified Quality Indicator}
\label{sec:res_uqi}

Finally, we  aim to find the best overall parameter combinations that produce the highest image quality while only depending on the receiver band and weather (i.e. phase corruption) scenario. 
For this purpose, the  Unified Quality Indicator is used, which considers the average and variations value for both the TDR and SPR indicators, i.e., TDRA, TDRV, SPRA, and SPRV together (see Sect.~\ref{sec:qiintro_uqi}). 
All of these indicators are combined with equal weight in the UQI (first with equal weights into TDR+ and SPR+ and then with equal weights into UQI).  
The resulting best parameter choices are again from the \param{niter}-\param{gain}-\param{robust} plots (see Figs.~\ref{fig:qiresults_UQI_b3}-\ref{fig:qiresults_UQI_b6}).  The exact UQI values are discussed in Sect.~\ref{sec:recommend_imaging}. 

We note that there are small differences in the best choices based on the TDR+ and SPR+ quality indicators, respectively. This outcome is expected as the two indicators are geared towards best reproduction of brightness temperature amplitudes on the one hand and towards best reproduction of spatial structure on the other hand. 
The differences in the optimum parameter combinations for Band~3 are smaller than for Band~6, reflecting the fact that Band~3 is less susceptible to the impact of weather conditions. 
In principle, one could choose the parameter combinations based on either one of the indicators depending on the requirements for a given scientific goal. 
For this study, however, we will now conclude on the overall best imaging approach. 

For Band~3, the better scenarios (excellent-moderate) favour a low \param{robust} value with high \param{niter} and moderate to high \param{gain}, whereas the more corrupted scenarios prefer a moderate \param{niter}, moderately high \param{gain}, and \param{robust}=0.0. These results are clearly a compromise as the TDR+ indicator favours medium to low \param{robust} and SPR+, on the other hand, rather implies higher \param{robust} values.  
 
For Band~6, the recommended choices for \param{robust} are similar, i.e. low values for excellent to moderate conditions and \param{robust}=0 for more difficult scenarios. In general, a combination of high \param{niter} and low \param{gain} produces imaging results with the best overall quality.

\begin{figure}[t!]
    \centering
    \vspace*{-1mm}
    \includegraphics{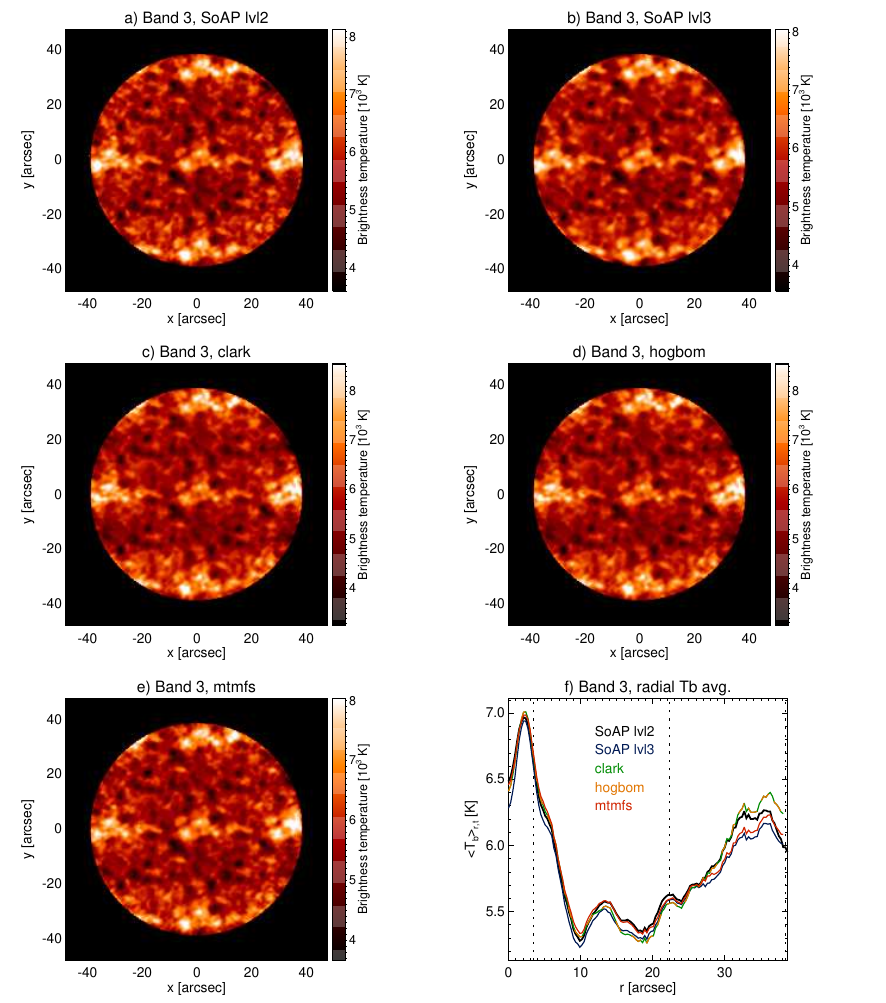}
    \vspace*{-2mm}
    \caption{Band~3 maps produced with different deconvolver algorithms.
    \textbf{a)}~MultiScale-CLEAN as used in \tool{SoAP}~level~2 (the default in this study), \textbf{b)}~MultiScale-CLEAN with self-calibration as used in \tool{SoAP}~level~3, 
    \textbf{c)}~Clark, \textbf{d)}~Hogbom, 
    \textbf{e)}~Multi-Term (Multi-Scale) Multi-Frequency Synthesis (MTMFS)
    \textbf{f)}~Brightness temperatures averages over the time series and radially for the five different cases. }
    \label{fig:algb3}
\end{figure}

\begin{figure}[t!]
    \centering
    \vspace*{-1mm}
    \includegraphics{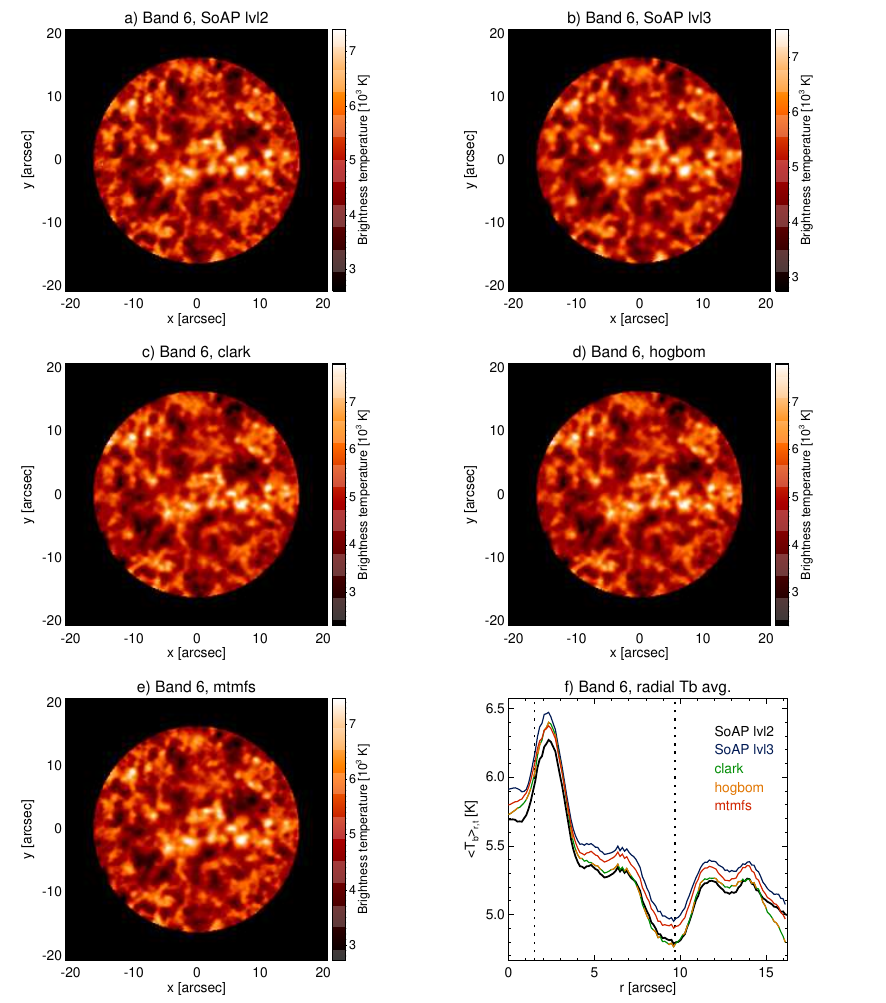}
    \vspace*{-2mm}
    \caption{Band~6 maps produced with different deconvolver algorithms. 
        \textbf{a)}~MultiScale-CLEAN as used in \tool{SoAP}~level~2 (the default in this study), \textbf{b)}~MultiScale-CLEAN with self-calibration as used in \tool{SoAP}~level~3, 
    \textbf{c)}~Clark, \textbf{d)}~Hogbom, 
    \textbf{e)}~Multi-Term (Multi-Scale) Multi-Frequency Synthesis (MTMFS)
    \textbf{f)}~Brightness temperatures averages over the time series and radially for the five different cases. 
    }
    \label{fig:algb6}
\end{figure}
 
\begin{figure}[t!]
    \centering
    \includegraphics{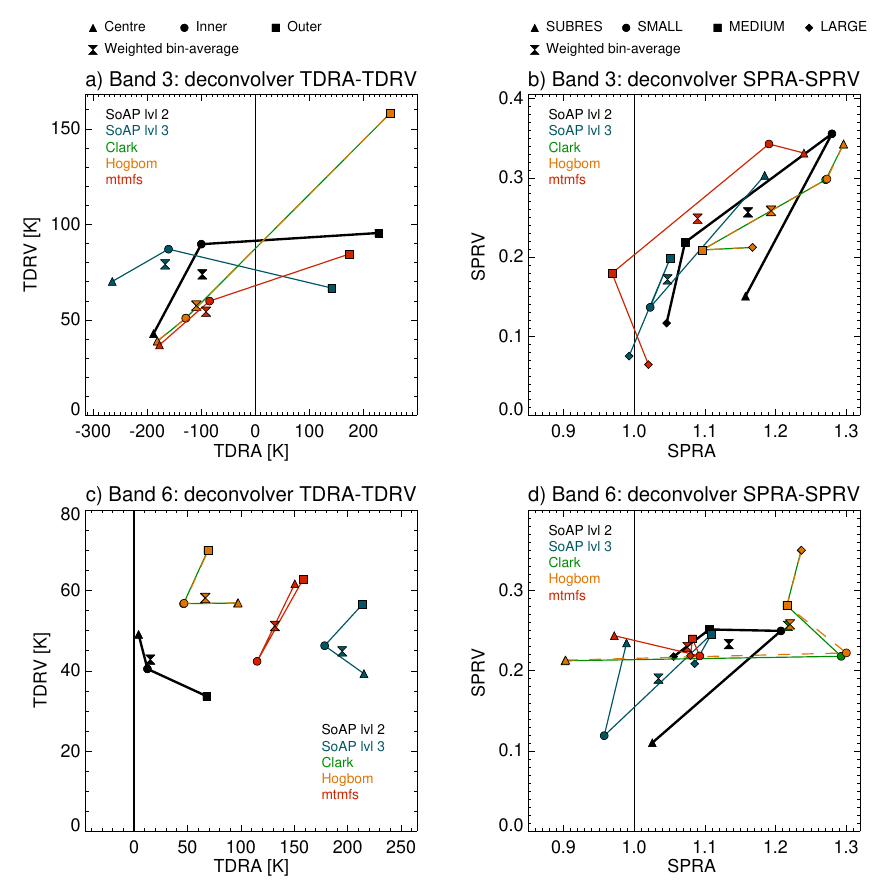}
    \caption{Comparison of the imaging quality indicators for the runs with different deconvolver algorithms for Band~3 (top) and Band~6 (bottom):  TDRA versus TDRV (left column) and SPRA versus SPRV (column). The different colours represent the different cases: 
    MultiScale Clean as used in \tool{SoAP}~level~2 (thick black, the default in this study),  
    MultiScale Clean with self-calibration as used in \tool{SoAP}~level~3 (blue), 
    Clark (green),  Hogbom (orange), Multi-Term (Multi-Scale) Multi-Frequency Synthesis (MTMFS, red).
    The values for the individual bins are represented by different symbols (see legend on top) and connected by lines in the respective colours. The weighted averages that are connected to the TDR+ and SPR+ indicators are marked with hourglass symbols in the  respective colours. The vertical lines mark TDRA=0\,K and SPRA=1.0, respectively. 
    }
    \label{fig:qialgbo}
\end{figure}

\subsection{Algorithm}

The results for the different deconvolver algorithms are compared for a selected time step in Figs.~\ref{fig:algb3} and \ref{fig:algb6}. The radially and temporally averaged brightness temperature profiles are plotted in the lower right panels. At first look, the maps look qualitatively similar. For both Band~3 and Band~6, there are minor differences in the form of  small shifts of the brightness temperature scale and minor features, especially in the outermost parts. This finding is supported by the radial brightness temperature profiles, which are quite similar for the tested cases, especially for Band~3. The deviations increase somewhat with radius but typically stay below $\sim$200\,K. There are more notable deviations for Band~6. \tool{SoAP} level~2 produces the lowest $T_\mathrm{b}$ values for a large range of radii out to $\sim 12$'' where the Clark and Hogbom cases drop to lower values. Like for Band~3, the deviations typically stay below $\sim$200\,K. 
Including self-calibration as in \tool{SoAP} level~3 leads to a systematic shift by on average -73\,K for Band~3 and +162\,K for Band~6, respectively, indicating a notable impact of self-calibration~on~the~results.  

A quantitative comparison of the results is facilitated by means of the time-averaged TDR and SPR quality indicators (see Sect.~\ref{sec:qi_desc}) for the different algorithms. See Table~\ref{tab:qi_algo} and  Fig.~\ref{fig:qialgbo}. 
Please note that the figure contains the quality indicators for all spatial bins but also the weighted averages (marked as hourglass symbols) that are calculated in the same way as in Sect.~\ref{sec:qi_desc} (TDR+, SPR+).

\noindent For Band~3, MTMFS, Hogbom, and Clark   reproduce brightness temperatures equally well although the latter two show a large TDRV value in the Outer region. 
The standard \tool{SoAP} level~2 MultiClean result is performs not quite as well for the Inner region but slightly better than \tool{SoAP} level~3. The situation is quite different for Band~6, for which \tool{SoAP} level~2 is clearly superior and here even produces higher quality than for Band~3. 
The SPRA and SPRV values for the Centre and Inner region reach very low values of 4-13\,K and 40-49\,K, respectively. 
It should be noted though that this test was performed for uncorrupted case and processing Band~6 for scenarios with phase corruption is expected to result in lower quality. 
Surprisingly, including self-calibration deteriorates the results mostly through higher TDRA deviations from the reference model for this case. 
In contrast, \tool{SoAP} level~3 performs best in reproducing the spatial structure for both receiver bands as measured in terms of the weighted cross-bin SPRA and SPRV  indicators (SPR+). 
It is not entirely clear why a slightly larger spread in the results is found for Band~6 but only standard parameters have been used for this experiment. It is highly likely that the spread can be much reduced when the parameters for the different algorithms would be optimised individually. However, that was computationally beyond the scope of this study. 

The results of the presented experiments imply that the imaging quality of \tool{SoAP} level~2 can be increased through the usage of self-calibration. Otherwise, using the MTMFS algorithm also leads to improved reproduction of spatial scale. At least for Band~3, this makes MTMFS an interesting alternative to MultiScale-CLEAN. It seems worth investigating this option in more detail for other PWV~levels in the future.

\subsection{Imaging pixel grid}

\begin{figure}[t!]
    \centering
    \includegraphics[width=15cm]{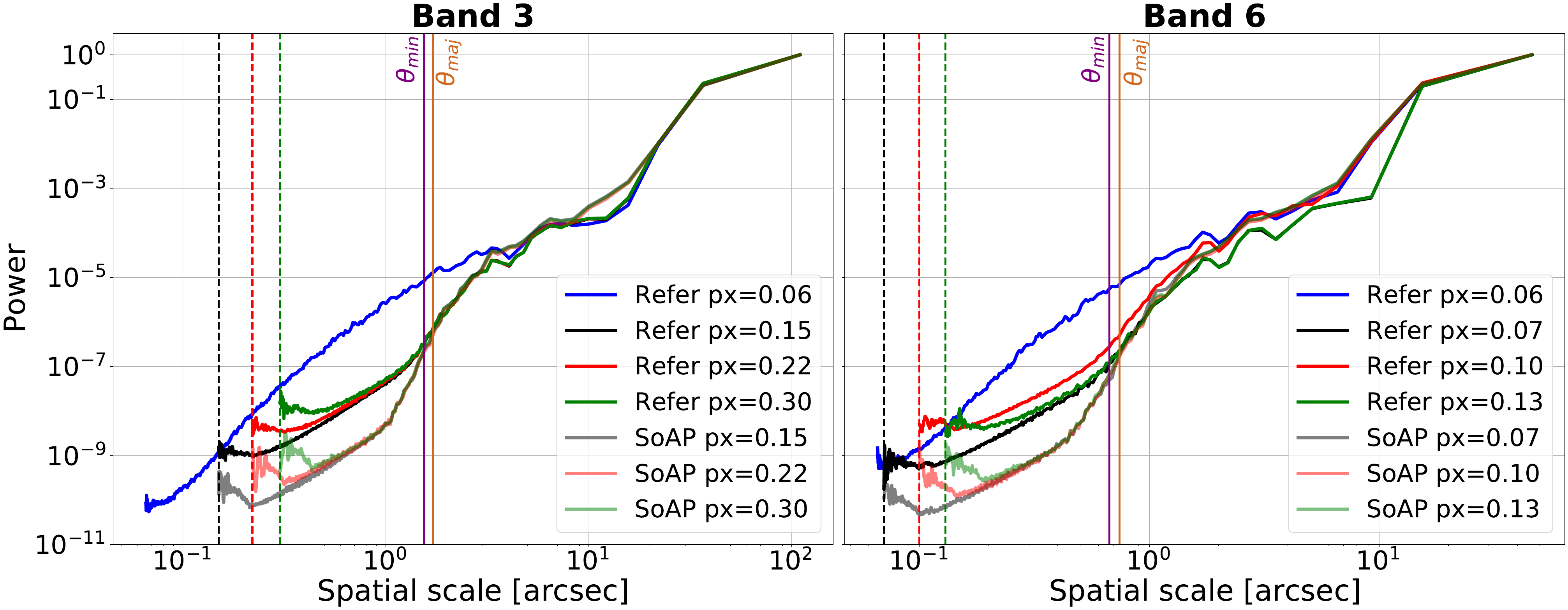}
    \caption{Influence of imaging grid pixel size on the spatial power spectra of the resulting \tool{SoAP} maps for Band~3 (left) and Band~6 (right). The respective reference model is shown at its original resolution (0.066'', blue line) and after convolution with the synthesised beam and resizing to pixel sizes of 0.15'' , 0.22''  and 0.30''  (black, red and green lines, respectively). The lighter shaded black, red and green lines represent the corresponding power spectra of the \tool{SoAP} output images for the same pixel sizes. Vertical dashed black, red and green lines mark the pixel sizes. The minor and major axis of the synthesised beam are represented by the purple and ochre vertical lines respectively. For Band~6 (right panel), the pixel sizes     of 0.07'' , 0.10''  and 0.13''  represented by black, red and green lines, respectively. 
 }
    \label{fig:pixelsize_power}
\end{figure}

Figure~\ref{fig:pixelsize_power} shows the spatial power spectra analysis for the runs with different pixel grids (see Sect.~\ref{sec:setup_pixelgrid}). 
For both Band~3 and 6, the original reference models show high power and a smooth slope for the covered spatial scale range. 
When the reference models are resized to the \tool{SoAP} pixel sizes and convolved with the synthesised  beams, their power decreases. This change is as expected most notable at smaller spatial scales and in particular for scales below the beam size and thus below the nominal angular resolution of the simulated array configuration. The \tool{SoAP} outputs behave similarly to the resized reference models for spatial scales larger than the beam size. However, for spatial scales smaller than the respective beam size, the power slope in the \tool{SoAP} output becomes  steeper, indicating the strong decrease of resolved structure towards smaller scales. 
It should be noted here that the interferometric images are reconstructed from sampling discrete points in the u-v space and therefore information at scales smaller than the beam size is not fully recovered. 
The power spectra for all test pixel sizes match and meet an inflection point at $\sim$1.0'' for Band~3 and $\sim$0.45'' for Band~6. There the power decreases with a shallower but still significant sizes until the different cases successively reach the spatial scale range, where additional noise is add from operations such interpolation on the discrete pixel grid. It is important to note here the power spectra for the different pixel sizes match essentially perfectly for the scales smaller than the synthesised beam size and even beyond the inflection point into the noise domain. 
In conclusion, there seems to be no notable gain from smaller pixel sizes, neither for Band~3 nor for Band~6, that would justify the increase in data volume  corresponding to the increase in number of pixels.

\clearpage
\subsection{Integration time}

\begin{figure}[th!]
    \centering
    \includegraphics{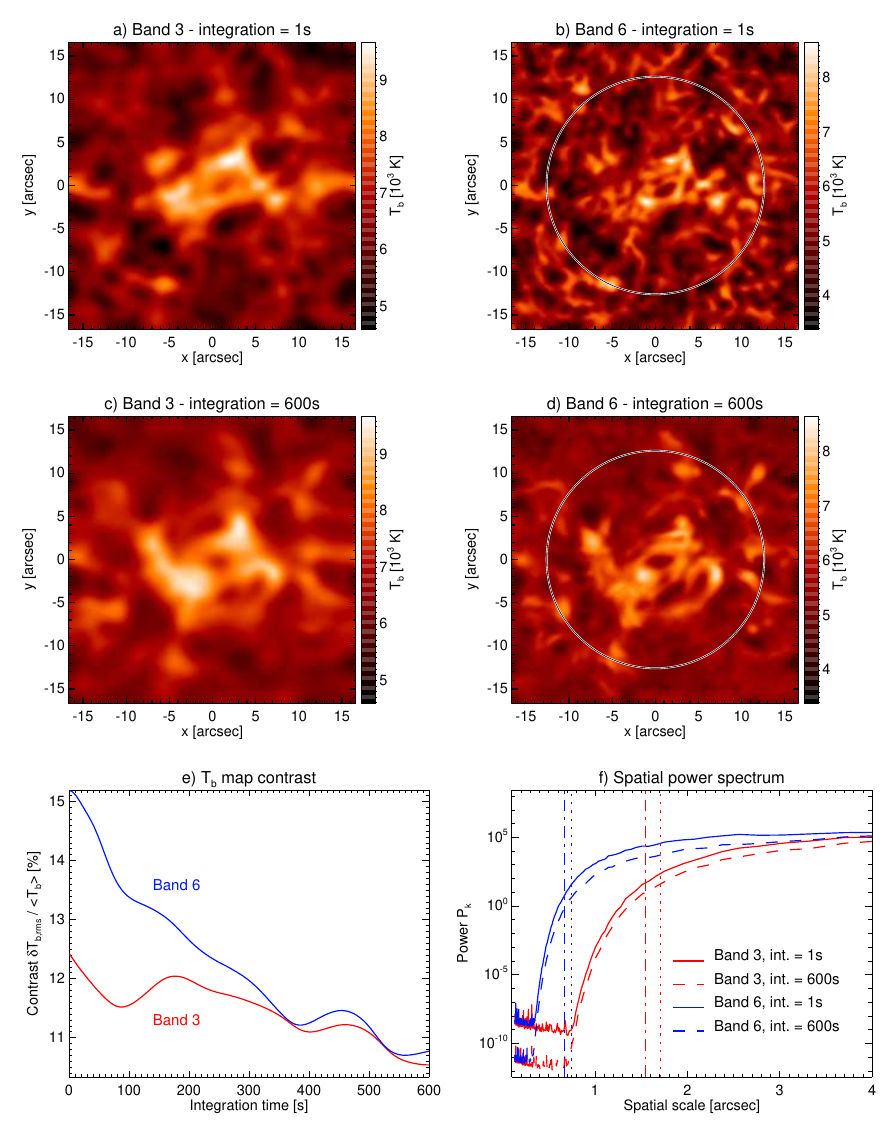}
    \caption{Simulated \tool{ART} maps for a) Band~3 and b) Band~6 for an integration time of 1s compared the maps integrated over 600s of the time series (panels c and d). The primary beam is marked for Band~6, while the Band~3 beam is larger than the size of the \tool{ART} map. e) The resulting brightness temperature contrast as function of integration time; f) Spatial power spectra for both bands and 1s and 600s integration time, respectively. The angular resolution given by the synthesised beam is marked for each band as vertical lines (minor axis: dot-dashed, major axis: dotted). The power spectra are limited to small scales of less than 4'' .}
    \label{fig:integrationtime}
\end{figure}

Observing a source over an extended time range while following its movement across the sky, as done for Earth rotation synthesis, results in better u-v coverage and thus better imaging results. 
However, this approach should only by applied for sources that do not change notably during the course of the observation. 
In this regard, the Sun is a particular case. Many scientific applications of solar observations with ALMA aim at the dynamics and structure at small spatial scales, which are ultimately connected to short timescales too. 
The \tool{ART} millimeter maps (see Sect.~\ref{sec:artcalc}) that are used as reference models for the parameter study are employed for illustrating the need for short integration times. In Fig.~\ref{fig:integrationtime}a-b, maps for Band~3 and Band~6 with 1\,s integration time are compared to corresponding maps that are integrated over 600\,s (Fig.~\ref{fig:integrationtime}c-d), which is the typical length of a solar ALMA scan. The displayed brightness temperature ranges are kept the same, although separately for Band~3 and 6. The maps integrated over 600s appear to have less contrast as a result of a narrower brightness temperature distribution. This effect is easily explained with the dynamics on small spatial scales that occur on short timescales. Consequently, increasing the integration times results in smearing out the small-scale dynamics. The resulting decrease in the contrast of a  brightness temperature map ($\delta T_\mathrm{b,rms}/<T_\mathrm{b}>$ ) as function of integration time is shown in Fig.~\ref{fig:integrationtime}e. For Band~3, the contrast drops from 12.4\,\% to 10.5\,\%, while the contrast changes from 15.2\,\% to 10.8\,\% for Band~6 when increasing the integration time from 1\,s to 600\,s. 
Interestingly though, the differences in the (radially averaged) spatial power spectrum for the maps with 1s and 600s integration times are not very pronounced (see Fig.~\ref{fig:integrationtime}f). A longer integration time does lead to a decrease in spatial power at small scales, including a turnoff point at a slightly larger spatial scale but otherwise closely resembles the power spectrum of the corresponding model for 1\,s integration time. The power spectrum of a map derived from observational data integrated over one scan or even over all related scans, as it is the case for the reference image produced with the imaging scripts distributed together with observational data to PIs, may thus look reasonable whereas much of the wanted small-scale structure and dynamics is lost, rendering that imaging product unusable with a majority of science cases. While good u-v coverage is always a concern, high cadence is crucial for the solar observing mode. 

\section{Results for ultra-high-cadence sequences}

\begin{figure}[t!]
    \centering
    \includegraphics[width=14.5cm]{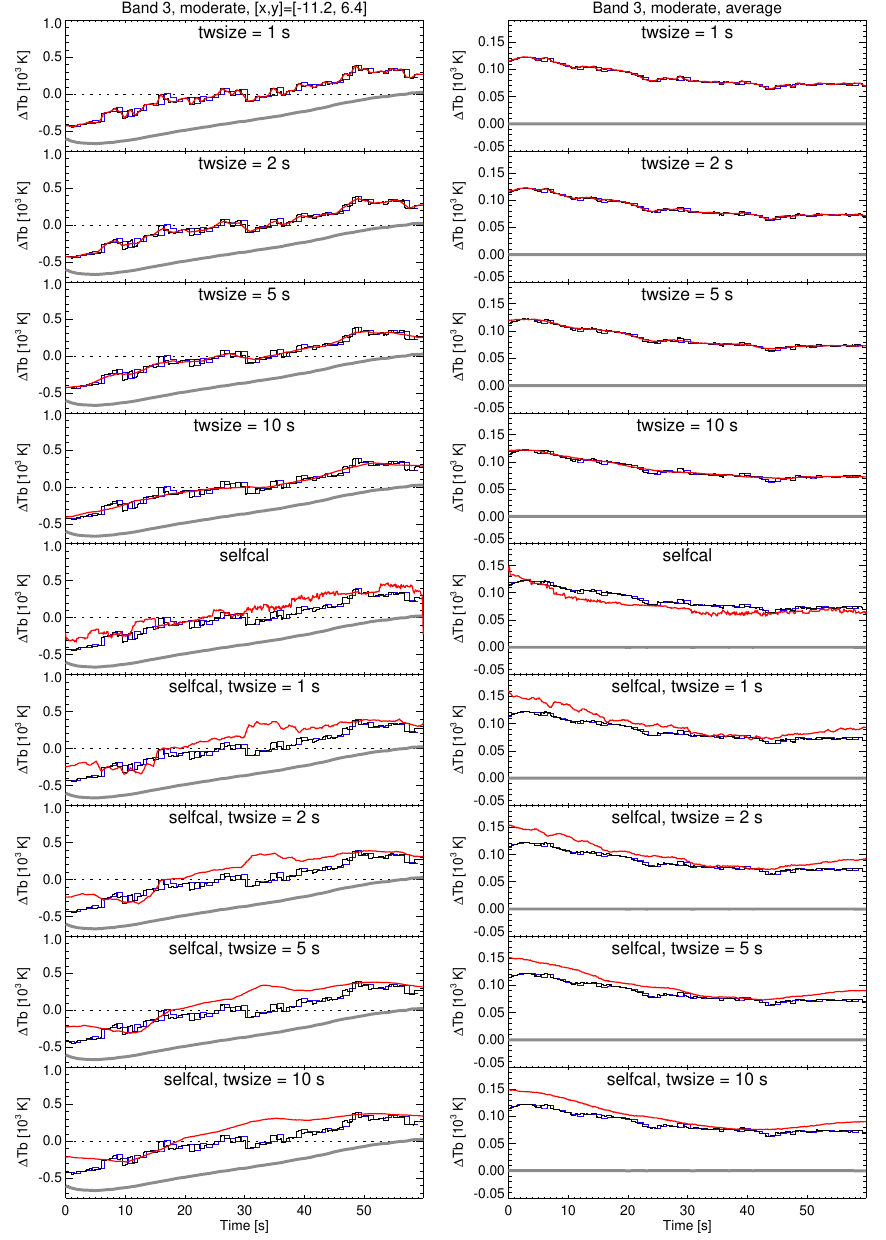}
    \caption{Brightness temperature as function time for Band~3 for the moderate weather scenario when exploiting the time domain to improve the data quality. 
    Left: Selected pixel with coordinates given above the topmost panel. Right: Horizontal average (for all pixels within the mask). 
    The rows show the results (red lines) for different choices of the twsize parameter in seconds with and without prior self-calibration. 
    For comparison, the results without applying a running window (twsize=0, no self-cal.) for the sequences with $\Delta t=0.1$\,s (black) and $\Delta t=1$\,s (blue) and the convolved reference model (grey) are shown. }
    \label{fig:tbcomp_twsizeselfcal_b3moderate}
\end{figure}

\begin{figure}[t!]
    \centering
    \includegraphics[width=14.5cm]{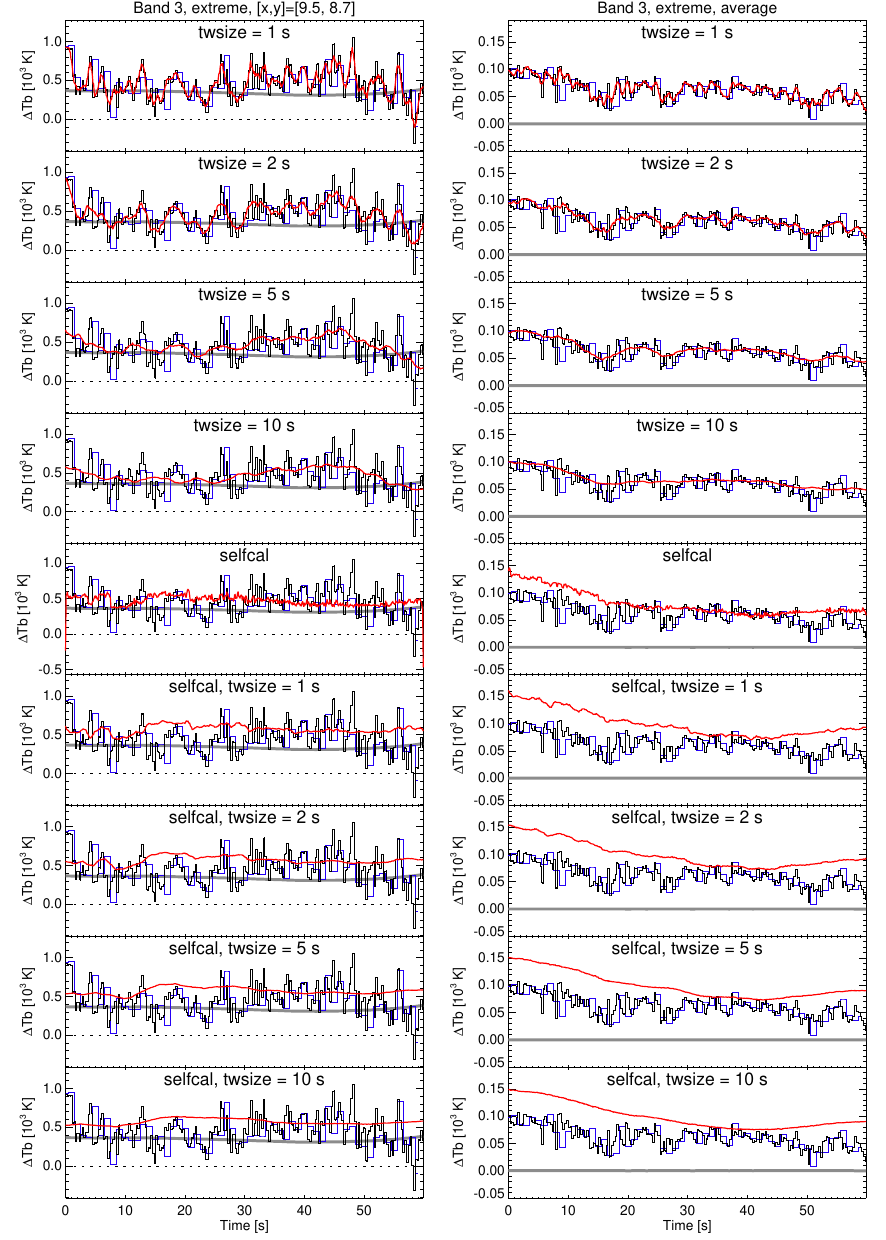}
    \caption{Brightness temperature as function time for Band~3 for the extreme weather scenario when exploiting the time domain to improve the data quality. 
    Left: Selected pixel with coordinates given above the topmost panel. Right: Horizontal average (for all pixels within the mask). 
    The rows show the results (red lines) for different choices of the twsize parameter in seconds with and without prior self-calibration. 
    For comparison, the results without applying a running window (twsize=0, no self-cal.) for the sequences with $\Delta t=0.1$\,s (black) and $\Delta t=1$\,s (blue) and the convolved reference model (grey) are shown. }
    \label{fig:tbcomp_twsizeselfcal_b3extreme}
\end{figure}

\begin{figure}[t!]
    \centering
    \includegraphics[width=14.5cm]{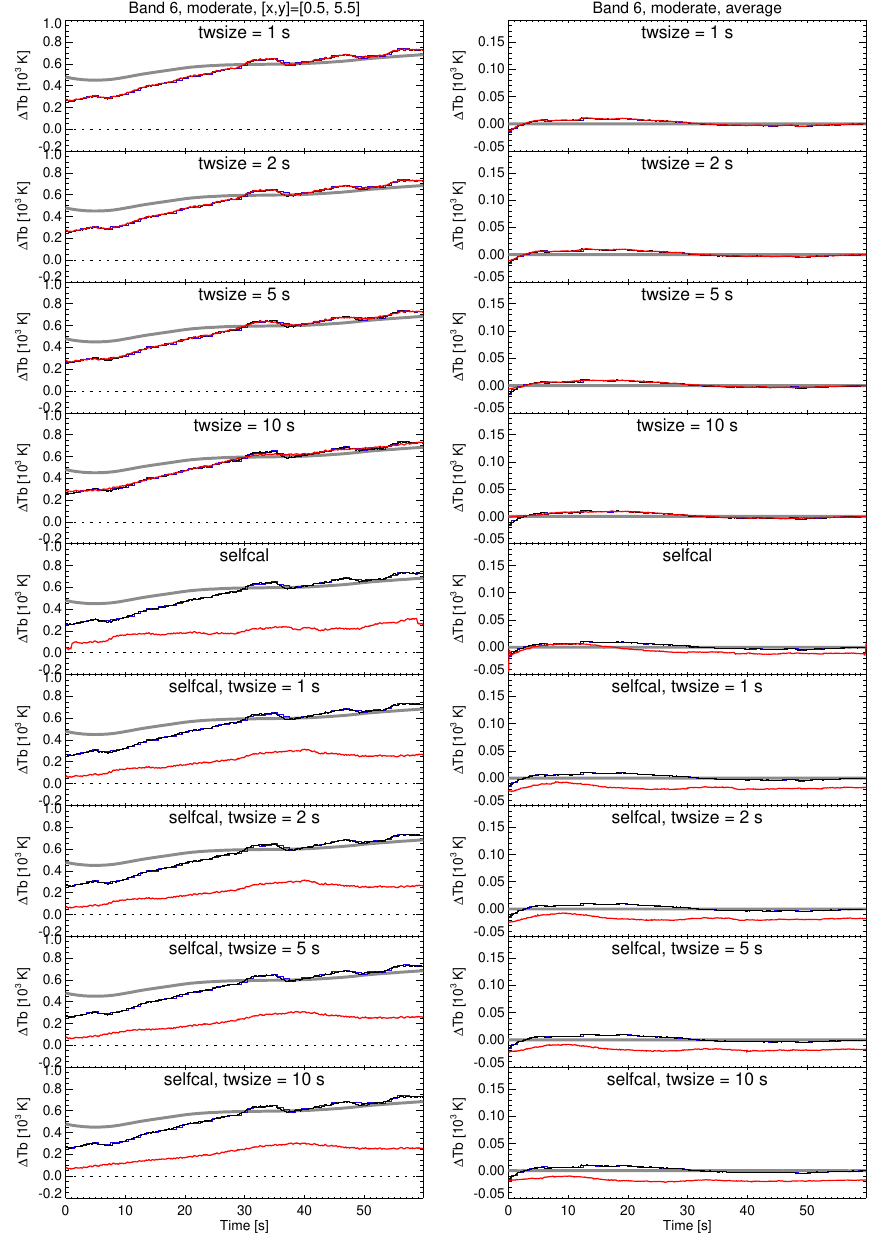}
    \caption{Brightness temperature as function time for Band~6 for the moderate weather scenario when exploiting the time domain to improve the data quality. 
    Left: Selected pixel with coordinates given above the topmost panel. Right: Horizontal average (for all pixels within the mask). 
    The rows show the results (red lines) for different choices of the twsize parameter in seconds with and without prior self-calibration. 
    For comparison, the results without applying a running window (twsize=0, no self-cal.) for the sequences with $\Delta t=0.1$\,s (black) and $\Delta t=1$\,s (blue) and the convolved reference model (grey) are shown. }
    \label{fig:tbcomp_twsizeselfcal_b6moderate}
\end{figure}

\begin{figure}[t!]
    \centering
    \includegraphics[width=14.5cm]{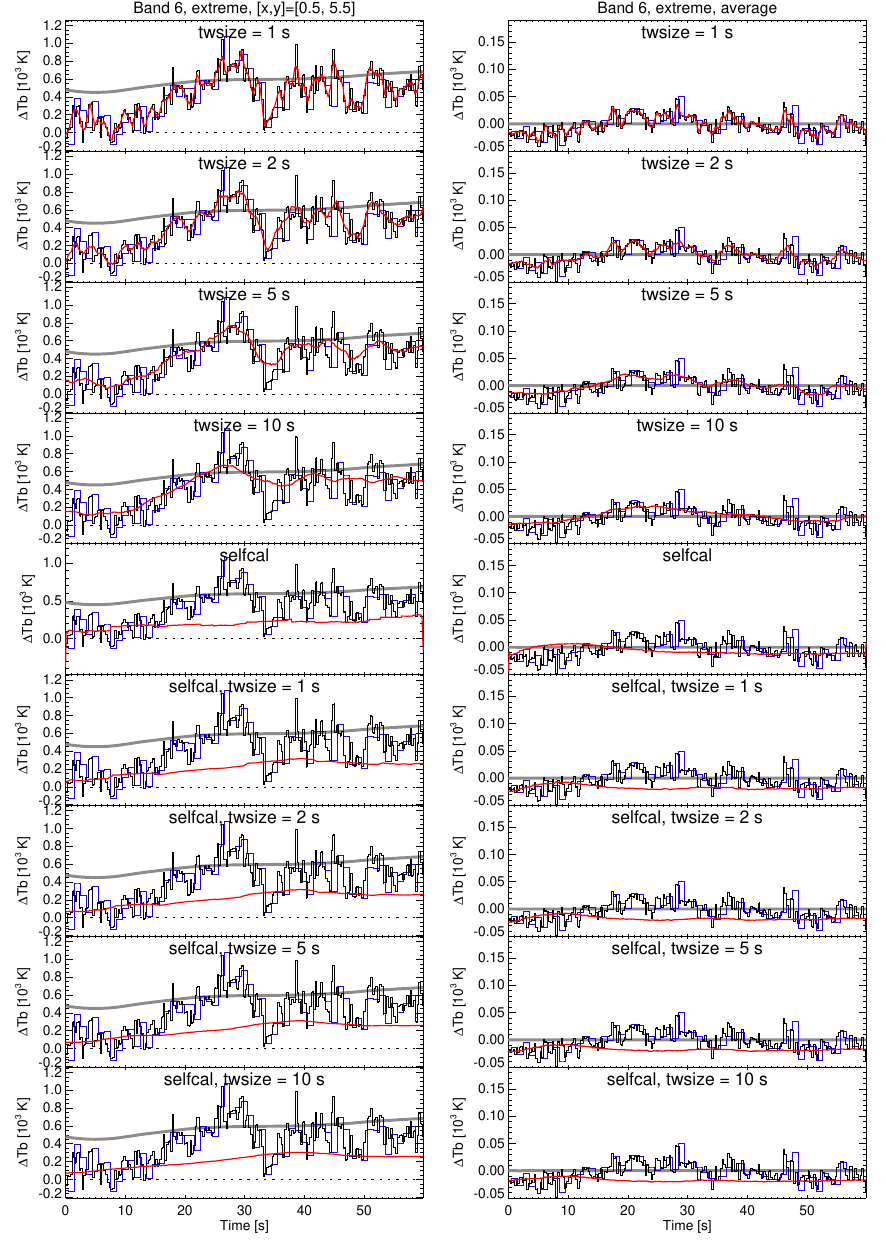}
    \caption{Brightness temperature as function time for Band~6 for the extreme weather scenario when exploiting the time domain to improve the data quality. 
    Left: Selected pixel with coordinates given above the topmost panel. Right: Horizontal average (for all pixels within the mask). 
    The rows show the results (red lines) for different choices of the twsize parameter in seconds with and without prior self-calibration. 
    For comparison, the results without applying a running window (twsize=0, no self-cal.) for the sequences with $\Delta t=0.1$\,s (black) and $\Delta t=1$\,s (blue) and the convolved reference model (grey) are shown. }
    \label{fig:tbcomp_twsizeselfcal_b6extreme}
\end{figure}

For comparison with the imaging runs with sliding time windows and/or self-calibration, first the results from the basic imaging runs are discussed in Sect.~\ref{sec:resbasic}. 
The resulting changes of the brightness temperature time series are visualised in Figs.~\ref{fig:tbcomp_twsizeselfcal_b3moderate}-\ref{fig:tbcomp_twsizeselfcal_b6extreme} for Band~3 and Band~6 for the moderate and extreme scenario.  
The results are compared to the (uncorrupted) reference case and the corresponding measurement sets for the original cadence of  1\,s for a selected pixel position in the inner field-of-view (FOV) and for the brightness temperature averaged over the FOV.  
The results for the sliding time windows, for self-calibration, and a combination thereof are discussed in \mbox{Sects.~\ref{sec:reswindow}
-\ref{sec:resselfcalplus}}, respectively, and then compared in Sect.~\ref{sec:rescomp}.

\subsection{Basic imaging}
\label{sec:resbasic}

Both the results of standard imaging for the data at 1\,s and at 0.1\,s cadence show a noise component with fluctuations on the order of 100\,K for the moderate scenario for Band~3 when considering individual positions in the FOV which strongly suppressed when averaging across the FOV (see Fig.~\ref{fig:tbcomp_twsizeselfcal_b3moderate}). In addition, there is typically an offset of a few 100\,K with respect to the reference model, which varies from position to position. An offset of $\sim 100$\,K  still persists in the spatially averaged brightness. While no exact match with the reference model is expected given that the simulated observations have by nature a limited $uv$-coverage, this potentially systematic offset should be further investigated and taken into account for the future development of imaging algorithms for solar ALMA data. This effect is also visible in the moderate scenario for Band~6 although much less pronounced (see Fig.\ref{fig:tbcomp_twsizeselfcal_b6moderate}). The deviation from the reference is even reduced to $\sim 10$\,K in the spatially averaged brightness temperature. 

By design, the extreme scenario does exhibit a higher level of remaining variations and thus deviation from the reference model 
(see Figs.~\ref{fig:tbcomp_twsizeselfcal_b3extreme} and \ref{fig:tbcomp_twsizeselfcal_b6extreme}). 
For Band~3, again an offset with respect to the reference model remains in the spatially averaged brightness temperature, amounting to 50-100\,K. 
For Band~6, the averaged brightness temperature fluctuates around the reference model with variations of only around a few 10\,K.

\subsection{Sliding time windows}
\label{sec:reswindow}

The top four rows of Figs.~\ref{fig:tbcomp_twsizeselfcal_b3moderate}-\ref{fig:tbcomp_twsizeselfcal_b6extreme} show the brightness temperature (red curves) that results from applying sliding time windows with widths from 1\,s to 10\,s for selected positions in the FOV and for the spatially averaged maps. 
The gain with a 1\,s window applied to the 0.1\,s-cadence data as compared to the unaltered 1\,s-cadence data is notable but small for moderate scenario for Band~3 and even smaller for Band~6. This outcome is expected as basic imaging is already performing reasonably well under these conditions, maybe except for the systematic offset for Band~3.
A longer time window does not result in any notable improvement for the tested Band~6 case but further suppresses the noise visible for Band~3 (Fig.~\ref{fig:tbcomp_twsizeselfcal_b3moderate}). 

The benefits of applying sliding time windows are clearly larger for the extreme scenario. Here, applying a  1\,s window  to the 0.1\,s-cadence data already shows less noise as compared to the unaltered 1\,s-cadence data (Figs.~\ref{fig:tbcomp_twsizeselfcal_b3extreme} and \ref{fig:tbcomp_twsizeselfcal_b6extreme}). 
Increasing the width of the time window then substantially reduces the noise and thus the deviation from the reference model for both considered receiver bands. The brightness temperatures deviations remain on a level of a few 100\,K for individual pixels over time so that scientific studies need to consider this error when analysing such time series. On the positive side, however, measurement sets that are obtained under such extreme conditions still can produce scientifically valuable data with the help of sliding time windows. 

\subsection{Self-calibration}
\label{sec:resselfcal}

The corrupted measurement sets for the moderate and extreme scenario for Band~3 and 6 were processed with \tool{SoAP} with self-calibration (\tool{SoAP} level 3, see Sect.~\ref{sec:slidingtimewindow} and the \textit{Tech.~Doc.}, Wedemeyer et al. 2023). 
The resulting time variation of brightness temperatures calculated with self-calibration are shown in the middle row of Figs.~\ref{fig:tbcomp_twsizeselfcal_b3extreme} and \ref{fig:tbcomp_twsizeselfcal_b6extreme}. 
There is only little improvement for the Band~3 moderate scenario with remaining small fluctuations on the level of what is produced with basic imaging for the 1\,s and 0.1\,s cadence data. 
In contrast, there is a notable reduction in noise for the Band~3 extreme scenario, clearly demonstrating the advantage of applying self-calibration under such conditions and thus salving otherwise strongly affected measurement sets. However, a small but systematic offset with respect to the reference model remains. In the spatially averaged brightness temperature maps, the offset at the beginning of the time series is almost 150\,K but decreases during the first 30\,s and then levels off to a value of $\sim 70$\,K. 
As before, there is not much gain for the Band~6 moderate scenario as the noise level was low to start with. 
In contrast, the Band~6 extreme scenario immediately benefits from a strong reduction in noise and a very low offset in the spatially averaged brightness temperature maps. For individual pixels, however, offsets of a few 100\,K with respect to the reference model must be expected. 
It is important to note that the deviations and thus the final systematic offsets are measured with respect to a reference model that in reality never can be matched by data obtained with an interferometric array with a finite number of baselines. The reference model corresponds to a single-dish telescope with the same angular resolution as the synthetised beam of the simulated array but with perfect $uv$-coverage and no atmospheric or instrumental degradation (see Sect.~2.6 in the \techdoc). We recommend therefore to review in a future study if more suitable reference models can be constructed and/or if an algorithm can be devised that corrects for the brightness temperature offsets that might likely be due to the nature of observing a complex source like the Sun with an interferometric array.

\subsection{Self-calibration plus sliding time window}
\label{sec:resselfcalplus}

The self-calibrated measurement set produced with \tool{SoAP} level~3 is now used as input for imaging with sliding time windows. 
The resulting time variation of brightness temperatures calculated with a combination of self-calibration and sliding time windows are shown in the lower four rows of Figs.~\ref{fig:tbcomp_twsizeselfcal_b3extreme} and \ref{fig:tbcomp_twsizeselfcal_b6extreme}. 
Applying a time window of 1\,s width to the Band~3 moderate data has no obvious benefits and can lead even result in slightly higher deviations from the reference model as compared to the self-calibrated data without sliding time windows.  
Larger windows again smooth the variations but result in a slightly larger offset than the data without self-calibration. In conclusion, there seem not to be strong benefits of combining self-calibration and sliding time windows for the Band~3 moderate scenario. The same is true and even more obvious for the Band~6 moderate scenario, which also remains with a larger offset although only a $\sim 20$\,K level. 
For the extreme scenario for Band~3, already a time window of 1-2\,s produces notably smoother time series both for individual pixels and for the spatially averaged maps. The offset from the reference model is larger on average but comparable to the basic imaging results for some pixels. In contrast, there is no notable gain from combining self-calibration with sliding time windows for the   extreme scenario for Band~3. 
Overall, additionally applying a time window with a width of 1-2\,s can improve the results for self-calibrated data but the improvements remain moderate in comparison to data that has been treated with either only sliding time windows or only with self-calibration. 

\begin{figure}[tp!]
    \centering
    \includegraphics[width=14.5cm]{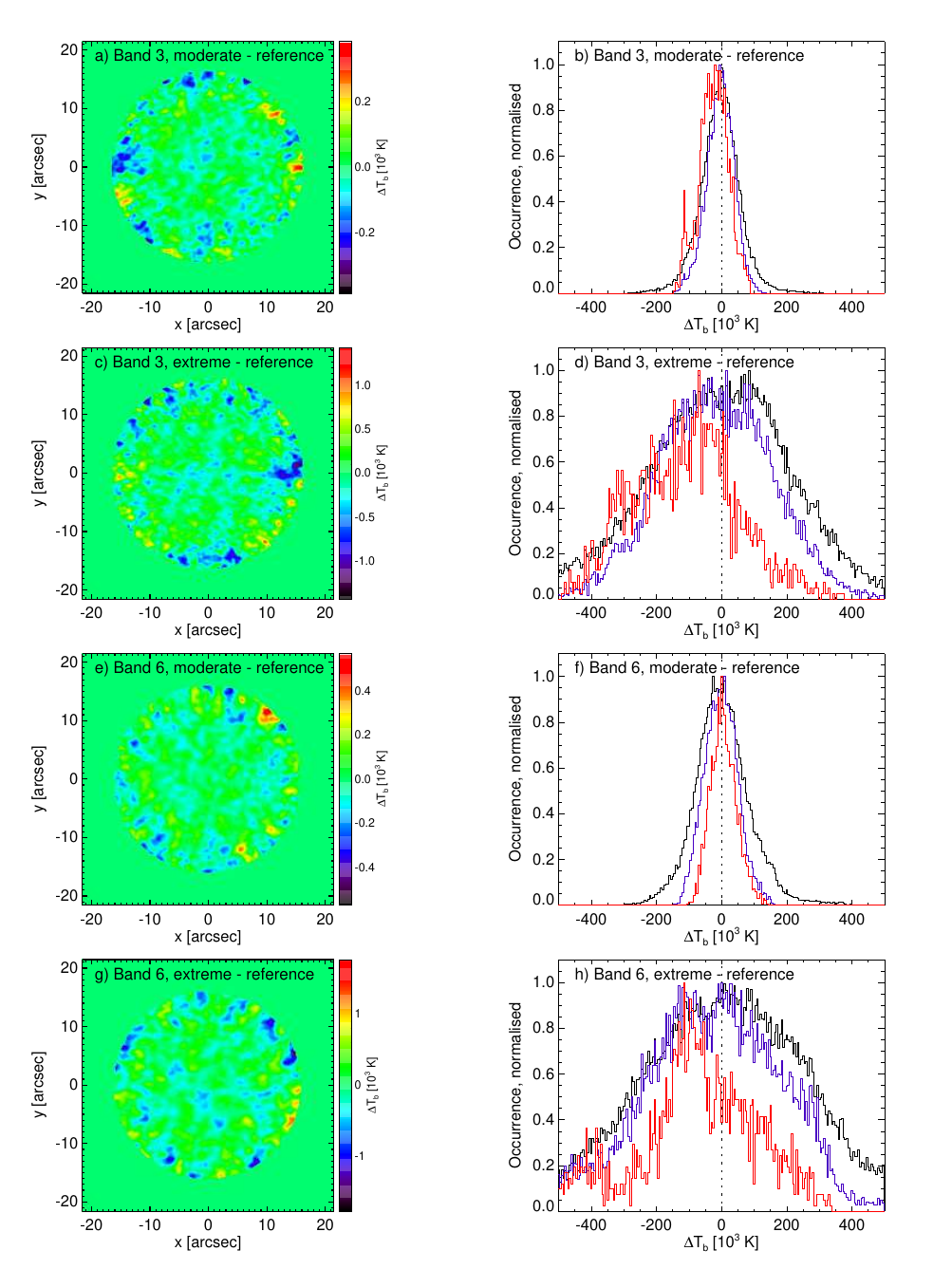}
    \vspace{-4mm}
    \caption{Brightness temperature differences for the first time step in the 0.1\,s cadence series as produced with the basic imaging mode of \tool{SoAP} (no sliding time windows, no self-calibration). 
    Each row shows the difference map (left) and the corresponding histograms (right) for the selected bands (3 and 6) and the two selected scenarios (moderate and extreme).   
    Each difference map was derived by subtracting the corresponding reference map (see Fig.~\ref{fig:Tbmaps_100ms_allcases}c-d). 
    Please note that the data range and the corresponding colour scales differ for each shown map. The histograms are calculated for the whole (circular) FOV (black), up to 50\,\% of the FOV radius (blue) and for 20\,\% of the radius~(red). 
    }
    \label{fig:Tbdiff_examples}    
\end{figure}

\begin{figure}[t!]
    \centering
    \includegraphics[width=14cm]{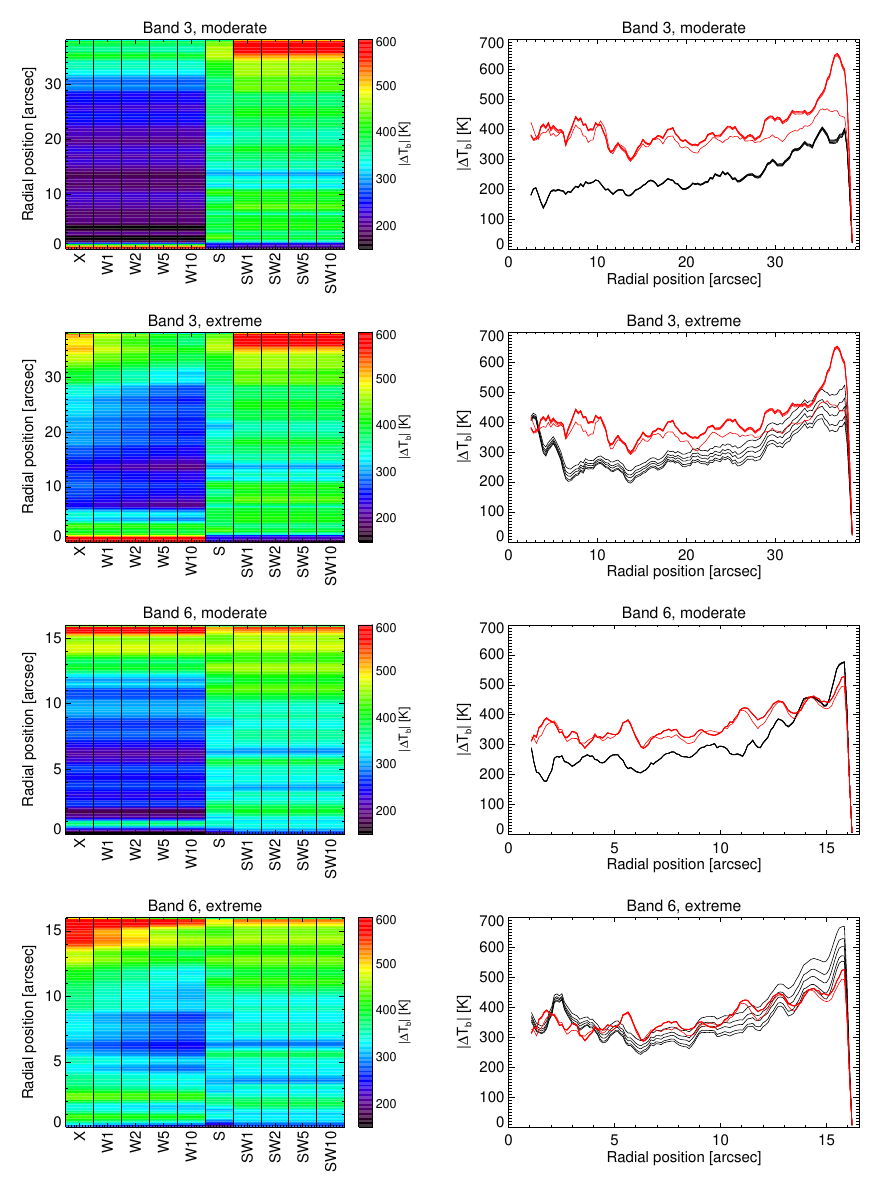}
    \caption{Absolute brightness temperature differences with respect to the reference model(s) for all considered imaging cases. Each row shows the different runs for Band~3 and 6 for the moderate and extreme scenario. In the right column, the radially and temporally averaged brightness temperature differences for the imaging runs without and with self-calibration are shown as black and red curves, respectively. As the curves are close to each other, the same data is shown as color-coded profiles in the left column with colours ranging from black ($\Delta T_\mathrm{b} = 150$\,K) to red ($\Delta T_\mathrm{b} = 600$\,K).}
    \label{fig:finalrad}
\end{figure}

\begin{figure}[t!]
    \centering
    \includegraphics[width=13.5cm]{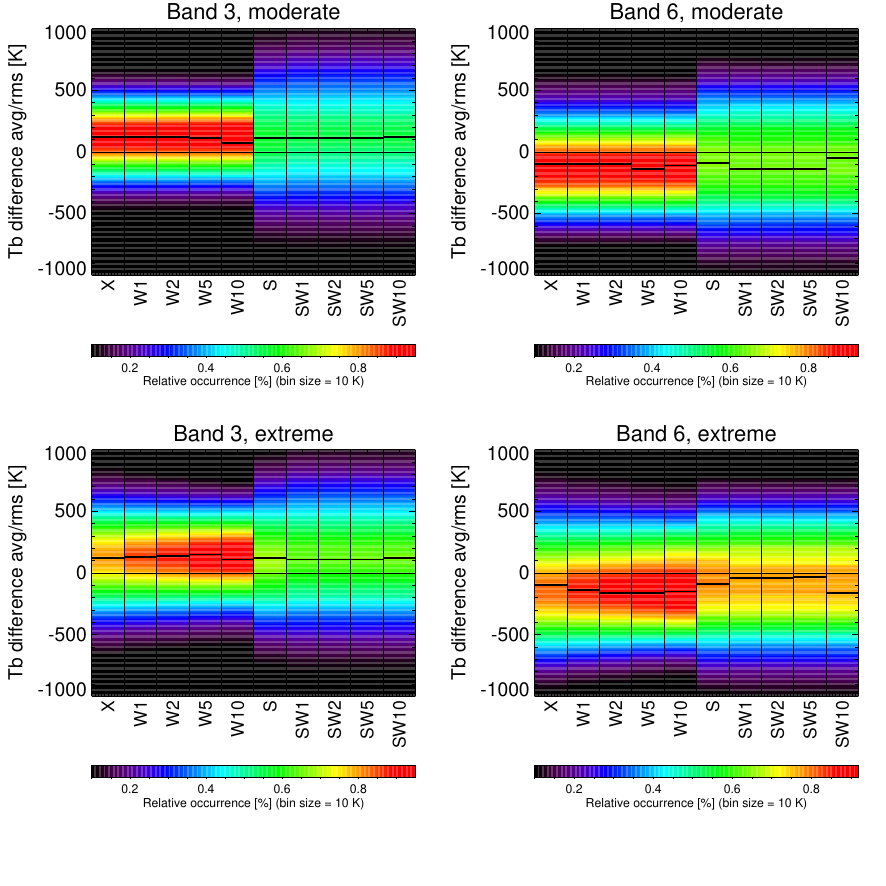}
  \vspace*{-9mm}
    \caption{Histograms for the brightness temperature difference of the individual imaging runs for Band~3 (left column) and Band 6 (right column) and the moderate (top) and extreme scenario (bottom) with respect to the corresponding reference model. Only pixels with a distance of up to 70\,\% of the FOV radius are considered. The first five columns in each panel show the data for different widths of the sliding time window. A window size of 0 means that this technique is not applied and that those brightness temperature maps are calculated with \tool{SoAP} frame by frame. The other five columns after equivalent but for the data with prior self-calibration (red). The histograms are scaled individually but presented on a common color scale for each panel. The data with prior self-calibration exhibit wider distributions and correspondingly lower values for the maximum relative occurrence of any Tb difference bin (notable as an apparent absence of red shades.)} 
    \label{fig:tbdiffhisto}
\end{figure}
\begin{figure}[t!]
    \centering
    \includegraphics[width=11cm]{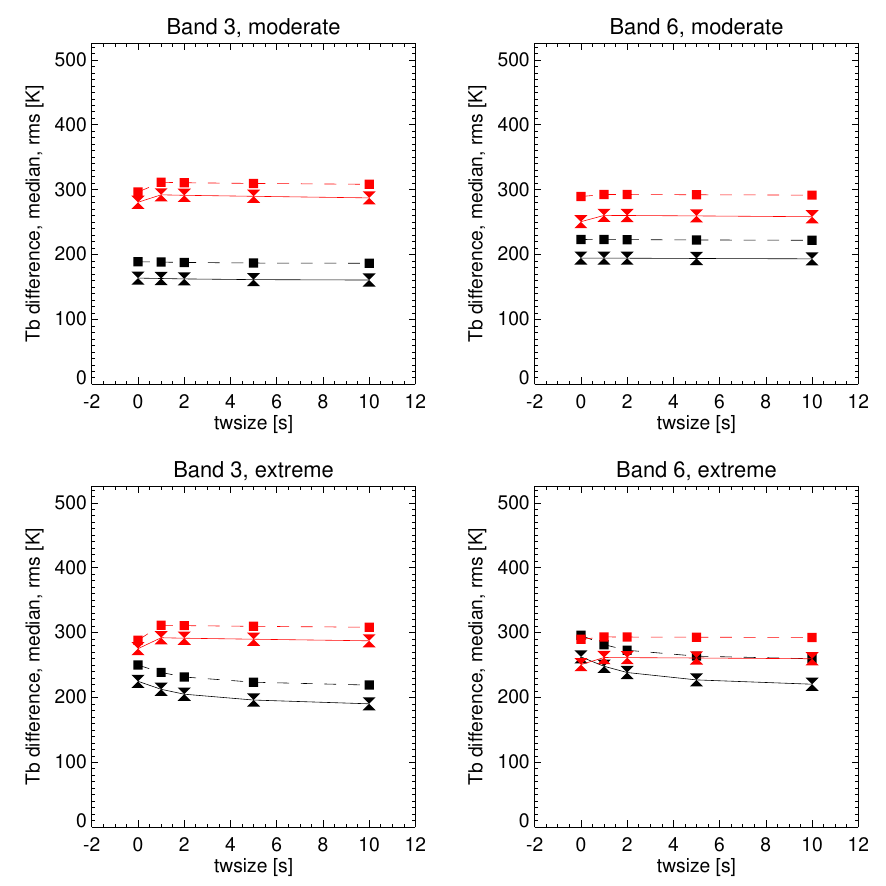}
    \caption{Data quality of the individual imaging runs for Band~3 (left column) and Band 6 (right column) and the moderate (top) and extreme scenario (bottom) without prior self-calibration (black) and with prior self-calibration (red). In each panel the average absolute brightness temperature with respect to the reference model (TDM, squares and dashed lines) and the its variation (TDV, double triangles and solid lines) are plotted as function of the size of the sliding time window. A window size of 0 means that this technique is not applied and that those brightness temperature maps are calculated with \tool{SoAP} frame by frame (basic imaging).    }
    \label{fig:finalquality}
\end{figure}

\begin{figure}[t!]
    \centering
    \includegraphics[width=11cm]{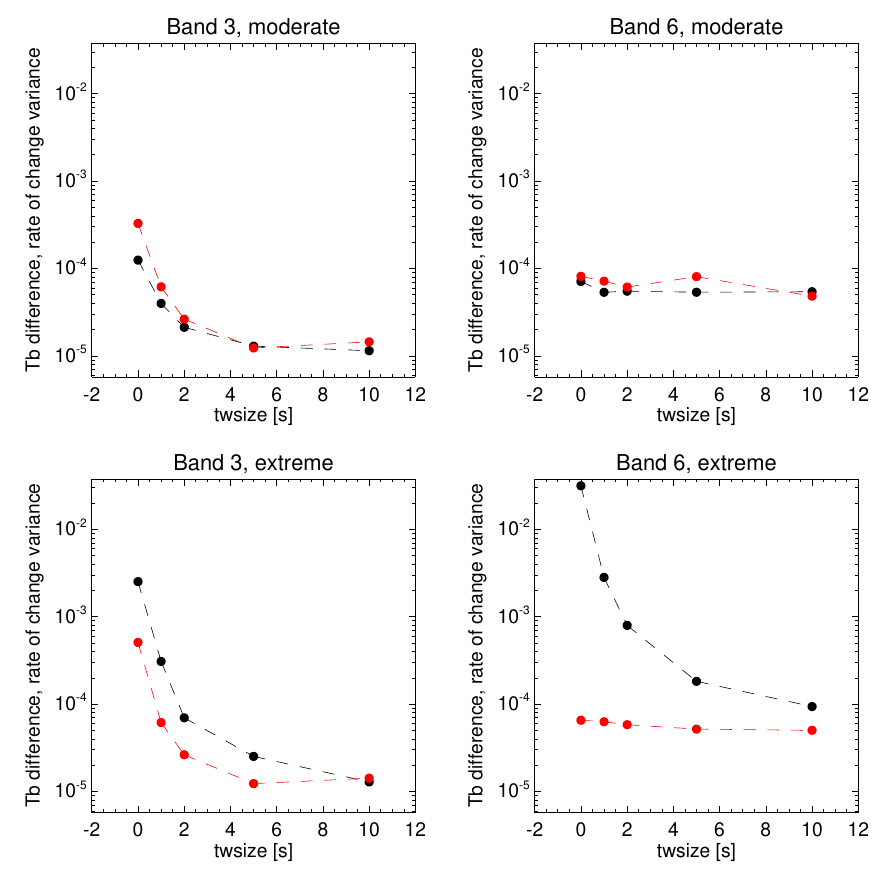}
    \vspace{-3mm}
    \caption{Variance of rate of change (VRoC) for the individual imaging runs for Band~3 (left column) and Band 6 (right column) and the moderate (top) and extreme scenario (bottom) without prior self-calibration (black) and with prior self-calibration (red).  A window size of 0 means that this technique is not applied and that those brightness temperature maps are calculated with \tool{SoAP} frame by fram (basic imaging).    }
    \label{fig:finalroc}
\end{figure}

\subsection{Comparison of the different approaches}
\label{sec:rescomp}

The quality of the imaging results is now quantified with respect to the corresponding reference model (see Sect.~\ref{sec:reference}). 
For each case shown in Table~\ref{tab:imagingcases} and each considered band and scenario, the difference between the corresponding brightness temperature maps ($T_\mathrm{b} (x,y,t)$) and the reference map  ($T_\mathrm{b,ref} (x,y,t)$) for all time steps is calculated: 
\begin{equation}
\Delta T_\mathrm{b} (x,y,t) = T_\mathrm{b} (x,y,t) - T_\mathrm{b,ref} (x,y,t)
\end{equation}
Only pixels within the circular mask are considered. 
Please see  Fig.~\ref{fig:Tbdiff_examples} for examples of the resulting difference maps as produced with basic imaging for both bands for the moderate and extreme scenario.  
The first time step of these time series is used for illustration. As expected, the moderate scenario for Band~3 shows the smallest brightness temperature differences and thus the smallest deviation from the reference case. In other words, the sky image of the solar target was reconstructed with reasonable accuracy. Except for a few positions the differences remain below 100\,K in the inner part of the field-of-view (see red histogram in Fig.~\ref{fig:Tbdiff_examples}b)  but increase to 200-300\,K at the outer parts of  the FOV. The other examples and all maps in general show a similar increase of the deviation with increasing distance from the centre of the FOV (i.e., the axis) as a result of the primary beam response and the resulting signal-to-noise ratio. 
The overall level of the brightness temperature differences, however, is notably higher for the moderate scenario for Band~6 as observations at the higher frequencies in this band are more severely impact by the tropospheric phase corruption as compared to Band~3. 
The deviations are typically on the order of up to a few 100\,K for the extreme scenario and can even exceed 500\,K in the outer parts of the FOV (see Fig.~\ref{fig:Tbdiff_examples}c-d).

\noindent\paragraph{Radial dependence} 
Due to the primary beam response and the resulting change of the signal-to-noise ratio, the deviation of the reconstructed brightness temperatures from the reference model changes with increasing radius outwards from the axis (centre of the FOV). 
This effect is clearly visible in the examples in  Fig.~\ref{fig:Tbdiff_examples} both in the maps and in the different histograms. The outer parts of the FOV clearly exhibits the highest deviations. In principle, the requirements of a given scientific application would determine if imaging with high accuracy is needed only for the innermost region of the FOV or rather for the whole FOV as considered here. 
 
The radial dependence of the deviations from the reference model(s) is therefore visualised in more detail for all considered imaging cases in Fig.~\ref{fig:finalrad}. 
In conclusion and as expected, the deviations are highest towards the edges of the considered FOV. On the other hand, the deviations remain at a smaller level for radii of $\sim 70$\,\% of the FOV radius. It is also notable that the remaining offsets discussed in the previous sections are lower for the sliding time window map series as compared to the sets with self-calibration.

\noindent\paragraph{Brightness temperature differences}
Histograms of all brightness temperature differences $\Delta T_\mathrm{b}$, i.e. the deviations from the corresponding reference model, are shown 
for all considered imaging cases in Fig.~\ref{fig:tbdiffhisto}. 
It is most obvious that the imaging runs with self-calibration produce much broader distributions than the basic imaging results and the runs with sliding time windows but no self-calibration. Also, as concluded in the previous sections, there is no obvious benefit from combining self-calibration with additional sliding time windows, at least in most situations. On the other hand, the extreme scenarios for Band~3 and 6 demonstrate that sliding time windows can result in a significant noise reduction and might thus have potential of salvaging otherwise highly problematic measurement sets.

\noindent\paragraph{Final comparison}
For each of the resulting data cubes of brightness temperature differences for the considered imaging cases, the average (TDM, see Eq.~\ref{eq:def_tdm}) and the standard variation (TDV, see Eq.~\ref{eq:def_tdv}) of the absolute differences are calculated. 
A pair of TDM and TDV values is therefore derived for each imaging case   and each considered band and scenario. 
The results, which are illustrated in Fig.~\ref {fig:finalquality}, confirm the conclusion of the previous sections that a sliding time window can reduce the deviations in  extreme situations but that some offsets with respects to the reference models and thus the true sky model remain. 
Overall, the median brightness temperature error in the reconstructed brightness temperature and also the offset are typically on the order of 200\,K for basic imaging and when applying sliding time windows but closer to 300\,K when applying self-calibration. 

However, brightness temperature differences as metric do  not fully capture the internal consistency of the reconstructed time series. For that reason, the rate of change (RoC) was calculated for the same data, again restricting the analysis to the inner 70\,\% of the brightness temperature maps. The corresponding variance of the RoC provides a good metric for the relative fluctuations of a time series: 
\begin{equation}
    \mathrm{VRoC} =  
  Var\left(\frac{|\Delta T_\mathrm{b}|(x,y,t_{i+1}) - |\Delta T_\mathrm{b}|(x,y,t_i)}{|\Delta T_\mathrm{b}|(x,y,t_i)} \right)
\end{equation}

The results are shown in Fig.~\ref{fig:finalroc} for all considered imaging cases. This metric reveals the benefits of applying sliding time windows for noise reduction. The moderate scenario for Band~6 again sticks out as it is less severe than the other cases and does not benefit from any further noise reduction.

\clearpage
\section{Conclusions and final recommendations} 

\subsection{Summary of the results}
Simulated time-dependent observations of the Sun with ALMA  in Band~3 and 6 for different weather (or ``seeing'') conditions, 
are used for testing different aspects of the processing of solar ALMA data.  The test data is produced with the Solar ALMA Simulator (\code{SASim}) and then processed with the Solar ALMA Pipeline (\code{SoAP}). Please refer to the \techdoc{}\citep{techdoc} for details of the tools. 
As a by-product of this study, it became clear that the level of precipitable water vapour (PWV) in the Earth's atmosphere at the ALMA site is not sufficient to describe the impact of Earth's atmosphere on the quality of ALMA data although the PWV level is commonly used as a primary quality indicator. For this study, the different tested seeing scenarios are instead defined based on the post-calibration phase variations that are seen in real (daytime) ALMA observations of the Sun.  

The focus is on the imaging parameters \param{niter}-\param{gain}-\param{robust} for the CLEAN algorithm as currently used for ALMA data but also the deconvolver algorithm and the imaging pixel grid size are investigated. The comparison of the resulting brightness temperature maps produced with \code{SoAP} with respect to reference models, which present perfect observations, is quantified in terms of quality indicators that are sensitive to the reproduction of the brightness temperature values on the one hand and to the reproduction of the spatial structure in the maps on the other hand. The optimization of these quality indicators then leads to the recommended best compromises for imaging parameters depending on receiver band and seeing scenario.  The resulting first recommendations for the processing of solar observations with ALMA are summarised in Sect.~\ref{sec:firstrecommend}.

\subsection{Final recommendations for optimal imaging of solar data}
\label{sec:firstrecommend}
\label{sec:recommend_imaging}

The following recommendations aim at the optimum choice of imaging parameters that result in the best overall reproduction of the true time-dependent properties of the solar source. For this purpose, the Unified Quality Indicator is used for which the results are described in more detail in Sect.~\ref{sec:res_uqi}. 
As a result of this study, we recommend the following approach for imaging of solar ALMA data: 
\begin{enumerate}
    \item Calibration of the measurement set (MS).
    \item Calculation of the Spatial Structure Function (SSF) for the calibrated MS.
    \item Determination of the median and standard deviation of the SSF. 
    \item Imaging with MultiScale-CLEAN with imaging parameters that can be chosen from Table~\ref{tab:finalrecomm_parameters} with respect to the SSF median and variation. 
\end{enumerate}

\begin{table}[h!]
\vspace{3mm}
    \centering
\begin{tabular}{|c|l|c|c|c|c|c|c|}
\hline
Band& Scenario& SSF med.&SSF rms&PWV$^*$&niter & gain & robust\\
    \hline
    \textbf{3}&
    Excellent-moderate &
    $\lesssim$~10$^\degree$& 
    $\lesssim$~2$^\degree$& 
    $\lesssim$~3\,mm& 
    25\,000 &
    0.20 &
     -0.5\\
    \cline{2-7}
    & 
    Challenging-extreme & 
    $\gtrsim$~10$^\degree$&  
    $\gtrsim$~2$^\degree$& 
    $\gtrsim$~3\,mm&
    10\,000 &
    0.20 &
    0.0\\
    \hline
    \multicolumn{8}{c}{}\\[-3mm]
       \hline
    \textbf{6}& 
    Excellent-moderate &
    $\lesssim$~10$^\degree$& 
    $\lesssim$~2$^\degree$& 
    $\lesssim$~3\,mm& 
    25\,000 &
    0.05&
    -2.0 -- -0.5\\
    \cline{2-8}
    & 
    Challenging-extreme & 
    $\gtrsim$~10$^\degree$&  
    $\gtrsim$~2$^\degree$& 
    $\gtrsim$~3\,mm&
    25\,000 &
    0.02 &
    0.0\\
    \hline
\end{tabular}
    \caption{Recommended imaging parameters with respect to the SSF median and variation.\\
    $^*$~Please note that we recommend not to use the PWV~value as sole criterion to judge the observing conditions but to rather at least review the SSF properties. }
    \label{tab:finalrecomm_parameters}
\end{table}

In addition, the following conclusions are drawn: 

\begin{itemize}
    \item In general, a higher \param{niter} value in combination with a lower \param{gain} is recommended for Band~6  as compared to Band~3. This finding is expected as Band~6 observations are more susceptible to weather conditions and a larger quality improvement can thus be gained from deeper CLEANing. In principle, Band~3 data can be handled with comparatively lower \param{niter} value and comparatively high \param{gain} value. 
    \item The recommendation to set the \param{robust} (sub)parameter not  to the default value of 0.5 is influenced by the aim of finding good compromises for parameter combinations that reproduce brightness temperature amplitudes on the one hand, and the spatial structure of the source image on the other hand. While these two different aspects are connected to opposite signs of the \param{robust} parameter, values around 0 are clearly best for an overall best imaging quality in most situations. 
    \item While the main recommendation is to process solar ALMA data with the determined parameter combinations for best overall imaging quality, the quality in terms of accurate  brightness temperature  values and spatial structure are analysed separately. It is therefore possible to rather choose the found parameter combinations that maximise the quality for either of those aspects if advantageous for the foreseen scientific application. The presented results for the TDR+ and SPR+ quality indicators might guide the appropriate parameter choice.
    \item The comparison of deconvolution algorithms revealed that MultiScale-CLEAN as used as default in the Solar ALMA Pipeline level~2 is an overall good choice. Yet, Multi-Term (Multi-Scale) Multi-Frequency synthesis performed well and should be investigated in more detail. To enable a fair comparison of the algorithms, comprehensive parameter grids should be explored for each algorithms. In that connection, alternatives such as the Maximum Entropy Method \citep[MEM,][]{mem} and Adaptive Scale Pixel \citep[ASP,][]{2004A&A...426..747B} deconvolution algorithm should be tested for the application to solar data but would likely require substantial development effort. New approaches based on machine learning might be suitable, too, but such a technical development would require substantial and dedicated effort.
    \item The investigation of different pixel sizes for the imaging grid revealed that the current choice (0.30''  for Band~3, 0.13'' for Band~6) is adequate and that there is no notable gain from smaller pixel sizes that justify the corresponding increase in data volume, which can be significant for solar ALMA data. 
    \item  While not discussed in detail in this report, the execution time for all \tool{SoAP} runs was measured. The overall execution time scales (mostly) linearly with \param{niter} and to a lesser extent with \param{gain} and \param{robust}. The parameter grid was calculated across a large number of nodes, most of them with AMD EPYC 7543 32-core processors (2.8\,GHz) with execution times of about 80-180\,s per time step. Clearly, a higher \param{niter} increases the execution time notably but the overall required time for a solar MS is still clearly feasible on a correspondingly large machine. Even high values for \param{niter} should not pose any problem given the very small number of solar observations carried out by ALMA. 
    \item Finally, the importance of high cadence  for the ALMA solar observing mode is stressed. The impact of highly time-dependent seeing conditions for day time observing of the Sun on the one hand and the scientific potential of high cadence make this aspect essential for solar observations. Strategies that lead to improved image quality at high cadence including self-calibration and the high-cadence mode (see Sect.~\ref{sec:finalrecomm_ultrahigh}) are therefore promising for a future increase in imaging quality for solar ALMA data and the resulting scientific applications. 
\end{itemize}

\subsection{Final recommendations for ultra-high-cadence observations}
\label{sec:finalrecomm_ultrahigh}

This study demonstrates that imaging of solar ALMA data is challenging due to the sparse $uv$-sampling in the required snapshot imaging approach but that exploiting the time domain can increase the overall quality of the resulting time series of brightness temperature maps.  
Introducing a sliding time window in the imaging process, as implemented in \tool{SoAP}, effectively reduces the noise already for short window sizes\footnote{Please note that these conclusions are based on the phase corruption according to the \function{settrop()} function. We recommend reviewing these conclusions once a more reliable phase corruption approach for solar observing conditions becomes available (see, e.g., Sects.~2.10-11 in the \techdoc).}.
Under extreme conditions, observations at sub-second cadence can give an extra advantage and might allow for producing scientifically useful data from measurement sets that otherwise would be challenging to process. 
As the approach resembles an implicit time-averaging, the choice of the optimal window length and the resulting data quality certainly depend on the exact priorities of the intended scientific application. While the approach is advantageous for most solar studies, only adequately short time windows should be chosen for science cases that focus on changes at extremely short time scales ($\lesssim 1$\,s). 

It must be emphasised that interferometric aperture synthesis observations of a highly dynamic and complex source like the Sun with a limited number of baselines can by nature never achieve a perfect reconstruction of the source image. The results of this study therefore reveal differences and variations in the reconstructed brightness temperature maps that remain for otherwise optimised imaging strategies. Based on the scenarios considered in this study, remaining brightness temperature errors on the order of 200\,K must be considered. 

As alternative to sliding time windows, applying self-calibration substantially suppresses fluctuations and thus produces much more reliable results, confirming the impression gained from processing real ALMA data with self-calibration. This study reveals, however, that notable deviations from the true source images, as represented by the reference models used here, remain and can even be larger than compared to the results of basic imaging. 
However, apart from the systematic offsets, the temporal evolution of the reconstructed brightness temperatures is well reproduced when using self-calibration. It should also be noted that the scans used in this study have only a duration of 60\,s while real data has scan durations of 600\,s, which may make a difference for self-calibration.   
The offsets might be due to the fact that the reference model is too idealised
\footnote{The reference model corresponds to a single-dish telescope with the same angular resolution as the synthetised beam of the simulated array but with perfect $uv$-coverage and no atmospheric or instrumental degradation. Please refer to  Sect.~2.6 in the \techdoc).}
and can never  be matched by data obtained with an interferometric array with a finite number of baselines. 
We recommend therefore to review in a future study if more suitable reference models can be constructed and/or if an algorithm can be devised that corrects for the brightness temperature offsets that might likely be due to the nature of observing a complex source like the Sun with an interferometric array. In conclusion, it is fair to say that the self-calibration technique and its current implementation for processing solar data have not yet been explored systematically so that the further development in this direction might lead to imaging with smaller remaining deviations from the true source. 

Combining self-calibration with a sliding time window  brings no substantial improvement in the cases considered in this study. 
This effect can be explained by self-calibration and sliding time windows being alternative ways of exploiting time domain information for noise reduction and their limitations for a highly dynamic source like the Sun. 

The model of the Sun used for this study is representative for a large part of the Sun as it includes an enhanced network structure with stronger magnetic fields and overarching coronal loops embedded in a Quiet Sun region. Consequently, the results and recommendations in this report apply to the majority of solar observations of typical targets (mostly Quiet Sun) that produce high-cadence time series. Other types of observations like observations of sunspots, at the solar limb, mosaics and single-dish mapping are not covered and would require a separate investigation.

Based on the cases studied here, it is recommended to further develop and test the techniques explored here.  
Moreover, although beyond the scope of this study, the inclusion of the Total-Power array in a full array combination imaging strategy might potentially help  to overcome the missing short baselines, which might be particular important for high-cadence imaging of the Sun with notoriously sparse $uv$-sampling.
It must also be noted that the imaging process, in particular the implementation of the CLEAN algorithm in itself is likely not ideal for solar data. It is therefore strongly recommended to evaluate and develop potentially more suitable methods, e.g. based on the Maximum Entropy Method (MEM).

\subsection{Concluding remarks} 

The presented results clearly show the potential improved imaging quality that can be reached by optimising the imaging parameters. The challenges of snapshot imaging of the Sun at high cadence with thus comparatively sparse sampling of the u-v space make the optimisation even more important. While the lack of a known reference makes this task difficult based on observational data alone, the here developed and employed usage of simulated measurement sets has large potential. 
Please note that the results may somewhat differ from the experience with processing real observational data. Improved imaging of the latter as e.g. measured with metrics such dynamic range (DNR) and signal-to-noise ratio (SNR) may indeed result in smoother looking images, which might thus be perceived as more reliable, but there is no way to judge how closely the real sky source is reproduced in those images. The strength of the approach used in this study is indeed that the imaging results can be systematically and quantitatively compared to a \textit{ground truth}, which is unavailable for real observations. 

The modelling approach can easily be applied to additional bands like Band 9 or 10, which are not yet offered for solar observations. It would require a small set of real observations for different seeing conditions for these bands in order to design realistic phase corruption scenarios, possibly with an updated phase corruption model, and running  large imaging parameter grids. However, seeing scenarios could even be designed without any actual observation sets if a more reliable phase corruption algorithm together with realistic atmospheric models for the ALMA site are developed first (see Sect.~2.10 in \techdoc). 

The potential of the here demonstrated modelling approach with the Solar ALMA Simulator goes even beyond the optimisation of imaging parameters. While this tool enables quantitative testing for development of alternative imaging approaches that are optimised for solar data, it can also be used to guide the selection and configuration of different instrumental set-ups like receiver bands and array configurations. The Solar ALMA Simulator  can thus provide valuable contributions to the future development of solar observations with ALMA. 

Finally, the results of this study can help to develop improved imaging strategies for solar ALMA data. As such, the results can potentially contribute to the future development of an official processing pipeline for solar ALMA observations. 
We recommend to extend the scriptForPIs.py so that it extracts the phase information of a calibrated measurement set and calculated the average and standard deviation of the resulting SSF. These parameters would allow a PI then to find corresponding cases as described in this report and tot choose the CLEAN parameters accordingly. This step could even be executed automatically as part of the scriptForPIs.py, possibly even generating a corresponding script for CLEANing.

\noindent\paragraph{Acknowledgments} 
This ALMA development study was supported as an ESO “Advanced Study for Upgrades of the Atacama Large Millimeter/submillimeter Array (ALMA)” (CFP/ESO/16/11115/OSZ). It also received support by the SolarALMA project, which has received funding from the European Research Council (ERC) under the European Union’s Horizon 2020 research and innovation programme (grant agreement No. 682462, until September 2021), and by the Research Council of Norway through its Centres of Excellence scheme, project number 262622. 
The study team acknowledges support from the Nordic ALMA Regional Centre (ARC) node based at Onsala Space Observatory. The Nordic ARC node is funded through Swedish Research Council grant No 2019-00208. 
The study team would like to thank C.~De~Breuck, N.~Phillips, E.~Villard, M.~Barta, D.~Petri, F.~Stoehr, L.~Maud, and the midterm and final study review panels for helpful discussions. 
\bibliographystyle{aa}
\bibliography{hcismemo}

\begin{thebibliography}{24}
\expandafter\ifx\csname natexlab\endcsname\relax\def\natexlab#1{#1}\fi

\bibitem[{{Bastian} {et~al.}(2018){Bastian}, {B{\'a}rta}, {Braj{\v s}a},
  {Chen}, {Pontieu}, {Gary}, {Fleishman}, {Hales}, {Iwai}, {Hudson}, {Kim},
  {Kobelski}, {Loukitcheva}, {Shimojo}, {Skoki{\'c}}, {Wedemeyer}, {White}, \&
  {Yan}}]{2018Msngr.171...25B}
{Bastian}, T.~S., {B{\'a}rta}, M., {Braj{\v s}a}, R., {et~al.} 2018, The
  Messenger, 171, 25

\bibitem[{{Bhatnagar} \& {Cornwell}(2004)}]{2004A&A...426..747B}
{Bhatnagar}, S. \& {Cornwell}, T.~J. 2004, \aap, 426, 747

\bibitem[{{Carlsson} {et~al.}(2016){Carlsson}, {Hansteen}, {Gudiksen},
  {Leenaarts}, \& {De Pontieu}}]{2016A&A...585A...4C}
{Carlsson}, M., {Hansteen}, V.~H., {Gudiksen}, B.~V., {Leenaarts}, J., \& {De
  Pontieu}, B. 2016, \aap, 585, A4

\bibitem[{{Clark}(1980)}]{clark}
{Clark}, B.~G. 1980, \aap, 89, 377

\bibitem[{Cornwell(2008)}]{multiscale}
Cornwell, T.~J. 2008, IEEE Journal of Selected Topics in Signal Processing, 2,
  793

\bibitem[{{Cornwell} \& {Evans}(1985)}]{mem}
{Cornwell}, T.~J. \& {Evans}, K.~F. 1985, \aap, 143, 77

\bibitem[{{Cornwell} {et~al.}(1993){Cornwell}, {Holdaway}, \&
  {Uson}}]{1993A&A...271..697C}
{Cornwell}, T.~J., {Holdaway}, M.~A., \& {Uson}, J.~M. 1993, \aap, 271, 697

\bibitem[{{De La Cruz Rodr{\'\i}guez} {et~al.}(2021){De La Cruz
  Rodr{\'\i}guez}, {Szydlarski}, \& {Wedemeyer}}]{2021zndo...4604825D}
{De La Cruz Rodr{\'\i}guez}, J., {Szydlarski}, M., \& {Wedemeyer}, S. 2021,
  {ART: Advanced (and fast!) Radiative Transfer code for Solar Physics.},
  Zenodo

\bibitem[{{Gudiksen} {et~al.}(2011){Gudiksen}, {Carlsson}, {Hansteen}, {Hayek},
  {Leenaarts}, \& {Mart{\'{\i}}nez-Sykora}}]{2011A&A...531A.154G}
{Gudiksen}, B.~V., {Carlsson}, M., {Hansteen}, V.~H., {et~al.} 2011, \aap, 531,
  A154+

\bibitem[{{H{\"o}gbom}(1974)}]{hogbom}
{H{\"o}gbom}, J.~A. 1974, \aaps, 15, 417

\bibitem[{{Ishizaki} \& {Sakamoto}(2005)}]{alma-memo-529}
{Ishizaki}, H. \& {Sakamoto}, S. 2005, ALMA Memo

\bibitem[{{Labeyrie}(1970)}]{1970A&A.....6...85L}
{Labeyrie}, A. 1970, \aap, 6, 85

\bibitem[{{McMullin} {et~al.}(2007){McMullin}, {Waters}, {Schiebel}, {Young},
  \& {Golap}}]{casa}
{McMullin}, J.~P., {Waters}, B., {Schiebel}, D., {Young}, W., \& {Golap}, K.
  2007, in Astronomical Society of the Pacific Conference Series, Vol. 376,
  Astronomical Data Analysis Software and Systems XVI, ed. R.~A. {Shaw},
  F.~{Hill}, \& D.~J. {Bell}, 127

\bibitem[{{Puschmann} \& {Beck}(2011)}]{2011A&A...533A..21P}
{Puschmann}, K.~G. \& {Beck}, C. 2011, \aap, 533, A21

\bibitem[{{Rau} \& {Cornwell}(2011{\natexlab{a}})}]{2011A&A...532A..71R}
{Rau}, U. \& {Cornwell}, T.~J. 2011{\natexlab{a}}, \aap, 532, A71

\bibitem[{{Rau} \& {Cornwell}(2011{\natexlab{b}})}]{mtmfs}
{Rau}, U. \& {Cornwell}, T.~J. 2011{\natexlab{b}}, \aap, 532, A71

\bibitem[{{Rau} {et~al.}(2019){Rau}, {Naik}, \& {Braun}}]{2019AJ....158....3R}
{Rau}, U., {Naik}, N., \& {Braun}, T. 2019, \aj, 158, 3

\bibitem[{{Shimojo} {et~al.}(2017){Shimojo}, {Bastian}, {Hales}, {White},
  {Iwai}, {Hills}, {Hirota}, {Phillips}, {Sawada}, {Yagoubov}, {Siringo},
  {Asayama}, {Sugimoto}, {Braj{\v s}a}, {Skoki{\'c}}, {B{\'a}rta}, {Kim}, {de
  Gregorio-Monsalvo}, {Corder}, {Hudson}, {Wedemeyer}, {Gary}, {De Pontieu},
  {Loukitcheva}, {Fleishman}, {Chen}, {Kobelski}, \&
  {Yan}}]{2017SoPh..292...87S}
{Shimojo}, M., {Bastian}, T.~S., {Hales}, A.~S., {et~al.} 2017, \solphys, 292,
  \#87

\bibitem[{{van Noort} {et~al.}(2005){van Noort}, {Rouppe van der Voort}, \&
  {L{\"o}fdahl}}]{2005SoPh..228..191V}
{van Noort}, M., {Rouppe van der Voort}, L., \& {L{\"o}fdahl}, M.~G. 2005,
  \solphys, 228, 191

\bibitem[{{Wedemeyer}(2023)}]{techdoc}
{Wedemeyer}, S. 2023

\bibitem[{{Wedemeyer} {et~al.}(2016){Wedemeyer}, {Bastian}, {Braj{\v s}a},
  {Hudson}, {Fleishman}, {Loukitcheva}, {Fleck}, {Kontar}, {De Pontieu},
  {Yagoubov}, {Tiwari}, {Soler}, {Black}, {Antolin}, {Scullion}, {Gun{\'a}r},
  {Labrosse}, {Ludwig}, {Benz}, {White}, {Hauschildt}, {Doyle}, {Nakariakov},
  {Ayres}, {Heinzel}, {Karlicky}, {Van Doorsselaere}, {Gary}, {Alissandrakis},
  {Nindos}, {Solanki}, {Rouppe van der Voort}, {Shimojo}, {Kato},
  {Zaqarashvili}, {Perez}, {Selhorst}, \& {Barta}}]{2016SSRv..200....1W}
{Wedemeyer}, S., {Bastian}, T., {Braj{\v s}a}, R., {et~al.} 2016, \ssr, 200, 1

\bibitem[{{Wedemeyer} {et~al.}(2022){Wedemeyer}, {Fleishman}, {de la Cruz
  Rodr{\'\i}guez}, {Gun{\'a}r}, {da Silva Santos}, {Antolin}, {Guevara
  G{\'o}mez}, {Szydlarski}, \& {Eklund}}]{2022FrASS...9.7878W}
{Wedemeyer}, S., {Fleishman}, G., {de la Cruz Rodr{\'\i}guez}, J., {et~al.}
  2022, Frontiers in Astronomy and Space Sciences, 9, 967878

\bibitem[{{White} {et~al.}(2017){White}, {Iwai}, {Phillips}, {Hills}, {Hirota},
  {Yagoubov}, {Siringo}, {Shimojo}, {Bastian}, {Hales}, {Sawada}, {Asayama},
  {Sugimoto}, {Marson}, {Kawasaki}, {Muller}, {Nakazato}, {Sugimoto}, {Braj{\v
  s}a}, {Skoki{\'c}}, {B{\'a}rta}, {Kim}, {Remijan}, {de Gregorio}, {Corder},
  {Hudson}, {Loukitcheva}, {Chen}, {De Pontieu}, {Fleishmann}, {Gary},
  {Kobelski}, {Wedemeyer}, \& {Yan}}]{2017SoPh..292...88W}
{White}, S.~M., {Iwai}, K., {Phillips}, N.~M., {et~al.} 2017, \solphys, 292,
  \#88

\bibitem[{{W{\"o}ger} {et~al.}(2008){W{\"o}ger}, {von der L{\"u}he}, \&
  {Reardon}}]{2008A&A...488..375W}
{W{\"o}ger}, F., {von der L{\"u}he}, O., \& {Reardon}, K. 2008, \aap, 488, 375

\end{thebibliography}
\end{document}